\documentclass{tesisUCH-jzv} \pdfoutput=1

\fancypagestyle{frontmatterstyle}{
\fancyhf{}
\pagenumbering{Roman}
\rfoot{\sffamily\thepage}

   \fancypagestyle{plain}{%
   \rfoot{\sffamily\thepage}
   }
}

\fancypagestyle{thesisfancy}{

\pagenumbering{arabic}
\lhead{{\normalsize\scshape\rightmark} } 
\chead{}
\rhead{\normalsize\scshape\leftmark}
\fancyfoot{} 
\rfoot{\sffamily\thepage}
\fancypagestyle{plain}{ 
	\fancyhf{} 
	\renewcommand{\chaptermark}{}

	\fancyfoot{} 
	\rhead{}
	}
}

\newcommand{\eref}[1]{(\ref{#1})}

\usepackage[english]{babel} 
\usepackage[T1]{fontenc}
\usepackage{fixltx2e} 
\usepackage{hyperref}

\begin{document} 

\baselineskip 18pt
\thispagestyle{empty}
\begin{wrapfigure}{l}{0.1\textwidth}
    \vspace{-0.72cm}
    \includegraphics[width=0.09\columnwidth]{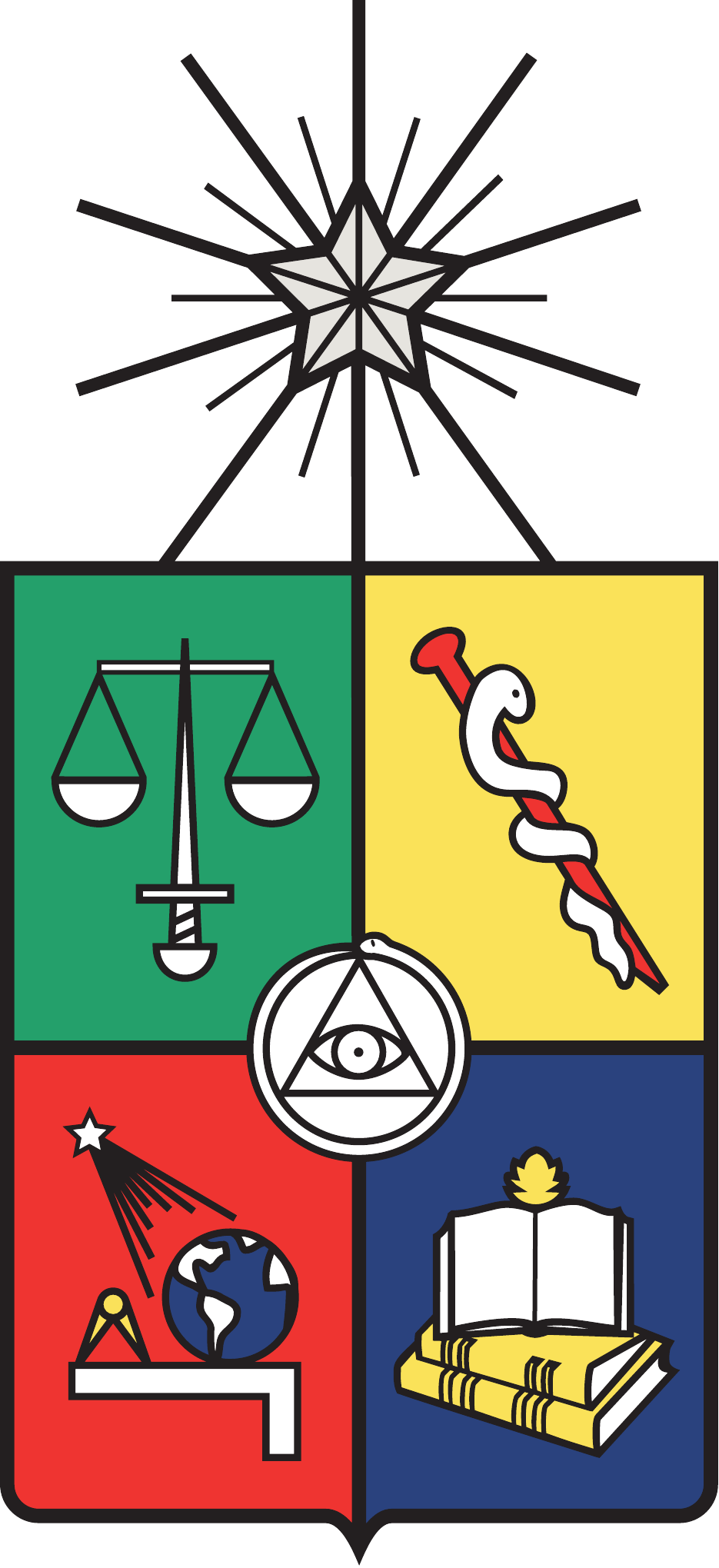} 
\end{wrapfigure}\hspace{1cm}
{\\
{\sc  Universidad de Chile\\
Facultad de Ciencias Físicas y Matemáticas\\
Departamento de Física} }
\vfill
{\centering
{\Large\bf Conductance in diffusive quasi-one-dimensional periodic waveguides: a semiclassical and random matrix study}
\vfill
{\sc TESIS PARA OPTAR AL GRADO DE DOCTOR CIENCIAS, MENCIÓN F\'ISICA}\\
\vspace{3cm}
{\bf\large Jaime Zuñiga Vukusich}\\
\vfill
{\sc Santiago de Chile\\
2011\\}
}
\newpage

\thispagestyle{empty}
\begin{wrapfigure}{l}{0.1\textwidth}
    \vspace{-0.72cm}
    \includegraphics[width=0.09\columnwidth]{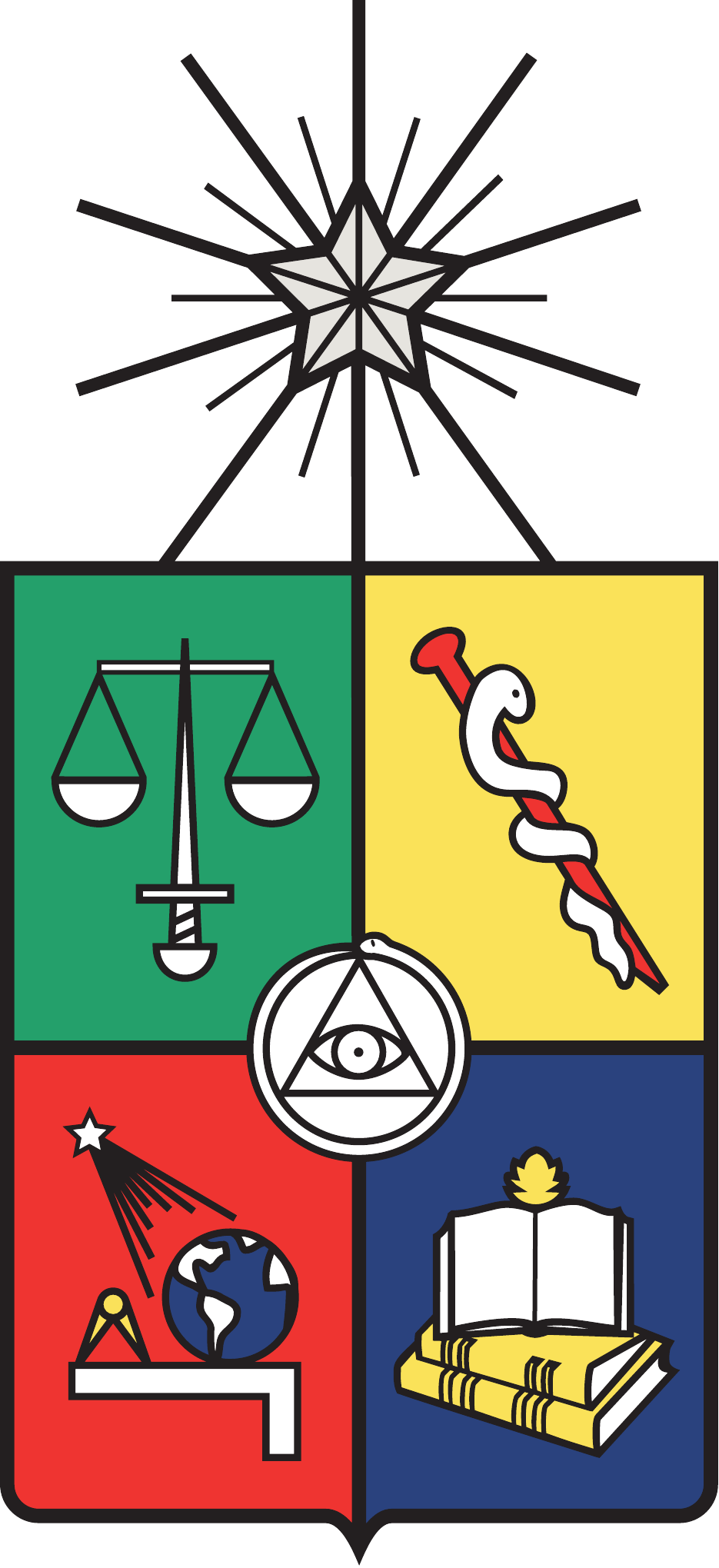} 
\end{wrapfigure}\hspace{1cm}
{\\
{\sc  Universidad de Chile\\
Facultad de Ciencias Físicas y Matemáticas\\
Departamento de Física} }
\vspace{2cm}\vfill
{\centering
{\Large\bf Conductance in diffusive quasi-one-dimensional periodic waveguides: a semiclassical and random matrix study}
\vspace{0.5cm}
\vfill
{\sc TESIS PARA OPTAR AL GRADO DE DOCTOR CIENCIAS, MENCIÓN F\'ISICA}\\
\vspace{2cm}
{\bf\large Jaime Miguel Zuñiga Vukusich}\\
\vspace{1cm}
{\sc PROFESOR GUÍA}\\
Felipe Barra de la Guarda\\
\vspace{1cm}
{\sc MIEMBROS DE LA COMISI\'ON}\\
Fernando Lund Plantat\\
Alvaro Nuñez Vásquez\\
Vincent Pagneux\\
Juan Carlos Retamal Abarzúa \\
\vfill
{\sc Santiago de Chile\\
Noviembre 2011\\}
}
\newpage

\pagestyle{frontmatterstyle}\setcounter{page}{2}
\baselineskip 14.5pt
\section*{Abstract}\addcontentsline{toc}{chapter}{Abstract/Resumen}

In this thesis we study quantum transport properties of finite periodic quasi-one-dimensional waveguides whose classical dynamics is diffusive. We focus in the semiclassical limit which enable us to employ a Random Matrix Theory (RMT) model to describe the system. The requirement of normal classical diffusive dynamics restricts the configuration of the unit cells to have finite horizon and the appropriate random matrix ensembles to be the Dyson circular ensembles. The system we consider is a scattering configuration, composed of a finite periodic chain of $L$ identical (classically chaotic and finite-horizon) unit cells, which is connected to semi-infinite plane leads at its extremes. Particles inside the cavity are free and only interact with the boundaries through elastic collisions; this means waves are described by the Helmholtz equation with Dirichlet boundary conditions on the waveguide walls. Therefore, there is no disorder in the system and all scattering is due to the geometry of the chain which is fixed. The equivalent to the disorder ensemble is an energy ensemble, defined over a classically small range but many mean level spacings wide. The number of propagative channels in the leads is $N$ and the semiclassical limit is achieved as $N\to\infty$. An important quantity for the transport properties of periodic chains is the number of propagating Bloch modes $N_B$ of the associated unfolded infinite periodic systems. It has been previously conjectured that for strongly diffusive systems in the semiclassical limit $\langle N_B\rangle \sim \sqrt{ND}$, where $D$ is the classical diffusion constant. We have checked numerically this result in a realistic cosine-shaped waveguide with excellent agreement. Then, by means of the Machta-Zwanzig approximation for $D$ we obtained the closed form expression $\langle N_B\rangle = \sqrt{N/\pi}$, which agrees perfectly with the circular ensembles. On the other hand, we have studied the (adimensional) Landauer conductance $g$ as a function of $L$ and $N$ in the cosine-shaped waveguide and by means of our RMT periodic chain model. We have found that $\langle g(L)\rangle$ exhibit two regimes. First, for chains of length $L\lesssim\sqrt{N}$ the dynamics is diffusive just like in the disordered wire in the metallic regime, where the typic ohmic scaling is observed with $\langle g(L)\rangle = N/(L+1)$. In this regime, the conductance distribution is a Gaussian with small variance (such that $\langle g^{-1} \rangle \approx 1/\langle g\rangle$) but which grows linearly with $L$. Then, in longer systems with $L\gg\sqrt{N}$, the periodic nature becomes relevant and the conductance reaches a constant asymptotic value $\langle g(L\to\infty)\rangle \sim \langle N_B\rangle$. In this case, the conductance distribution loses its Gaussian shape becoming a multimodal distribution due to the discrete integer values $N_B$ can take. The variance approaches a constant value $\sim\sqrt{N}$ as $L\to\infty$. Comparing the conductance using the unitary and orthogonal circular ensembles we observed that a weak localization effect is present in the two regimes. Finally, we study the non-propagating part of the conductance in the Bloch-ballistic regime, which is dominated by the mode with largest decay length $\ell$ which goes to zero as $g_{\rm np} = 4e^{-2L/\ell}$ as $L\to\infty$. Using our RMT model we obtained that under appropriate scaling the pdf $P(\ell)$ converge, as $N\to\infty$, to a limit distribution with an algebraic tail $\hat{P}(\ell)\sim\ell^{-3}$ for $\ell\to\infty$; this allowed us to conjecture the decay $\langle g_{\rm np}\rangle\sim L^{-2}$ which was observed in the cosine waveguide.

\newpage
\section*{Resumen}
En esta tesis estudiamos propiedades de transporte cuántico en guías de onda finitas periódicas quasi-unidimensionales, cuya dinámica clásica asociada es difusiva. Nos enfocamos en el límite semiclásico el cual nos permite emplear un modelo de \textit{Teoria de Matrices Aleatorias} (TMA) para describir el sistema. El requisito de difusión normal de la dinámica clásica restringe la configuración de la celda unitaria a tener horizonte finito, y significa que los ensembles apropiados de TMA son los ensembles circulares de Dyson. El  sistema que consideramos corresponde a una configuración de scattering, compuesto de una cadena finita de $L$ celdas unitarias (clásicamente caóticas y con horizonte finito) la cual esta conectada a dos guías planas semi-infinitas en sus extremos. Las partículas dentro de esta cavidad son libres y solo interactúan con los bordes a través de choques elásticos; esto significa que las ondas son descritas por una ecuación de Helmholtz con condiciones de borde tipo Dirichlet en las paredes la guía.  Por lo tanto, no hay desorden en el sistema y el scattering es debido a la geometría de la cadena la cual es estática. El análogo al ensemble de desorden es un ensemble de energía, definido sobre un intervalo clásicamente pequeño pero cuyo ancho es varias veces un \textit{espaciamiento de niveles promedio} (mean level spacing). El número de canales propagativos en las guías planas es $N$ y el límite semiclásico se alcanza cuando $N\to\infty$. Un número importante para las propiedades de transporte en cadenas periódicas es el número de modos de Bloch $N_B$ del sistema extendido infinito asociado. Previamente, ha sido conjeturado que en sistemas fuertemente difusivos en el límite semiclásico $\langle N_B\rangle\sim\sqrt{ND}$, donde $D$ es la constante de difusión clásica. Hemos comprobado numéricamente este resultado en una guía de ondas con forma de coseno obteniendo excelente concordancia. Luego, mediante la aproximación de Machta-Zwanzig para $D$ obtuvimos la expresión analítica $\langle N_B\rangle = \sqrt{N/\pi}$, la cual concuerda perfectamente con los ensembles circulares. Por otro lado, hemos estudiado la conductancia (adimensional) de Landauer $g$ como función de $L$ y $N$ en la guía-coseno y mediante nuestro modelo RMT para cadenas periódicas. Hemos encontrado que $\langle g(L)\rangle$ muestra dos regímenes. Primero, para cadenas de largo $L\lesssim\sqrt{N}$ la dinámica es difusiva tal como en un cable desordenado en el régimen metálico, donde se observa el escalamiento ohmnico típico con $\langle g(L)\rangle = N/(L+1)$. En este régimen, la distribución de conductancias es Gaussiana con una varianza pequeña (tal que $\langle g^{-1} \rangle \approx 1/\langle g\rangle$) pero que crece linealmente con $L$. Luego, para sistemas más largos con $L\gg\sqrt{N}$, su naturaleza periódica se hace relevante y la conductancia alcanza un valor asintótico constante $\langle g(L\to\infty)\rangle \sim \langle N_B\rangle$. En este caso, la distribución de la conductancia pierde su forma Gaussiana convirtiéndose en una distribución multimodal debido a los valores discretos (enteros) que $N_B$ puede tomar. La varianza alcanza un valor constante $\sim\sqrt{N}$ cuando $L\to\infty$. Comparando la conductancia para los ensembles circulares unitario y ortogonal, mostramos que un efecto de localización débil esta presente en ambos regímenes. Finalmente, estudiamos la parte no-propagativa de la conductancia en el régimen Bloch-balístico, la cual esta dominada por el modo con la longitud de decaimiento mayor $\ell$ que va a cero como $g_{\rm np} = 4e^{-2L/\ell}$ cuando $L\to\infty$. Usando nuestro modelo de TMA obtuvimos que bajo un escalamiento apropiado la pdf $P(\ell)$ converge, cuando $N\to\infty$, a una distribución límite con cola algebraica $\hat{P}(\ell)\sim\ell^{-3}$ para $\ell\to\infty$; esto nos permitió conjeturar el decaimiento $\langle g_{\rm np}\rangle\sim L^{-2}$, el cual fue observado en nuestra guía de ondas coseno.    

\section*{Agradecimientos}\addcontentsline{toc}{chapter}{Agradecimientos}
\vspace{1cm}
\begin{tabular*}{0.75\textwidth}{ ll }
  \hspace{2cm} &
  
\begin{minipage}{12cm}

Esta tesis no habría sido posible sin el apoyo de Felipe Barra, quien durante estos años siempre mostró ge\-nu\-ino interés en mi trabajo y estuvo permanentemente disponible como guía y colaborador. Además, quiero agradecer a Agnes Maurel y Vincent Pagneux por la excelente acogida que me brindaron durante mis días en París y por su importante aporte al desarrollo de este proyecto. 

Quiero dar también las gracias a mis amigos del pregrado que me acompañaron dando los apasionantes primeros pasos en Física, por las innumerables horas de estudio y debate que compartimos. 

Por último, quiero agradecer a mi familia por todo el soporte y cariño que me entregaron, en particular a Maricarmen por su compañía y aliento durante los momentos más difíciles.\\

Este proyecto contó con apoyo de una beca doctoral CONICYT.

\end{minipage}

\end{tabular*}

\tableofcontents
\listoffigures

\mainmatter 
\pagestyle{thesisfancy}
\renewcommand{\sectionmark}[1]{\markright{\S\thesection \ #1}{}}
\renewcommand{\chaptermark}[1]{\markboth{#1}{}}

\onehalfspacing

\chapter{Introduction}\label{chapter-introduction}

The study of wave transport phenomena is quite an old subject dating back at least to the days of Newton, d'Alembert, Euler, Laplace and Bernoulli among others, when the wave equation itself was originally derived. Its first applications in the 1700's were to mechanical (elastic) waves such as vibrating strings and membranes, sound, water waves and later to light. In 1862, Maxwell showed that electromagnetic radiation follows the same equation. In 1927, Schrodinger proposed his equation to describe non-relativistic particles, which is a generalized wave equation with quadratic dispersion relation, where the field takes complex values and the potential energy plays a role similar to the internal index of refraction in dielectric media. The ubiquity of the wave equation in so many different physical contexts can be explained by the fact that it can be derived from a Lagrangian variational principle, where the system consists of a continuum of coupled harmonic oscillators. Since generic systems perturbed near an equilibrium point can be described to first order by a harmonic oscillator, the wave equation must hold at least to first order in the field amplitude. 

With the introduction of electromagnetic resonant cavities, waveguides and optical fibers, the properties of spatially constricted waves gained attention. For instance, microwave resonators are closed metallic structures that confine electromagnetic waves with wavelength in the microwave part of the spectrum, this is from around one meter to one millimeter. Optical cavities, where light between the infrared and ultraviolet part of the spectrum is confined with high reflective mirrors, are key components of lasers. Elastic resonators and waveguides have also been studied. In the realm of condensed matter, technological advances in the last decades have opened the possibility to manufacture devices at the mesoscopic scale, usually composed of metallic or semiconductors two-dimensional layers at the nanometer scale. For instance, the study of electric conduction in thin and highly disordered metallic wires in this scale brought about new physical phenomena such as weak localization and universal conductance fluctuacions\cite{beenakker1997}, which are a consequence of the coherent propagation of waves in a disordered medium. Other notable examples are GaAs--AlGaAs heterojunctions, where a thin conducting layer is formed at the interface between GaAs and AlGaAs\cite{book_datta}, within which the electron's dynamic is well approximated by an effective mass description forming a two-dimensional degenerate electron gas (2DEG). What distinguishes a 2DEG in GaAs is its low scattering rate due to impurities, meaning that at temperatures low enough to suppress phonon scattering the electron wavefunction coherence can be maintained at scales larger than the device size and therefore its dynamic depends only on scattering due to boundary conditions in the confining two-dimensional layer, which can be designed at will in the laboratory. 

Spatially confined waves can be decomposed in a finite number of stationary modes which are associated to the solution of an eigenvalues problem where the operator is the time-independent wave equation. The shape of the cavity where the field is confined plays a major role in the statistical properties of this spectrum as well as in the solvability of the wave equation. In the small wavelength limit, waves are well described by geometric optics, this is by the propagation of \emph{rays} which follow the classical trajectory of free particles (with possible position dependent speed if the potential or refraction index is not constant). Thus, its behavior depends crucially on whether the underlying classical mechanics is integrable or chaotic. A dynamical system with $n$ degrees of freedom is integrable when it has $n$ constants of motion; in this case the action-angle variables exist and in principle an analytical solution in term of quadratures can be found. When an integrable system is separable\cite{book_gutzwiller}, the associated wave equation is also separable. Only systems with a high degree of symmetry are integrable; on the other hand, generic systems are non-integrable. Such systems are called \emph{chaotic} and are characterized by the exponential divergence of nearby initial conditions in phase space. Therefore, closed analytical solutions do not exist for the classical dynamics in chaotic systems and the wave equation is not separable. Besides non-separability, one of the signatures of chaos in the wave equation is found on its spectrum. In chaotic systems, the statistical properties of the energy spectrum are universal after appropriate system-dependent scalings. This was first proposed by Wigner when studying the energy levels of heavy nuclei and later Bohigas, Giannoni, and Schmit\cite{bohigas1984} conjectured the universality for general chaotic systems. Note, however, that the quantum (wave) dynamics in a classically chaotic system does not displays a property like exponential divergence of similar initial fields, because the wave equation (and Schrodinger equation) is linear in the field. Nevertheless, if one considers the evolution of an initial field under the action of two slightly perturbed propagators they diverge exponentially in the sense of expectation values\cite{prosen2002}. It is also possible to consider other characterizations of quantum chaos, for instance, in terms of the wave equation eigenstates\cite{rudnick1994, backer1998}. However, probably the most general and profound characterization of quantum chaos is the non-separability of the wave equation, which is shared with the Hamilton-Jacobi equation\cite{book_gutzwiller}, allowing a direct connection between the quantum and classical domains. 

\section{Quantum dots}

One of the most common systems where quantum chaos has been studied, both in the theoretical and experimental literature, is the quantum dot. Quantum dots are confining cavities on the nanometer scale, usually carved in semiconductive material, were waves are only scattered by the geometric features of its enclosing walls; electrons in the cavity form a 2DEG. The underlying classical dynamics corresponds to a \emph{billiard} where free non-interacting particles travel with constant velocity between elastic bounces against the quantum dot boundaries. Typically, the billiard dynamics is chaotic, so an ensemble of initial conditions will cover the full phase space volume after the ergodic time $\tau_e$. In order to perform transport measurements, the quantum dot is attached to two electron reservoirs called contacts. If this coupling is weak so that the residence time in the quantum dot $\tau_r \gg \tau_e$, the spectral statistical properties are insensitive to microscopic details (in particular, to the geometry of the cavity boundaries). By applying a small potential difference to the contacts it is possible to measure the quantum dot conductance $G$, which at low temperatures and in the absence of impurities can be calculated using the Landauer formula\cite{landauer1957, buttiker1985},
\begin{equation}\label{landauer-intro}
G = \frac{e^2}{h} \mbox{Tr}[\bm t\bm t^\dagger],
\end{equation}
where $e$ is the electron charge, $h$ is Plank's constant and $\bm t$ is the transmission matrix. Landauer conductance is not related to an intensive resistance as is usually defined for macroscopic bulk conductors because it does not arise from phonon or impurity scattering; instead, one must think of a mesoscopic conductor (in particular quantum dots) as a complete phase-coherent unit, where resistance and transmission are a consequence of wave scattering and interference effects. All measurable scattering and transport properties of a quantum dot are encoded in the scattering matrix. 

Quantum dots display features similar to disordered wires shorter than its Anderson localization length, in particular weak localization and universal conductance fluctuations. Weak localization, which is the enhancement of reflection probability in time-reversal-symmetric systems, was first considered in disordered wires by Abrahams \textit{et al.}\cite{anderson1979} and followed by extensive theoretical and experimental work\cite{bergmann1984}. Then, the universal conductance fluctuations (reproducible conductance fluctuations of order $e^2/h$ as a function of fermi energy or magnetic field) were theoretically predicted in disordered wires by Altshuler\cite{altshuler1985} and Stone\cite{stone1985}, and observed experimentally for the first time by Webb and Washburn\cite{webb1985} in metallic rings. Several years later, Baranger, Jalabert and Stone\cite{baranger1993} developed a semiclassical theory to predict these effects in quantum dots which were also observed experimentally\cite{marcus1992, bird1994}. 

\section{Semiclassical theory}

The first efforts to understand the effects of an underlying chaotic classical dynamics in the quantum level were focused on closed systems, in particular to the study of the energy spectrum's statistical properties. In this regard, Gutzwiller put forward his celebrated trace formula\cite{gutzwiller1971, book_gutzwiller}, a semiclassical expression for the density of energy states in terms of a sum over classical periodic orbits. Subsequently, Gutzwiller theory was generalized by Eckhardt \textit{et al.}\cite{eckhardt1992} to express arbitrary quantum observables matrix elements in terms of classical periodic orbits. On the other hand, the work of Berry and Tabor\cite{berry1977} and Bohigas, Giannoni and Schmit\cite{bohigas1984} focused on the universal character that many empirical and numerical evidence attributed to the spectrum of chaotic systems. They pointed out the fact that classically integrable and chaotic systems had quite different level statistics in the semiclassical limit; in particular they proposed that the nearest-neighbor level spacing distribution of integrable systems follows a Poisson law whereas chaotic systems show a higher degree of repulsion with a distribution similar to the one implied from ensembles of random hamiltonian with Gaussian elements. This is the foundation of modern Wigner-Dyson random matrix theory for Hamiltonians in chaotic systems and one of the better understood manifestations of quantum chaos. Recently, starting from the trace formula, it was proved that the Wigner-Dyson random matrix theory correctly describes the spectral statistics of quantum chaotic systems in the semiclassical limit\cite{haake2004, haake2007}, i.e. the Bohigas, Giannoni, Schmit conjecture is fulfilled.  

In closed chaotic systems there is only one dimensionless parameter relevant for its asymptotic semiclassical properties which can be taken as the wavelength in units of the system typical size, $\lambda/L$. In the limit $\lambda/L \to 0$ the universal semiclassical regime is reached and energy fluctuations are described by random matrix theory. Note that this is formally equivalent to the limit $\hbar\to0$ which is a common way to take the semiclassical limit. On the other hand, the semiclassical treatment of open chaotic systems is more complex due to the existence of an additional scale given by the mean residence time $\tau_r$, which in case of closed systems is infinite. The exponential divergence of nearby initial conditions in chaotic systems implies that the quantum evolution of an initial wave packet follows closely the classical dynamics up to the Ehrenfest time, $t_E \sim \lambda^{-1}\ln{S/\hbar}$ \cite{berry1979}, where $\lambda$ is the sum of positive Lyapunov exponents and $S$ is a characteristic classical action such as that of the shortest periodic orbit. Thus, a universal statistical description of open chaotic systems in terms of random matrix theory only holds for $t_E \ll \tau_r$ and a system-dependent regime is observed for $\tau_r \ll t_E$. Note that for quantum dots, the requirement of satisfying the limit $\tau_e\ll\tau_r$ means that they fall in the random matrix universal regime for fixed $\hbar$. A semiclassical theory for quantum scattering in open chaotic systems was first worked out by Baranger, Jalabert and Stone\cite{baranger1993} in terms of a semiclassical propagator, similar to which had been used before by Gutzwiller for the trace formula. They wrote the $\bm t$ matrix elements $t_{nm}$ for a two-terminal quantum dot as a sum over trajectories connecting mode $m$ on the left entry to mode $n$ on the right. This theory correctly predicts the weak localization and universal conductance fluctuations observed in quantum dots. A random matrix theory for transport was then introduced by Jalabert, Pichard and Beenakker\cite{jalabert1994} using the Dyson circular ensembles to model the quantum dot scattering matrix. 

\section{Diffusion in extended chaotic systems}

It is well known that in extended classical chaotic systems, under general assumptions, the dynamics relax by a diffusion process\cite{book_gaspard}. This means that an initially confined ensemble of initial conditions will spread as $\langle \bm x_t^2 \rangle \to D t$ for $t\to\infty$, where $\bm x_t$ is the position vector at time $t$, $D$ is the diffusion coefficient and the average $\langle \cdot\rangle$ is taken over the equilibrium measure. One of the earliest theories of transport is due to Drude, who tried to explain electron conduction in metals from a purely classical perspective. An equivalent, more contemporary formalism, is the Green-Kubo linear response theory\cite{book_dorfman}. Drude's model treats electrons as non-interacting (but electrically charged) particles traveling in a lattice of infinite mass spheres representing the ions; the only interaction are elastic collisions between particles and spheres. When a small electric field $\bm E$ is applied to the metal, the electric current is given by Ohm's law $\bm J = \sigma \bm E$, where the conductivity can be written as a function of the electron density of states at the Fermi energy, $\rho(E_F)$, by
\begin{equation}\label{einstein-relation}
\sigma = e^2 \rho(E_F) D. 
\end{equation}
This is known as the Einstein relation for degenerate conductors. The quantum property of electrons being distributed according to a Fermi-Dirac distribution is implicit in equation \eref{einstein-relation}; all other quantum effects are neglected. In this sense, \eref{einstein-relation} is a semiclassical result. Regardless of its simplicity, Drude's model provides a good approximation for the electric conductivity, Hall effect and thermal conductivity in metals at room temperature. 

The quantity which can be really measured experimentally is not the conductivity but the conductance $G=I/V$, with $I$ the electric current and $V$ the applied voltage. In a macroscopic homogenous two-dimensional conductor, Ohm's law implies that 
\begin{equation}
G= \frac{W\sigma}{L} ,
\end{equation}
where $W$ and $L$ are the conductor width and length, respectively. Obviously this scaling must break down when we go to the mesoscopic scale and quantum phase coherence starts playing a role. In particular, the conductivity loses its meaning for systems smaller than the phase-relaxation length, and in this case, the conductance is given by the Landauer formula \eref{landauer-intro} in terms of transmission probability. One of the most striking novel features observed in mesoscopic conductors is  Anderson localization in disordered wires, which consists in the exponential suppression of transmission (and hence of conductance) in a disordered wire longer than the localization length $\xi$. Notably, however, for wires of length in the range $\ell <L <\xi$, with $\ell$ the mean free path, the conductance displays ohmic scaling with $G \sim L^{-1}$; this is the metallic or \emph{quantum diffusive} regime, in which weak localization and universal conductance fluctuations are observed. Edwards and Thouless noticed that conductivity in disordered systems is closely related to the sensitivity of its eigenvalues on an external perturbation\cite{thouless1972} and argued that the metal-insulator transition could be described by a single parameter, namely the ratio between diffusion energy $E_c =  \hbar D/ L^2$ (which is the energy associated to the time it takes a classical particle to escape the systems by diffusion) and the mean level spacing $\Delta E$. In fact, the quantity $g_c = E_c/\Delta E$ has been shown to be the average conductance in disordered systems in the metallic regime\cite{simons1993}. For short systems in the semiclassical limit, the conductance starts being $g_c\gg 1$ and decays as $L^{-1}$ until $g_c \sim 1$ when Anderson localization kicks in. 

Localization is a purely quantum (wave) effect, caused by the destructive interference of multi-scattered waves inside disordered media. In the opposite extreme is the case of \emph{periodic} media, where there is other well known quantum result, namely Bloch theorem, which states that in infinite periodic systems the energy spectrum is organized in bands and eigenstates are ballistic, i.e. they travel through the system with constant velocity. This means that transport in extended periodic systems is ballistic even if the underlying classical dynamics is chaotic. The signatures of classical diffusion in band spectra where studied by Dittrich \textit{et al.}\cite{dittrich1997, dittrich1998} using the semiclassical trace formula and by Simons and Altshuler\cite{simons1993} in a more general context using the non-linear $\sigma$ model. Simons and Altshuler showed that the energy levels of chaotic and disordered systems subject to an external perturbation (which for periodic systems is identified with an Aharonov-Bohm flux) are universal up to two system dependent parameters. Later, Dittrich \textit{et al.} studied the two-point correlation function in finite periodic systems using semiclassical methods and found agreement with the results of Simons and Altshuler in the large system limit. 

In this thesis we are interested in the semiclassical transport properties of periodic systems possessing an underling chaotic or diffusive classical dynamics. We focus in two related quantities, the number of propagating Bloch modes $N_B$ and the (adimensional) Landauer conductance $g$. We show that in periodic chains, $\langle g\rangle \sim \langle N_B\rangle$ in the long-chain limit. An analytical expression for $\langle N_B\rangle$ was calculated by Faure\cite{faure2002} in the semiclassical limit. We test this result in a physically realistic cosine waveguide model, show its connection to random matrix theory and obtain an universal expression for $\langle N_B\rangle$ in a infinite chain of quantum dots with and without time-reversal symmetry, revealing a weak-localization-like effect in this quantity. On the other hand, we study the conductance properties of diffusive periodic systems employing a cosine shaped waveguide and a random matrix quantum dot chain model. We find that an ohmic regime is observed in such systems, but over a shorter length range than in disordered systems. The conductance can be separated in a Bloch-ballistic term, which remains non-null in the long system limit, and a decaying part related to closed modes. We show that the non-propagating part of the conductance decays with a power-law as $L^{-2}$. Finally, we consider the conductance distribution and fluctuations as a function of length and how the diffusive to Bloch-ballistic transition is signaled in them.

\section{Outline}

This thesis is organized as follows. In chapter \ref{chapter-waveguides} we review the scattering formalism for two-dimensional waveguides and Bloch theorem in such systems. By means of Oseledets theorem we show that the (adimensional) Landauer conductance is bounded above by $N_B$. Additionally, we show formally that the conductance is quasi-periodic as a function of length in the Bloch-ballistic regime. In chaper \ref{chapter-billiards}, we present relevant aspects of chaotic classical dynamics in billiards; in particular, we write an explicit result for the diffusion coefficient in the Machta-Zwanzig approximation for arbitrarily shaped quantum dots. The billiard model we use in the next chapters, namely the cosine billiard, is studied numerically. Chapter \ref{chapter-rmt} is a digression on random matrix theory and its applications to study chaotic systems. We define Dyson circular ensembles and present the quantum dot periodic chain model. In chapter \ref{chapter-nb} we review Faure's result for $\langle N_B\rangle$ and show numerical calculations done with the cosine waveguide. A connection with the universal parametric correlation function of Simons and Altshuler is presented. We also derive a parameterless expression for $\langle N_B\rangle$ in a quantum dots chain, and using this model we characterize some of the features of the $N_B$ probability distribution. The main results regarding transport properties are in chapter \ref{chapter-conductance-props}, where we study the conductance in the periodic cosine waveguide and random matrix models. First,  we focus on the Bloch-ballistic regime, and then on the ohmic regime in periodic diffusive systems. We also show the existence of a weak localization correction in both regimes and study the conductance fluctuations. Finally, in chapter \ref{chapter-conclusion}, we conclude and present future work prospects.

\chapter{Waveguides}\label{chapter-waveguides}

In this chapter we give an account of the quantum and classical dynamics of a particle inside a waveguide. The most general definition of a \emph{waveguide} is a structure able to confine and guide waves. They can be found in a number of contexts such as electromagnetism -- with different application on different parts of the spectrum-- and acoustics. More recently, mesoscopic waveguides have been constructed where electrons are constricted in coherent metallic (or semiconductor) rings and wires with  transversal section a few effective electron wavelengths wide. Depending on the underlying physical system, the wave equation and boundary conditions defining the waveguide mathematically can vary. We consider a two dimensional system where the wave dynamics is governed by the Helmholtz equation
\begin{equation}\label{helmholtz}
(\nabla^2  +k^2) \Psi = 0 \;,
\end{equation} 
for the field $\Psi=\Psi(x,y)$, which is equivalent to the Schrodinger equation for a free particle with energy $E=\hbar^2 k^2/2m$. The boundary conditions are Dirichlet $\Psi=0$ on the waveguide borders. This means that waves propagate coherently in the guide and are only scattered by the geometry of the boundaries. The associated classical dynamics to this Schrodinger equation is given by free non-interacting point particles colliding elastically with the waveguide walls; such systems are usually called billiards and are discussed in more detail in chapter \ref{chapter-billiards}. In this work we are interested in waveguides whose classical dynamics is strongly chaotic, so an initially bounded ensemble of particles will spread diffusively in the system such that the average square displacement  $\langle \Delta x^2 \rangle_t \sim t$ with time $t\to\infty$. For the wave equation (and Hamilton-Jacobi equation), classical chaotic dynamics translates into non-separability and hence, in general, the lack of analytic solution. Note that it is also possible to describe charged particles subject to a magnetic field with equation \eref{helmholtz} imposing minimal coupling, $\nabla \to \nabla - ieA/\hbar$ where $A$ is the vector potential and $e$ the particle charge. Since the wave equation we study is the Helmholtz equation, the results we obtain are also valid for classical waves.

In section \ref{section-s-matrix}, we review how to describe the scattering properties of a waveguide by means of the scattering and transfer matrices, and how to obtain its Landauer conductance from the transfer matrix spectrum using the polar decomposition. In addition, we define the Bloch spectrum of periodic waveguides using the scattering approach. Finally, in section \ref{section-landauer-conductance} we obtain an expression for the conductance of a long periodic waveguide using Oseledets theorem.

\section{Scattering and transfer matrices}
\subsection{Definition and properties}\label{section-s-matrix}

A natural way to describe the wavefunction $\Psi(x,y)$ in the waveguide is to project it on the local transverse basis, this is, writing the wavefunction as
\begin{equation}\label{LTM-general}
 \Psi(x,y) = \sum_{n=1}^\infty{\left(c_n^{+}(x) + c_n^{-}(x)\right)\, g_n(x,y)},
\end{equation}
where $g_n(x,y)$ are the local transverse modes which satisfy the boundary conditions on each $x$, and $c^{+}_n(x)$ ($c^{-}_n(x)$) is the right-going (left-going) longitudinal mode. In the particular case of a hard-wall waveguide, 
\begin{equation}
 g_n(x,y)= \sqrt{\frac{2}{h(x)}}\,\sin{\left(n\pi \frac{y-h_1(x)}{h_2(x)-h_1(x)}\right)} \;,\quad n=1,\ldots,\infty,
\end{equation}
where $h_1(x) < h_2(x)$ are the wall heights as a function of the longitudinal coordinate $x$. This set of functions satisfies the null boundary conditions $g_n(x,h_1(x)) = g_n(x,h_2(x))=0$ everywhere in the guide. The longitudinal modes are obtained by inserting (\ref{LTM-general}) in (\ref{helmholtz}), which transforms the original partial differential equation into a system of coupled ordinary differential equations which can be efficiently solved numerically. 

In a plane lead, $h_1(x)$ and $h_2(x)$ are constant, so (\ref{helmholtz}) is separable since $g_n(x,y)=g_n(y)$ is independent of $x$. In this region, the longitudinal modes are given by 
\begin{equation}
e^{\pm}_n(x) = \frac{e^{\pm i k_n x}}{\sqrt{k_n}},
\end{equation}
where $k_n^2 = k^2 - \left(n\pi/h_0\right)^2$ is the longitudinal wavenumber, $h_0=h_2-h_1$ is the lead width and the normalization is to impose unit flux. There are 
\begin{equation}\label{N-open-modes}
N = \left\lfloor \frac{h_0 k}{\pi} \right\rfloor
\end{equation}
propagating modes because for $n>N$ the longitudinal momentum $\hbar k_n$ is imaginary, implying null energy flux. These are called evanescent modes and decay exponentially with $x$. 
\begin{figure}[t]
 \centering
 \includegraphics[width=0.5\columnwidth]{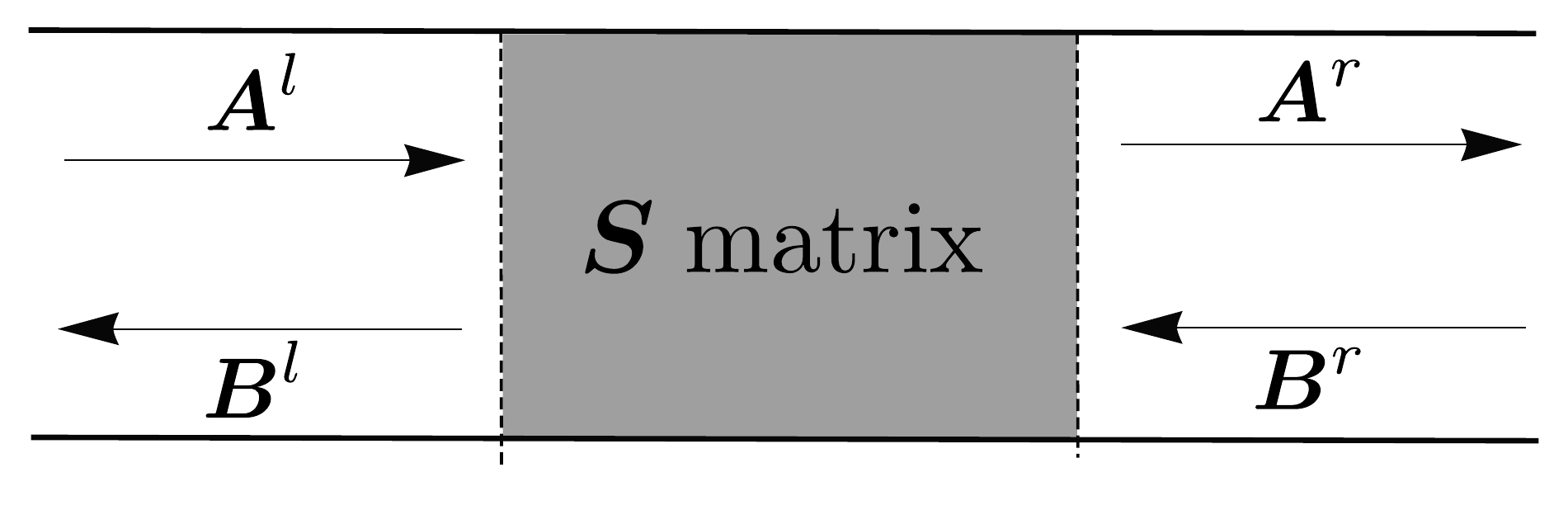}
 \caption[Diagram of the wavefunction decomposition in the plane leads]{Schematic diagram of the wavefunction decomposition in the plane leads. The gray area represents the scattering zone with scattering matrix $\bm S$. The far field is completelly represented by the vectors $\bm A^l$, $\bm B^l$, $\bm A^r$ and $\bm B^r$.}
 \label{fig-scat-scheme} 
\end{figure}

We will consider a scattering geometry [see figure \ref{fig-scat-scheme}], i.e. a waveguide composed of an interaction or scattering region connected to two semi-infinite plane leads at its extremes. The far field wavefunction in the leads can be described with a $2 N$ dimensional complex vector composed of coefficients $A_n$ and $B_n$ for $n=1,\ldots,N$ and we can write the field as
\begin{equation}\label{LTM}
 \Psi(x,y) = \sum_{n=1}^{N}{\left(A_n e^{+}_n(x) + B_n e^{-}_n(x)\right)\, g_n(x,y)} .
\end{equation}
Let $\bm A^{\rm r}$ ($\bm B^{\rm r}$) be the $N$ dimensional complex vector of right-going (left-going) amplitudes in the right lead and $\bm A^{\rm l}$ ($\bm B^{\rm l}$) the same on the left lead. We denote the incoming and outgoing (or incident and scattered) fields in vector notation as 
\begin{equation}
    \psi_{\rm in}=\left(\begin{array}{c}
   \bm A^{\rm l}\\
   \bm B^{\rm r}
 \end{array}\right) \quad \mbox{and}\quad
 \psi_{\rm out} =\left(\begin{array}{c}
   \bm B^{\rm l}\\
   \bm A^{\rm r}
 \end{array}\right).
\end{equation}
The scattering matrix $\bf S$ is defined as the linear transformation that maps incoming to outgoing fields,
\begin{equation}\label{s-matrix-def}
\psi_{\rm out} = \bm{S} \psi_{\rm in}.
\end{equation}
If $\bm{t}$ and $\bm{r}$ ($\bm{t}'$ and $\bm{r}'$) are the left-to-right (right-to-left) transmission and reflexion matrices, we have
\begin{equation}
 \label{S-system}
 \begin{array}{lll}
   \bm{r} \bm{A}^{\rm l} + \bm{t}' \bm{B}^{\rm r} &=& \bm{B}^{\rm l} \\
   \bm{t} \bm{A}^{\rm l} + \bm{r}' \bm{B}^{\rm r} &=& \bm{A}^{\rm r} ,
 \end{array}
\end{equation}
thus, we can write $\bm S$ as 
\begin{equation}
 \label{s-matrix}
   \bm S =\left(\begin{array}{cc}
   \bm{r}&  \bm{t}'\\
   \bm{t} &  \bm{r}'\\
 \end{array}\right)
\end{equation}
which is a $2N \times 2N$ dimensional matrix. The conservation of probability flux (or energy flux for classical waves) $\psi_{\rm in}^\dagger\, \psi_{\rm in} = \psi_{\rm out}^\dagger\, \psi_{\rm out}$ implies that $\bf{S}$ is a unitary matrix, i.e. 
\begin{equation}\label{s-unitarity}
\bm{S}^\dagger \bm{S} = \bm{1}, 
\end{equation}
with $\bm{A}^\dagger$ the hermitian conjugate of $\bm A$. On the other hand, if time reversal symmetry is not broken by a magnetic field the scattering matrix is symmetric,
\begin{equation}\label{s-symmetry}
\bm{S} = \bm{S}^T ,
\end{equation}
with $\bm{A}^T$ the transpose of $\bm A$.

The archetypal waveguide configuration we consider in this thesis will be a periodic chain of scatterers. Given the unit cell scattering matrix $\bm{S}_1$ and assuming the coupling of evanescent modes between cells in the chain is negligible, it is possible to obtain the $(n+1)$-cells chain $\bm S$ matrix by the usual Feynman paths sum rules\cite{book_datta},
\begin{equation}\label{t-r-comp}
\begin{split}
\bm{r}_{n+1} &= \bm{r}_n + \bm{t}'_n \bm{r}_1 (\bm{1} - \bm{r}'_n\bm{r}_1)^{-1} \bm{t}_n \\
\bm{t}_{n+1} &= \bm{t}_1  (\bm{1} - \bm{r}'_n\bm{r}_1)^{-1} \bm{t}_n\\
\bm{r}'_{n+1} &= \bm{r}'_1 + \bm{t}_1 \bm{r}'_n(\bm{1}-\bm{r}_1\bm{r}'_n)^{-1}\bm{r}'_1 \\
\bm{t}'_{n+1} &=\bm{t}'_n(\bm{1}-\bm{r}_1\bm{r}'_n)^{-1}\bm{t}'_1 ,
\end{split}
\end{equation}
where the subscripts denotes the number of cells in the chain. If the coupling of evanescent modes between cells cannot be neglected, transmission and reflexion matrices are in general not invertible and other methods to relate unit cell and chain scattering matrices is necessary [see appendix \ref{appendix-s-matrix-comp}]. 

We have seen that the $\bm S$ matrix links incident and scattered waves. A complementary picture is given by the transfer matrix which connects the wavefunction on the right and left leads,
\begin{equation}
\psi_{\rm right} =\left(\begin{array}{l}
   \bm A ^r\\
   \bm B^r
 \end{array}\right) \quad \mbox{and}\quad
\psi_{\rm left}  =\left(\begin{array}{l}
   \bm A ^l\\
   \bm B^l
 \end{array}\right).
\end{equation}
Then the transfer matrix $\bm M$ is defined by
\begin{equation}\label{m-matrix-def}
\psi_{\rm right} = \bm{M}\psi_{\rm left}.
\end{equation}
In order to obtain an explicit form for $\bm M$ we recast \eref{S-system} to the form
\begin{equation}\label{m-matrix-def-gen}
\bm{M}_{\rm l} \psi_{\rm right} =  \bm{M}_{\rm r} \psi_{\rm left}
\end{equation}
with
\begin{equation}
 \label{gen_transfer_matrix}
   \bm{M_r} =\left(\begin{array}{cc}
   \bm{t} & \bm{0}\\
   -\bm{r} & \bm{1}\\
 \end{array}\right)
\quad \mbox{and} \qquad
\bm{M_l} = \left(\begin{array}{cc}
   \bm{1} & -\bm{r}'\\
   \bm{0} & \bm{t}'\\
 \end{array}\right).
\end{equation}
Hence, the transfer matrix can be written in terms of transmission and reflexion matrices as
\begin{equation}\label{M(S)}
\bm{M} = \bm{M_l}^{-1} \bm{M_r} .
\end{equation}
From the unitarity of $\bm S$ \eref{s-unitarity} follows that the $\bm{M}$ matrix satisfies
\begin{equation}\label{M-identity}
\bm{M}\bm{\Sigma}\bm{M}^\dagger = \bm{\Sigma},
\end{equation}
where 
\begin{equation}
\bm{\Sigma} = \left(
\begin{array}{cc}
\bm{1}_N & \bm{0}_N\\
\bm{0}_N & -\bm{1}_N
\end{array}\right)
\end{equation}
with $\bm{1}_N$ and $\bm{0}_N$ the $N\times N$ identity and zero matrices. Thus, $\bm M$ is a pseudo-unitary matrix\footnote{ The pseudo-unitary group defined by a metric with a signature equivalent to $\bm{\Sigma}$ is usually denoted \mbox{U}(N,N).} with metric $\bm\Sigma$ whose induced norm $\psi^\dagger\bm{\Sigma}\psi$ can be interpreted as the probability flux of a field $\psi$ through the system. In addition, the scattering matrix symmetry \eref{s-symmetry} in time-reversal symmetric systems implies\cite{bachmann2010} a further symmetry of $\bm M$ given by
\begin{equation}\label{M-identity-2}
\bm{K}\bm{M}\bm{K}=\bm{M}^*,
\end{equation}
where  
\begin{equation}
\bm{K} = \left(
\begin{array}{cc}
\bm{0}_N & \bm{1}_N\\
\bm{1}_N & \bm{0}_N
\end{array}\right).
\end{equation}

The assumption of negligible coupling between adjacent cells required for \eref{t-r-comp} implies that it is possible  to truncate the unit cell reflexion and transmission matrices keeping only open channels without loosing significant accuracy. When this is the case we can relate the unit cell transfer matrix $\bm{M}_1$ to the length $L$ periodic chain transfer matrix $\bm{M}$ by 
  \begin{equation}
\bm{M}  = (\bm{M}_1)^L. 
\end{equation}
This is formally equivalent to the $\bm S$ matrix composition equations (\ref{t-r-comp}) but numerically unstable since $\bm M_1$ contains unbounded elements. In appendix \ref{appendix-s-matrix-comp} we show an alternative method to obtain the length $L$ chain $\bm S$ matrix using the unit cell $\bm M$ matrix spectrum. The numerical integration procedure we employ to obtain the $\bm S$ matrix is due to V. Pagneux and is discussed in detail in \cite{pagneux2010}.

\subsection{Bloch spectrum of periodic waveguides}\label{waveguide-bloch-spec}

Lets considered the unfolded periodic waveguide, i.e. the periodic scatterers chain of infinite length. As is well known, Bloch theorem states that in a system invariant to (discrete) translations in the $x$ direction Schrodinger equation solutions take the form $\phi_{n,q}(x,y) = e^{i q x}\, u_{n,q}(x,y)$, with $u_{n,q}(x,y)$ a periodic function of $x$ with a period equal to the translation length (or equivalently the potential periodicity or unit cell length). Moreover, the energy levels of the system form continuos bands parametrized by the quasi-momentum $q$ and the integer index $n$. One method to obtain the set of Bloch eigenfunctions and eigenvalues is solving the hamiltonian eigenvalue problem directly in the unit cell domain for arbitrary quasi-momentum $q$. This means imposing periodic boundary conditions such that 
\begin{equation}\label{bloch-condition}
\phi_{n,q}(x+Z, y) = e^{iqZ}\,\phi_{n,q}(x, y) \;\quad\mbox{for all $y$},
\end{equation}
with $Z=x_1 - x_0$ the unit cell length, $x_0 < x < x_1$ and $x_0$ and $x_1$ the unit cell borders. This produces a discrete number of solutions $E=E_n(q)$ and $\phi_{n,q}(x,y)$ labeled by $n$. On the other hand, we see from \eref{m-matrix-def} that we can impose Bloch condition \eref{bloch-condition} for a fixed energy $E$ using the unit cell transfer matrix $\bm{M}_1$ (calculated for that energy) in the eigenvalue problem
\begin{equation}\label{m-bloch-problem}
\bm{M}_1 \psi = \lambda \psi .
\end{equation}
The advantage of this scattering approach is that once we have computed the $\bm S$ matrix it only takes the solution of a matrix eigenvalue problem to obtain the Bloch basis. In addition, using \eref{m-matrix-def-gen} we can rewrite \eref{m-bloch-problem} as the generalized eigenvalue problem
\begin{equation}
\bm{M}_{\rm l} \psi = \lambda \bm{M}_{\rm r}  \psi ,
\end{equation}
which is numerically more robust since it avoids taking the inverse of $\bm{M}_{\rm l}$. 

Solving \eref{m-bloch-problem} produces $2N$ eigenvalues $\lambda_i = e^{iq_iZ}$ and associated eigenstates $\psi_i$, with $q_i$ in general complex numbers. Eigenvectors $\psi_i$ associated to eigenvalues $\lambda_i$ with unit modulus (or equivalently real quasi-momentum $q_i$)  correspond to propagating quasi-periodic Bloch-Floquet modes whereas states with eigenvalues such that $|\lambda_i|\neq1$ are closed channels in the unfolded infinite periodic system and have null flux $f_{\psi}=\psi^\dagger\bm{\Sigma}\psi$. Propagating Bloch modes are right-going or left-going if $f_\psi>0$ or $f_\psi<0$ respectively and have a group velocity given by
\begin{equation}\label{vh_def2}
v_{n}(\theta) = \frac{Z}{\hbar}\left.\frac{dE_n}{d\theta'}\right|_{\theta'=\theta}
\end{equation}
with $\theta = qZ$. Note that the velocity $v_n(\theta)$ is proportional to the energy band $E_n(\theta)$ slope at the intersections $E=E_n(\theta)$. Hence, since the bands must be $2\pi$-periodic, the number of right-going and left-going Bloch modes must be the same. 

From the transfer matrix pseudo-unitarity (\ref{M-identity}) follows\cite{horvat2007} that if $\psi$ is a right eigenvector of \eref{m-bloch-problem} with associated eigenvalue $\lambda$, then $(\bm{\Sigma} \psi)^\dagger$ is a left eigenvector with eigenvalue $1/\lambda^*$. This means closed channels Bloch quasi-momentum comes in pairs $(q, q^*)$, which correspond to states decaying in both chain directions. In addition, for time-reversal symmetric waveguides we have from \eref{M-identity-2} that if $\psi$, $\lambda$ is an solution of \eref{m-bloch-problem} then $\bm{K}\psi$, $\lambda^*$ is also a solution, so in this case the $\bm M$ matrix eigenvalues come in quadruples $(\lambda, \lambda^*, 1/\lambda, 1/\lambda^*)$. This means that open channel Bloch quasi-momentum comes in pairs $(q, -q)$, i.e. the bands are symmetric with respect to $q=0$ when time-reversal symmetry is preserved. 

In finite periodic waveguides, the Bloch basis is also important since transmission in a long chain will only be allowed for modes that are linear sums of the elements of this subspace $\{\psi_i \;:\; |\lambda_i|=1\}$. Since $\bm{M} = (\bm{M}_1)^L$ the length $L$ chain transfer matrix spectrum is given by $\{\lambda_i ^L\}_{i=1}^{N}$, hence closed Bloch modes will decay exponentially in the chain.  In section \ref{section-oseledet} we will see this in more detail.

\subsection{Polar decomposition}\label{section-polar-decomp}

The unitarity of the $\bm{S}$ matrix (\ref{s-matrix}) implies that the Hermitian matrices $\bm{t}\bm{t}^\dagger$, $\bm{t}'\bm{t}'^\dagger$, $1-\bm{r}\bm{r}^\dagger$ and $1-\bm{r}'\bm{r}'^\dagger$ have the same espectrum $\{T_i\}_{i=1}^N$, with all elements real and bounded in the $[0,1]$ interval. The scattering matrix can be written in terms of this set of eigenvalues by means of the polar decomposition\cite{mello1988},
\begin{equation}\label{s-matrix-polar}
\bm{S} = \left(\begin{array}{cc}
   \bm{U} & \bm{0}\\
   \bm{0} & \bm{V}\\
 \end{array}\right)
 \left(\begin{array}{cc}
   -\sqrt{\bm{1-\mathcal{T}}} & \sqrt{\bm{\mathcal{T}}}\\
   \sqrt{\bm{\mathcal{T}}} & \sqrt{\bm{1-\mathcal{T}}} \\
 \end{array}\right)
 \left(\begin{array}{cc}
   \bm{U}' & \bm{0}\\
   \bm{0} & \bm{V}'\\
 \end{array}\right),
\end{equation}
where $\bm U$, $\bm V$, $\bm{U}'$ and $\bm{V}'$ are unitary $N \times N$ matrices and $\bm{\mathcal{T}}$ is a diagonal matrix with elements $T_1, T_2, \ldots, T_{N}$. The transfer matrix can also be written using this decomposition, 
\begin{equation}\label{m-matrix-polar}
\bm{M} = \left(\begin{array}{cc}
   \bm{V} & \bm{0}\\
   \bm{0} & \bm{V}'^\dagger\\
 \end{array}\right)
  \left(\begin{array}{cc}
   \sqrt{\bm{\mathcal{T}}^{-1}} & \sqrt{\bm{\mathcal{T}}^{-1} - \bm{1}}\\
   \sqrt{\bm{\mathcal{T}}^{-1} - \bm{1}} & \sqrt{\bm{\mathcal{T}}^{-1}} \\
 \end{array}\right)
 \left(\begin{array}{cc}
   \bm{U}' & \bm{0}\\
   \bm{0} & \bm{U}^\dagger\\
 \end{array}\right).
\end{equation}
From  (\ref{M-identity}), we have that the spectrum of $\bm{M}\bm{M}^\dagger$ consists of inverse pairs $\{\Lambda_i, \Lambda_i^{-1}\}_{i=1}^{N}$, which are real and positive. From (\ref{s-matrix-polar}) and (\ref{m-matrix-polar}) one obtains 
\begin{equation}\label{M-T-relation}
[2\bm{I} + \bm{M}\bm{M}^\dagger + (\bm{M}\bm{M}^\dagger)^{-1}]^{-1} = \frac{1}{4}\left(\begin{array}{cc}
   \bm{t}\bm{t}^\dagger & 0\\
   0 & \bm{t}'\bm{t}'^\dagger\\
 \end{array}\right),
\end{equation}
which is a 2-to-1 relationship between the spectrum of $\bm{M}\bm{M}^\dagger$,  $\{\Lambda_i^{1}, \Lambda_i^{-1}\}_{i=1}^{N}$, and the spectrum of $\bm{t}\bm{t}^\dagger$, $\{ T_{i} \}_{i=1}^{N}$, given by
\begin{equation}\label{polar-rel}
T_i = \frac{4}{2 + \Lambda_i + \Lambda_i^{-1} } \,,\quad i=1,\ldots,N.
\end{equation}

\begin{figure}[h]
 \centering
 \includegraphics[width=0.75\columnwidth]{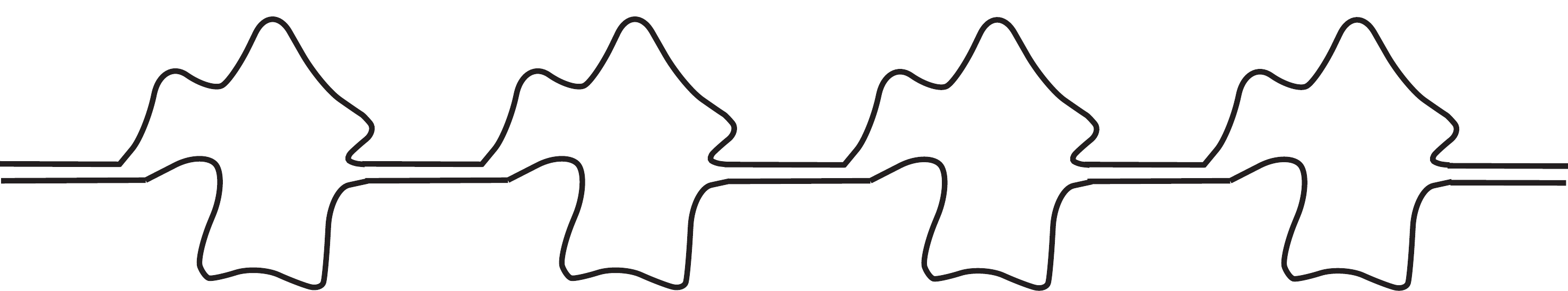}
 \caption[Example of a periodic chain]{Example of a periodic chain with four unit cells. We denote the unit cell length $Z$ and the leads width $h_0$.}
 \label{fig-waveguide} 
\end{figure}

\section[Asymptotic conductance]{Asymptotic conductance of a periodic {\nobreak waveguide}}

\subsection{Landauer conductance}\label{section-landauer-conductance}

Landauer scattering theory of electron conduction \cite{fisher1981,buttiker1985} provides a complete description of transport in mesoscopic devices at low temperatures, voltages and negligible electron-electron interaction. The typical system consists of a phase-coherent region where scattering is elastic (the periodic chain in our case) connected by ideal leads to two electron reservoirs at zero temperature and energies $E$ and $E+\delta E$, respectively. The small potential difference $\delta E \ll E$ generates an non-equilibrium net current between the reservoirs,
\begin{equation}
I = \frac{e}{\pi \hbar}\, g\, \delta E,
\end{equation}
where $e$ is the electron charge and $g=\mbox{tr} [\mathbf{t}\mathbf{t^\dagger}]$ is the $N$-channels transmission probability through the scattering zone. The conductance $G$ is defined by $I = G\, \delta E/e$, hence  
\begin{equation}\label{landauer}
G = \frac{e^2}{\pi\hbar}\;g =  \frac{e^2}{\pi\hbar}\; \mbox{tr} [\mathbf{t}\mathbf{t^\dagger}].
\end{equation}
Note that if transmission were perfect $\mbox{tr} [\mathbf{t}\mathbf{t^\dagger}] = N$, thus Landauer's conductance implies a non-null resistance $G^{-1}=\pi\hbar/(N e^2)$ even for reflectionless samples, known as contact resistance, which arise from non-elastic scattering when the discrete $N$ channels propagating in the leads couple to the reservoirs continuous spectrum and thermalize at the equilibrium potential. This is a purely quantum effect since it vanishes when $N\rightarrow\infty$. 

Conductance seen as transmission is not only relevant for conduction in mesoscopic degenerate electron gases but is an important quantity in other contexts when we are interested in understanding transport properties of waves, for example in electromagnetic and photonic devices, in molecular nanowires and in acoustic waveguides. In addition, is possible to define other transport quantities as a function of transmission eigenvalues $T_i$ such as shot noise power and thermopower \cite{buttiker1992, kumar1996, vanlangen1999}. For the sake of broadness and notation simplicity, in what follows we consider the adimensional conductance 
\begin{equation}\label{adim-g}
g = \mbox{tr} [\mathbf{t}\mathbf{t^\dagger}] = \sum_{i=1}^N T_i .
\end{equation}

\subsection{Oseledets theorem for transmission eigenvalues}\label{section-oseledet}

The $\bm{M}$ matrix spectrum of a periodic chain is trivially obtained from the unit cell Bloch spectrum since as we have seen $\lambda_i(L)=\lambda_i^L$. However, the conductance \eref{adim-g} --and in general any other transport property dependent of the transmission eigenvalues $T_i$-- is a function of $\bm{M}\bm{M}^\dagger$ eigenvalues as shown by \eref{polar-rel}. Thus, $g$ depends on the chain length $L$ in a non trivial way because of composition relations  (\ref{t-r-comp}). Nevertheless, it is possible to obtain an asymptotic approximation for $g$ from the unit cell transfer matrix and Oseledets theorem\cite{oseledets1968}. Let $\{\lambda_i\}_{i=1}^{2N}$ be the set of $\bm{M}_1$ eigenvalues, then Oseledets theorem implies that
\begin{equation}\label{oseledet-limit}
\lim_{L\rightarrow\infty} (\bm{M}_1^L\bm{M}_1^{L\dagger})^{\frac{1}{2L}} = 
\left(
\begin{array}{cccc}
|\lambda_1| 	&   	0		& \cdots	& 0\\
 	0		& |\lambda_2| 	&		& \vdots \\
  	\vdots	&  			& \ddots  	& 0 \\
	0		&	\cdots	&	0	& |\lambda_{2N}| 
\end{array}
\right),
\end{equation}
this is, $\Lambda_i(L)^{1/(2L)} \rightarrow |\lambda_i|$ where $\{\Lambda_i(L), \Lambda_i(L)^{-1}\}_{i=1}^{N}$ is the spectrum of $\bm{M}\bm{M}^\dagger = \bm{M}_1^L\bm{M}_1^{L\dagger}$. Hence, we have that
\begin{equation}\label{oseledets} 
\Lambda_i(L) \underset{L\rightarrow\infty}{\longrightarrow} a_i(L) e^{2L \log{|\lambda_i|}},
\end{equation}
where $a_i(L)$ is a positive and  (generically) bounded function of $L$. Then, using relation (\ref{polar-rel}), we can decompose $g$ in two terms, one with the sum of transmission modes related to the $2 N_B$ propagating Bloch modes $|\lambda_i|=1$ and another with the sum of modes related to evanescent Bloch states $|\lambda_i| \neq 1$, which have a decay length 
\begin{equation}\label{T-decay-length}
\ell = \left(\min_{|\lambda_i|>1}\{\log|\lambda_i|\}\right)^{-1}  
\end{equation}
determined by the slowest to decay non-propagating state. From  (\ref{oseledets}) and  (\ref{polar-rel}) we deduce that the transmission eigenvalues $T_i$ associated to Bloch modes is of order one, so for chains of length $L\gtrsim \ell$,
\begin{equation}\label{g_long_L}
g(L) \lesssim N_B + 4 a_{m}(L)^{-1}\, e^{-2L/\ell} .
\end{equation}
where $a_m$ is the Oseledet function $a_i$ associated to $\ell$. The equality in \eref{g_long_L} is non-generic and occurs if $\bm M_1$ is \textit{normal}, in which case $T_i(L)=1$ for all Bloch modes. In appendix \ref{appendix_perturbation}, a perturbative calculation for the Bloch transmission eigenvalues $T_i(L)$ is performed in the limit $\bm M_1$ is near a normal matrix. For the generic case, $g(L\to\infty) \sim N_B$ and only once we had defined the appropriate ensemble averages for $g$ and $N_B$ in chapter \ref{chapter-rmt} we will be in condition to establish an exact relation between this quantities. This is done in chapter \ref{chapter-conductance-props}.

\subsection{Conductance quasi-periodicity}\label{section-long-chain-conductance}

Not unexpectedly, \eref{g_long_L} contrasts with quasi-one-dimensional disordered systems where waves localize leading to zero conductance in a long wire. In a periodic chain, if the associated unfolded periodic system has a non-null number of propagating Bloch modes, the conductance do not decay to zero and instead show a quasi-periodic behavior in the big $L$ limit. We can see this as follows. From the transfer matrix definition  (\ref{gen_transfer_matrix}) follows that $\bm{t} = (\bm{X}_4)^{-1}$ where
\begin{equation}\label{transfer_matrix_decomp}
   \bm{M} =\left(\begin{array}{cc}
   \bm{X}_1 & \bm{X}_2\\
   \bm{X}_3 & \bm{X}_4\\
 \end{array}\right) = (\bm{M}_1)^L = \bm{P} \bm{D}^L \bm{P}^{-1}.
\end{equation}
The matrix $\bm{P}=[\bm{v}_i]_{i=1}^{2N}$ has the set of $\bm{M}_1$ right eigenvectors in its columns and $\bm{P}^{-1}=[\bm{u}_i^T]_{i=1}^{2N}$ the set of $\bm{M}_1$ left eigenvectors in its rows. Then, 
\begin{equation}\label{t-inv}
\bm{t}^{-1}  = \sum_{i=1}^{2N} \lambda_i^L \bm{\beta}_i \bm{\eta}_i^T
\end{equation}
where $\bm{\beta}_i$ and $\bm{\eta}_i$ are $N$ dimensional vectors composed of $\bm{v}_i$ and $\bm{u}_i$ second halfes respectively. We separate the sum in  (\ref{t-inv}) in two parts, one with the $2N_B$ unit modulus eigenvalues $\lambda_i$ and another with the rest, where in the long chain limit all terms with $|\lambda_i|<1$ can be neglected so $N-N_B$ elements survive. Then,
\begin{equation}\label{t-inv-a+ucv}
\bm{t}^{-1} = \bm{A} + \bm{U}\bm{G}^L\bm{V} ,
\end{equation}
 where 
\begin{eqnarray}
      \bm{A}=&\sum_{i \,:\, |\lambda_i|=1}{ \lambda_i^L\bm{\beta}_i\bm{\eta}_i^T}\\
      \bm{U}=&[\bm{\beta}_i]_{i \,:\, |\lambda_i|>1} \\
      \bm{V}=&[\bm{\eta}^T_i]_{i \,:\, |\lambda_i|>1} \\
      \bm{G}=&\mbox{diag}(\{\lambda_i\}_{i \,:\, |\lambda_i|>1}),
\end{eqnarray}
i.e. $\bm{A}$ is a $N \times N$ matrix, $\bm{U}$ is a $N \times (N-N_B)$ matrix with the vectors $\{\bm{\beta}_i \,:\,|\lambda_i|>1\}$ in its columns,  $\bm{V}$ is a $ (N-N_B) \times N$ matrix with the vectors $\{\bm{\eta}_i \,:\,|\lambda_i|>1\}$ in its rows, and $\bm{G}$ is a diagonal matrix with $N-N_B$ elements $\{\lambda_i\}_{i \,:\, |\lambda_i|>1}$. Note that $\bm A$ may not be invertible since, at best, it has rank $2N_B$ which can be less than $N$. To obtain $\bm{t}$ for $L\gg \ell$ using this decomposition, we invert  (\ref{t-inv-a+ucv}) using a generalized Woodbury Identity\cite{riedel1992}, which leads to
\begin{equation}\label{t_periodic_decomp}
\bm{t} = \bm{A}^{+} - \bm{A}^{+}\bm{U}_\parallel \bm{C}_1^\dagger  - \bm{C}_2 \bm{V}_\parallel^\dagger \bm{A}^{+} + \bm{C}_2\left( \bm{G}^{-L} + \bm{V}_\parallel^{\dagger} \bm{A}^{+}\bm{U}_\parallel \right)\bm{C}_1^\dagger,
\end{equation}
where $\bm{C}_1 = \bm{U}_\perp \left(\bm{U}_\perp^\dagger \bm{U}_\perp \right)^{-1}$, $\bm{C}_2 = \bm{V}_\perp \left(\bm{V}_\perp^\dagger \bm{V}_\perp \right)^{-1}$ and $\bm{A}^{+}$ is the Moore-Penrose generalized inverse of $\bm{A}$. For any matrix $\bm{Z}$, $\bm{Z}_\parallel = \bm{P}_{A} \bm{Z}$ and $\bm{Z}_\perp = (\bm{1}-\bm{P}_{A})\bm{Z}$ with $\bm{P}_A$ the projector to the column space of $\bm{A}$. Note that $\bm{G}^{-L}$ biggest element decay as $e^{-2L/\ell}$ and $\bm{U}, \bm{V}$ are independent of $L$. Finally, we have that for long chains $g=\mbox{tr}[\bm{t}\bm{t}^\dagger]$ depends only on $L$ quasi-periodically through the propagating Bloch modes in $\bm{A}^{+}$ given by $\bm{M}_1$ eigendecomposition. 

The quasi-periodicity of $g(L)$ can be understood as a consequence of the coupling between Bloch modes in the periodic chain and the plane leads propagating channels. The amplitude of this coupling, i.e. of the transmission between periodic chain and plain leads, oscillates as a function of $L$ because of the $\bm t$ matrix resonances due to interior reflection.

\chapter{Classical dynamics in chaotic billiards}\label{chapter-billiards}

The classical analog of the waveguide system presented in chapter \ref{chapter-waveguides} is an open billiard. An open (closed) billiard is a bidimensional system of non-interacting point particles enclosed in a infinite (finite) domain delimited by hard elastic walls. In open billiards, one is usually interested in the transport properties (relaxation) of an initially spatially confined ensemble of particles or in the scattering properties of an isolated interaction region. Of course, both problems are connected in the sense that transport properties can be thought to be a result of local scattering phenomena\cite{book_lax}. In chapter \ref{chapter-waveguides}, we have defined the conductance of a periodic chain --which is a measure of transport probability--  as a function of the scattering matrix of a single cell. This scattering approach to transport is also natural to experimentalists because, in the laboratory, finite pieces of materials are proved and then its properties extrapolated to the bulk.

The classical dynamics of a billiard, depending on the shape of its boundaries, can be integrable, chaotic or mixed. Billiards have been extensively studied in the context of classical chaos --both in the physical and mathematical literature--  because their simplicity has allowed several analytical results for particular billiards such as the Lorentz gas (Sinai billiard) and the Bunimovich stadium where, for instance, hyperbolicity, ergodicity and the central limit theorem have been proved. 

In this chapter we define a billiard in more detail and give a brief account of properties relevant for this thesis. In section \ref{section-billiard-dyn} we define the billiard collision map and present the concepts of chaos and hyperbolicity. Then, in section \ref{section-billiard-stat-props}, some statistical properties of deterministic chaotic systems are discussed. A good treatment of these topics can be found in \cite{book_gaspard}. We describe the Machta-Zwanzig approximation in section \ref{section-MZ-approx}, which will be used in chapter \ref{chapter-rmt} to obtain an analytical expression for the number of propagating Bloch modes in a random matrix periodic chain model. Finally, the cosine billiard is defined; this is the system employed in our numerical calculations of the conductance properties discussed in chapter \ref{chapter-conductance-props}.

\section{Billiard dynamics}\label{section-billiard-dyn}

\subsection{Billard flow and collision map}

\begin{figure}[h]
 \centering
 \includegraphics[width=0.65\columnwidth]{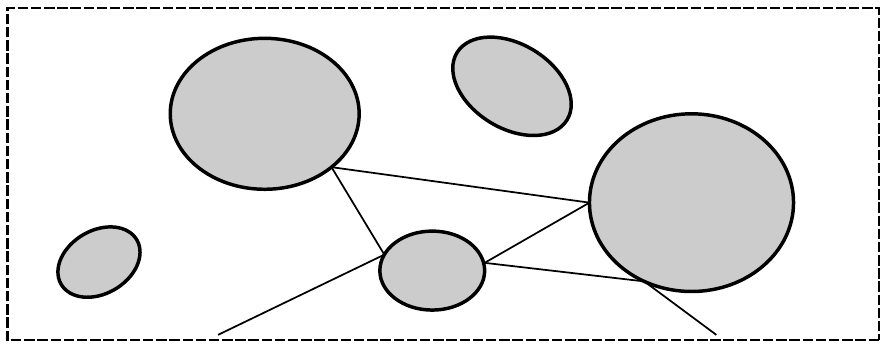}
 \caption[Generic billiard with convex obstacles]{Generic billiard with convex obstacles (gray objects) and a particle trajectory. If all the dashed lines were part of the boundary $\partial D_0$ this would be a closed billiard. Otherwise, taking periodic boundary conditions on the dashed lines we could construct a periodic billiard with this unit cell.}
 \label{fig-billiard-example} 
\end{figure}

A billiard is a hamiltonian system of non-interacting free particles bouncing between obstacles where they collide elastically [see figure \ref{fig-billiard-example}]. We consider bidimensional billiards only; their hamiltonian is simply $H=(p_x^2 + p_y^2)/2m$ so particles travel with constant speed. Without loss of generality we can set the particle speeds and mass to one and rescale later if necessary. Let $\mathcal{D}\subseteq \mathbb{R}^2$ be the position space where the particles can move. The billiard phase space is given by $\mathcal{M} = \mathcal{D} \otimes S^1$ where $S^1$ is the unit circle which represents all possible directions of motion. The continuos time evolution in the billiard consists of linear segments of free flights in $\mathcal{D}$ and specular reflection at the boundaries $\partial\mathcal{D}$. We can denote a phase space point in the billiard as $\bm X = (\bm q, \bm{\hat{v}})$ with $\bm q = (x,\,y)\in \mathcal{D}$ and $\bm{\hat{v}} \in S^1$. Formally, the dynamical system is defined by the system of equations
\begin{equation}\label{dyn-system}
\bm{\dot{X}} = \bm F(\bm X),
\end{equation}
where 
\begin{equation}\label{hamiltonian-field}
\bm F(\bm X) = \bm\Sigma \cdot \partial_{\bm X} H 
\end{equation}
with $\bm \Sigma$ the symplectic fundamental matrix. Equation \eref{dyn-system} induces a continuous time evolution or flow denoted by 
\begin{equation}
\bm X_t = \bm \Phi^t \bm X_0,
\end{equation}
where $\bm X_0$ and $\bm X_t$ are the phase space coordinates at time zero and $t$ respectively. 

\begin{figure}[h]
 \centering
 \includegraphics[width=0.65\columnwidth]{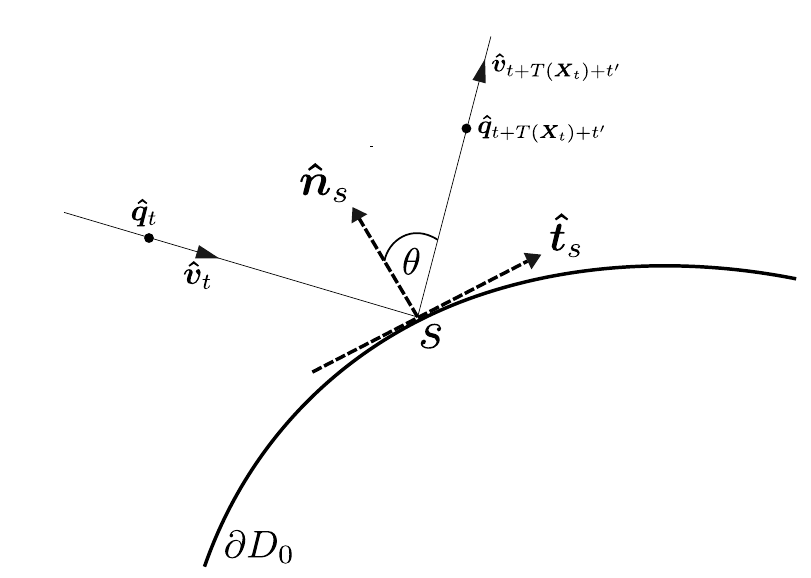}
 \caption[Collision map scheme]{Schematics of the collision map over a convex scatterer. A particle initially at $\bm X_t = (\bm{\hat{q}}_t,\bm{\hat{v}}_t)$ collides with an obstacle boundary $\partial D_0$ at $s=\bm{\hat{q}}_{t+T(\bm X_t)}$. The collision occurs at time $t+T(\bm X_0)$ and change the particle velocity direction according to \eref{billiard-flow} and \eref{billiard-map}. }
 \label{fig-collision-map} 
\end{figure}

Let $T(\bm X)$ be the time of first collision of a phase space point $\bm X$, i.e. the time $\bm X$ takes to reach the first intersection with $\partial \mathcal{D}$. Given an initial condition $\bm X_{t_0} = (\bm q_0,\, \bm{\hat{v}}_0)$ we can iteratively define the billiard flow as
\begin{equation}\label{billiard-flow}
\begin{split}
\bm\Phi^t \bm X_{t_0}  &= \bm\Phi^{t - T(\bm X_{t_0})} \bm X_{t_1}  \\
\bm X_{t_1} & = \left(\bm q_{t_1},\; \bm{\hat{v}}_0 -2(\bm{\hat{n}}_{\bm q_{t_1}} \cdot \bm{\hat{v}}_0)\bm{\hat{n}}_{\bm q_{t_1}} \right ) \\
\bm q_{t_1} &= \bm q_0 + \bm{\hat{v}}_0 \,T(\bm X_{t_0}) \\
t_1 &= t_0 + T(\bm X_{t_0}),
\end{split}
\end{equation}
where $\bm{\hat{n}}_{\bm q} \in S^1$ is the vector normal to the billiard boundary at $\bm q \in \partial \mathcal{D}$. The piecewise continuos nature of the billiard flow induces naturally the billiard map, which is the discrete time version of $\bm\Phi$ that iterates particles from collision to collision, i.e. is the Poincare map for the surface of section given by the billiard boundary. We define the billiard map $\bm\phi$ for elements $\bm\chi = (s,\,\sin\theta) \in \partial\mathcal{D} \otimes [-1,1]$ where $\sin\theta$ is the tangential momentum at the collision point $\bm r = \bm r(s)$ with $s$ the boundary curve natural parametrization; these are called Birkhoff coordinates. The map,
\begin{equation}\label{billiard-map}
\begin{split}
\bm\chi_{n+1} &= \bm\phi \bm\chi_n  \\
t_{n+1} &= t_n + T(\bm \chi_n),
\end{split}
\end{equation}
is defined using \eref{billiard-flow} taking $\bm X_{t_0} = (\bm r_n(s),\, \cos{\theta}\,\bm{\hat{n}}_{\bm r_n} + \sin{\theta}\,\bm{\hat{t}}_{\bm r_n})$ with $\bm{\hat{t}}_{\bm r}$ the tangencial boundary vector $\bm{\hat{n}}_{\bm r} \cdot \bm{\hat{t}}_{\bm r} = 0$. Note that there is ambiguity in the choice of $\bm{\hat{t}}_{\bm r}$ direction; for instance, we can define it such that $\bm{\hat{t}}_{\bm r} \cdot \bm{\hat{v}} > 0 $ for $\theta>0$. See figure \ref{fig-collision-map}.

For open periodic billiards, since the dynamic flow and map can be reduced to the unit cell with periodic boundary conditions plus an integer value to label its position, the phase space can be decomposed as $\mathcal{M} = \mathcal{D}_0  \otimes S^1 \otimes \mathbb{Z}^d$ where $\mathcal{D}_0$ is the unit cell fundamental domain and $d=1$ ($d=2$) for a chain (bidimensional lattice) configuration. \\

\subsection{Linear stability and chaos}

The most common characterization of chaos is \emph{sensibility to initial conditions}, which means that, given two initially infinitesimally close initial conditions, the local dynamical map will separate them exponentially. This is characterized by the Lyapunov exponents defined as follows. Let $\bm X_0$ and $\bm Y_0 = \bm X_0 + \delta\bm X_0$ be two initially close phase space point, i.e. $\| \delta\bm X_0 \| \ll 1$ for some norm $\|\cdot\|$, then
\begin{equation}
\delta \bm X_t = \partial_{\bm{X}} \bm\Phi^t(\bm X)|_{\bm X = \bm X_0} \cdot \delta \bm X_0 + \mathcal{O}(\delta\bm X_0^2)
\end{equation}
is the separation at time $t$, $\delta\bm X_t = \Phi^t\bm Y_0 - \Phi^t\bm X_0$. The Lyapunov exponent at $\bm X_0$ for direction $\bm{\hat{e}} = \delta \bm X_0 / \| \delta \bm X_0  \| $  is given by
\begin{equation}\label{lyapunov-exp}
\lambda(\bm X_0, \bm{\hat{e}}) = \lim_{t\to\infty}\, \lim_{\|\delta\bm X_0\| \to0}\, \frac{1}{t} \ln{ \frac{\|\delta \bm X_t\| }{ \| \delta \bm X_0  \|}  },
\end{equation}
hence $\| \delta \bm X_t \| \sim e^{\lambda t} \|\delta \bm X_0\|$. In general dynamical systems, according to Oseledets theorem, \eref{lyapunov-exp} takes values from a discrete set $\lambda_1(\bm X_0) > \lambda_2(\bm X_0) > \cdots >\lambda_r(\bm X_0)$ called the Lyapunov spectrum, whose multiplicities $m_1, m_2, \ldots, m_r$ sum up to the dimension of the phase space $\mathcal{M}$. Since Hamiltonian systems define symplectic systems [see \eref{hamiltonian-field}], it can be shown that the fundamental matrix $\partial_{\bm{X}} \bm\Phi^t(\bm X)$ is in fact symplectic, therefore its eigenvalues comes in pairs $\pm\lambda$. Hence, in a general Hamiltonian system with $d$ degrees of freedom there are $d$ independent pairs of Lyapunov exponents $\pm \lambda_i$ with $i=1,\ldots,d$. From this follows that the sum of all Lyapunov exponents is null, a property satisfied by any conservative system (not necessary Hamiltonian). Additionally, it can be easily seen that the Lyapunov exponent parallel to the flow direction in phase space is null since local divergence in this direction can not be exponential; the Lyapunov exponent paired with the latter corresponds to the direction orthogonal to the energy surface $H=E$ (this can be generalized to any constant of motion present in the system). These restrictions leave $d-1$ not null independent Lyapunov exponents pairs. Thus, in a bidimensional billiard there is only one pair $\pm \lambda(\bm X)$ of non-null exponents. Basically, a billiard is called chaotic when $\lambda(\bm X)>0$ for all $\bm X \in \mathcal{M}$. In case this holds in a subset of $\mathcal{M}$ the dynamics is called mixed and otherwise is integrable. The latter occurs only for highly symmetrical billiards, the mixed case being the more common for an arbitrarily chosen billiard shape. 

The union of all local tangent spaces in phase space is called tangent (bundle) space and is where $\delta\bm X$ lives. According to the Oseledets theorem, the Lyapunov spectrum is associated to a vector basis in this tangent space at each $\bm X$, which can be obtained by the solution to the linear problem 
\begin{equation}
\partial_{\bm{X}} \bm\Phi^t(\bm X) \cdot \bm{\hat{e}}_i(\bm X) = \Lambda_i(t, \bm X)\,  \bm{\hat{e}}_i(\bm\Phi^t \bm X).
\end{equation}
Then, the Lyapunov exponent
\begin{equation}
\lambda_i(\bm X) = \lim_{t\to\infty} \frac{1}{t}\, \ln|\Lambda_i(t,\bm X)|
\end{equation}
is associated to the vector $\bm{\hat{e}}_i(\bm X)$. This association allows a separation of the tangent space into a local stable subspace, spanned by $\mathcal{E}_s(\bm X)=\{ \bm{\hat{e}}_i(\bm X) \}_{\lambda_i<0}$, a local unstable subspace, spanned by $\mathcal{E}_u(\bm X)=\{ \bm{\hat{e}}_i(\bm X) \}_{\lambda_i>0}$, and a neutral subspace $\mathcal{E}_0(\bm X)=\{ \bm{\hat{e}}_i(\bm X) \}_{\lambda_i=0}$. Initial conditions in $\mathcal{E}_u(\bm X)$ ($\mathcal{E}_s(\bm X)$) will diverge (converge) exponentially to $\bm X$. 

\subsection{Hyperbolicity}

A stronger and more formal definition of chaotic systems is given by hyperbolicity. Given an invariant set $\mathcal{A} =\bm\Phi^t \mathcal{A} \subseteq \mathcal{M}$, we say it is hyperbolic if for each $\bm X\in \mathcal{A}$
\begin{enumerate}
\item the local neutral subspace $\mathcal{E}_0(\bm X)$ contains only the direction of the flow, and
\item the angle between $\mathcal{E}_s(\bm X)$ and $\mathcal{E}_u(\bm X)$ is always different to zero.
\end{enumerate}
A system is hyperbolic if it contains a single invariant hyperbolic subset. If the Lyapunov spectrum is the same for all trajectories of the invariant subset then the system is uniformly hyperbolic and nonuniformly hyperbolic otherwise. The latter is the most common case and holds in particular for hyperbolic billiards. If the flow is hyperbolic then the same holds for the associated map. The definition of hyperbolicity for a map is the same as before but the neutral direction can be ignored choosing an appropriate Poincar\'e surface. 

In hyperbolic systems all periodic orbits have positive Lyapunov exponents which imply they are unstable. This is the key property to have fast correlation decay and good mixing properties as will be discussed bellow. Nonhyperbolic systems have stable periodic orbits which live in KAM tori surrounded by unstable chaotic trajectories. The presence of these stable periodic orbits in phase space usually cause correlations to decay slower than in hyperbolic systems. 

\begin{figure}[h]
 \centering
 \includegraphics[width=0.85\columnwidth]{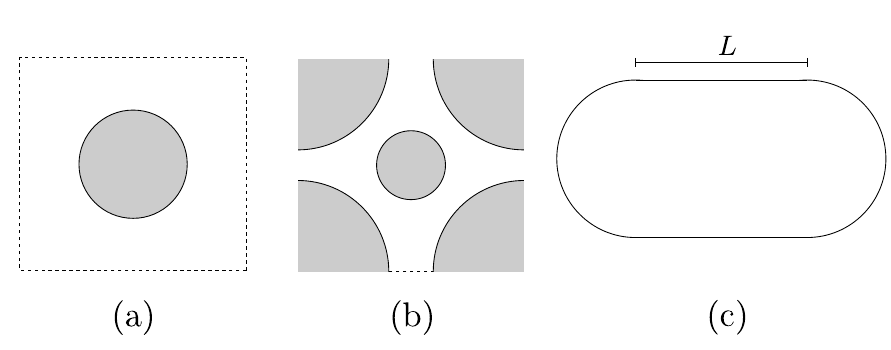}
 \caption[Sinai billiard and Bunimovich Stadium]{Lorentz Gas unit cell of an infinite-horizon (a) and finite-horizon (b) configuration. The most common form of the Sinai Billiard is given by (a). The Bunimovich Stadium (b) is composed of two semi-circles connected by parallel segments of length $L$.}
 \label{fig-sinai-buni} 
\end{figure}

\subsection{Sinai Billiard and Bunimovich Stadium}\label{section-lorentz-gas}

Probably the most studied and well known billiard systems are the Lorentz Gas (which is an unfolded Sinai Billiard) and the Bunimovich Stadium. Sinai Billiards corresponds to a class of closed billiards composed of smooth convex scatterers (usually circles); the Lorentz Gas is generated by placing this unit cell periodically or randomly on the plane. The Sinai Billiard dynamics is completely dispersing or \emph{defocusing} because an incident beam of particles gets defocused --the beam spreads--  after colliding with the convex obstacles. It has been rigorously proved that Sinai Billiards are hyperbolic\cite{sinai1970}, which allows to establish strong mixing and ergodic properties as we discuss in the next section. 

The Lorentz Gas can be divided into two different classes depending on its geometry. When the obstacles are arranged in such a way that unbounded free flight trajectories can exist in the system, the billiard is said to have \emph{infinite-horizon}. In the opposite case, it is said to have \emph{finite-horizon}. The speed of correlations decay and hence statistical properties of hyperbolic billiards crucially depends on this distinction. In particular, in a finite-horizon Lorentz Gas the dynamics can be shown to be diffusive whereas in the infinite-horizon case the diffusion is anomalous [see section \ref{corr-decay-and-diff}].

On the other hand, the Stadium is defined as a closed billiard composed of two parallel line segments and two circle arcs as shown in figure \ref{fig-sinai-buni}. The Stadium is a mixture of \emph{focusing} elements given by the arcs (beams of particles shrink after colliding with them) and neutral parallel segments where there is no dispersion. In the limit case of null length parallel segments the billiard reduces to a circle, which is integrable. When the circle is deformed making these segments sufficiently large the system becomes hyperbolic, as was proved by Bunimovich\cite{bunimovich1979}. This is explained by the fact that the focusing effect caused by the arcs is transformed into defocusing when the distance between them is sufficiently large.

The focusing and defocusing mechanisms are the two basic ingredients for chaos in dynamical billiards. 

\section{Statistical properties}\label{section-billiard-stat-props}

\subsection{Statistical ensembles}

We now consider an ensemble of $N$ trajectories in phase space
\begin{equation}
\bm X_t^{(i)} = \bm\Phi^t \bm X_0^{(i)} \,\in\, \mathcal{M}\;, \quad i=1,\ldots,N.
\end{equation}
These trajectories, which form a cloud of points in phase space at each time $t$, do not interact in any way since they are different realizations of the dynamical system in question. Given an observable $A(\bm X)$, its average over the ensemble $\{\bm X_t^{(i)}\}$ is given by
\begin{equation}\label{obs-average}
\langle A \rangle_t = \lim_{N\to\infty}\, \frac{1}{N} \sum_{i=1}^N A\left(\bm X_t^{(i)}\right).
\end{equation}
In the limit $N\to\infty$, the ensemble can be represented by a probability density distribution
\begin{equation}
f_t(\bm X) = \lim_{N\to\infty}\, \frac{1}{N} \sum_{i=1}^N \delta\left(\bm X - \bm X_t^{(i)}\right),
\end{equation}
hence, the average \eref{obs-average} may be expressed as
\begin{equation}
\langle A \rangle_t = \int_\mathcal{M} A(\bm X) f_t(\bm X)\, d\bm X.
\end{equation}

Since probability must be conserved, from the continuity equation in phase space we obtain
\begin{equation}\label{liouville-eq}
\partial_t f_t(\bm X) = - \partial_{\bm X}\cdot (\bm F(\bm{X})\, f_t(\bm X)),
\end{equation}
which is known as the Liouville equation. Assuming the vector field $\bm F(\bm X)$ is time-independent, we can write its solution formally as
\begin{equation}\label{liouville-eq-sol}
f_t(\bm X) = e^{\hat{L} \,t}f_0(\bm X),
\end{equation}
where $\hat{L}(f) = - \partial_{\bm X}\cdot (\bm F(\bm{X})\, f) $ is called the Liouville operator. In an invertible conservative system, the solution to \eref{liouville-eq-sol} may be expressed as
\begin{equation}
f_t(\bm X) = f_0(\bm\Phi^{-t}\bm X).
\end{equation}

Every phase space probability density $f(\bm X)$ induces a measure $\mu(U)=\int_U f(\bm X)\,d\bm X$ for $U \subseteq \mathcal{M}$. An important class of measures is induced by the stationary solutions of the Liouville equation \eref{liouville-eq}, $\hat{L} f(\bm X)=0$. Invariant measures satisfy
\begin{equation}
\mu(\bm\Phi^{-t} U) = \mu(U)
\end{equation}
for any phase space subset $U$. In general there are many invariant measures, for instance a density defined over any set of periodic orbits will induce an invariant measure. However, if the system is chaotic and we take a generic initial distribution in phase space, we might expect that the dynamical instabilities will produce the ensemble to be distributed over the whole phase space after enough time. Intuitively, the chaotic dynamic will blur any particular initial distribution features, and correlations between observables evaluated at the initial and long time densities will be null. This is a property called asymptotic stationarity, and can be formalized as
\begin{equation}\label{asymp-stat}
\lim_{t\to\infty}\langle A\rangle_t = \int_\mathcal{M} \mu_\infty(d\bm X) A(\bm X),
\end{equation}
for some observable $A(\bm X)$ with $\mu_\infty$ the invariant measure. In bounded systems (with a regular invariant measure), the previous property is equivalent to the mixing condition, namely 
\begin{equation}
\lim_{t\to\infty} \langle A(\bm\Phi^t\bm X) B(\bm X)\rangle_\infty = \langle A\rangle_\infty \langle B\rangle_\infty ,
\end{equation}
where $B(\bm X) = f_0(\bm X)/f_\infty(\bm X)$. The mixing condition implies ergodicity, which is the equivalence of time average and phase-space average,
\begin{equation}\label{ergodicity}
\lim_{t\to\infty} \frac{1}{t} \int_0^t A(\bm\Phi^t\bm X_0)\,dt = \int_\mathcal{M} A(\bm X) \mu_\infty(d\bm X) 
\end{equation}
for any initial condition $\bm X_0$ except for a set of null measure. 

In hamiltonian systems, volumes in phase space are preserved, thus the Liouville measure $d\bm X = d^f\bm q\, d^f\bm p$, with $(\bm q, \bm p)$ the canonical variables and $f$ the number of degrees of freedom, is an invariant measure. For systems with conserved quantities such as energy, the invariant measure can be defined in each energy shell $H=E$ (or equivalent) creating the microcanonical invariant measure. In billiards, this is given by the natural (Lebesgue) measure in Birkhoff coordinates $\bm\chi = (s, \sin\theta)$ [see discussion around \eref{billiard-map}],
\begin{equation}
d^2\bm\chi = d s\, d\sin\theta.
\end{equation}

\subsection{Correlation decay and diffusion}\label{corr-decay-and-diff}

The time-correlation function of classical observables plays an important role characterizing the dynamics of a system when no analytical results are known. Given two square integrable observables $A(\bm X)$ and $B(\bm X)$, their time-correlation function is given by
\begin{equation}\label{dyn-correlation}
C_{AB}(t) = \langle A(\bm\Phi^t\bm X) B(\bm X)\rangle - \langle A\rangle \langle B\rangle .
\end{equation}
In this section we always take averages using the relevant invariant measure, $\langle \cdot\rangle = \langle \cdot\rangle_\infty$, unless stated differently. In systems satisfying the mixing condition, it holds that $C_{AB}(t) \to 0$ as $t\to\infty$ for all square integrable observables $A$ and $B$. The decay rate of $C_{AB}(t)$ depends on how strongly the system mixes, i.e., it is conditional to the strength of the dynamical instability or chaoticity and to the smoothness of the observables. For instance, in an hyperbolic billiard with finite horizon (such as the Lorentz Gas) the decay of the velocity (and other smooth observables) is exponential. On the other hand, in systems with mixed phase space where integrable tori live immersed in an ergodic sea the correlation decay may be subexponential and in general polynomial.

The discrete-time-correlation function can be defied replacing the flow with the collision map in \eref{dyn-correlation}. It is worth noting that the decay rate may be of different character for the continuos and discrete time correlation functions. A good example of this is the infinite horizon Lorentz Gas where the discrete time correlations decay exponentially as in the finite horizon case but the continuos time correlation decay algebraically as $1/t$ as a consequence of the existence of unbounded free flights in the system\cite{bleher1992}. 

In systems where correlations decay sufficiently fast, the dynamical fluctuations of observables converging to the ergodic limit are Gaussian in the sense of the central limit theorem (CLT). Before giving a precise definition for the CLT lets consider $\Delta A^2 = \langle (S_A(n) - n \langle A\rangle)^2\rangle$ where $S_A(n)=\sum^n_{i=0} A_i$ and $A_n=A(\bm \phi^n \bm\chi_0)$. According to the discrete version of the ergodic theorem \eref{ergodicity}, in an ergodic system
\begin{equation}
\lim_{n\to\infty}\frac{1}{n} S_A(n) = \langle A\rangle. 
\end{equation}
Then, $\Delta A^2$ is the average squared of $S_A(n)$ fluctuations around its ergodic limit. It is easy to see that,
\begin{equation}
\Delta A^2 = n C_{AA}(0) + 2\sum_{i=1}^{n-1} (n-i)\, C_{AA}(i).
\end{equation}
Therefore, if the correlation function decays sufficiently rapid so that the sum\footnote{According to Chernov\cite{chernov2008}, the weaker condition $\sum_{n=0}^\infty | C_{AA}(n)|  <\infty$ also guaranties \eref{SLLN}.} 
\begin{equation}\label{LLN_condition1}
\sum_{n=0}^\infty n\, C_{AA}(n)  <\infty,
\end{equation}
we obtain that
\begin{equation}\label{SLLN}
\langle (S_A(n) - n \langle A\rangle)^2\rangle = n \sigma^2 + o(n) \quad \mbox{for large $n$,}
\end{equation}
where 
\begin{equation}\label{discrete-green-kubo}
\sigma^2 = C_{AA}(0) + 2\sum_{n=1}^\infty C_{AA}(n) .
\end{equation}
Hence, $S_A(n)  = n \langle A\rangle + \mathcal{O}(\sqrt{n})$ for $n\to\infty$, i.e., the average squared deviation of $S_A(n)$ from its mean grows as $\sigma\sqrt{n}$. Taking the observable as the $n$-th displacement $A(\bm \chi_n) =\Delta\bm r_n = \bm r_{n} - \bm r_{n-1}$  and assuming $\langle \Delta\bm r_n  \rangle = 0$, we have that if the system is strongly mixing (in the sense discussed above) from \eref{SLLN},
\begin{equation}
\langle \bm r_n^2\rangle = n\sigma^2 \quad \mbox{for large $n$},
\end{equation}
which is the weakest characterization of deterministic diffusion in a dynamical system. We can identify the discrete time diffusion coefficient as $\tilde{D}=\sigma^2$.  

Equation \eref{SLLN}, which corresponds to the Law of Large numbers in probability theory, gives us the average size of an observable fluctuations around its ergodic limit. A stronger property is given by the CLT and concerns the distribution of these fluctuations. Given an ergodic dynamical system, the CLT holds if
\begin{equation}\label{clt}
\lim_{t\to\infty} \mu\left\{ \bm X \;:\; \frac{\int_0^t A(\bm\Phi^t\bm X)\,dt - t\, \langle A\rangle}{\sqrt{D_A t}} < y\right\} = \frac{1}{\sqrt{2\pi}} \int_{-\infty}^y e^{-z^2/2}\,dz , 
\end{equation}
where $D_A$ is the (generalized) diffusion coefficient which can be written in continuos time as
\begin{equation}
D_A = \lim_{t\to\infty} \frac{1}{t} \left\langle \left( \int_0^t A(\bm\Phi^t\bm X)\,dt -t\, \langle A\rangle \right)^2 \right\rangle
\end{equation}
or, in analogy to \eref{discrete-green-kubo},
\begin{equation}\label{green-kubo}
D_A = \int_{-\infty}^\infty C_{AA}(t)\,dt ,
\end{equation}
assuming that $C_{AA}(t)$ decay faster than $t^{-1}$. When this condition does not hold the system is not diffusive but may be super-diffusive. Equation \eref{green-kubo} is the celebrated Green-Kubo formula which is fundamental in non-equilibrium statistical mechanics\cite{book_dorfman}.

If we choose the observable to be the instantaneous velocity in direction $x$, $A(\bm X)=  v_x$, then the CLT expressed in \eref{clt} means that the limit
\begin{equation}
\lim_{t\to\infty} \frac{x_t-x_0}{\sqrt{Dt}} 
\end{equation}
exists and converge in distribution to a standard Gaussian random variable. This implies that $\langle x_t^2\rangle = Dt$ for large $t$. If the random variable
\begin{equation}
y(s) = \lim_{t\to\infty} \frac{x_{st}-x_0}{\sqrt{Dt}} 
\end{equation}
is Gaussian with variance $\sqrt{s}$ for all $s$, then $y(s)$ is called a Brownian process. This implies that particle ensembles in the billiard spread according to a diffusion equation when the billiard is scaled such that the mean free path and obstacles are small and the number of particles is large. 

The CLT and  convergence to a Brownian motion has been proved for general finite-horizon hyperbolic billiards by Bunimovich, Sinai and Chernov\cite{bunimovich1991}. An analogous result was obtained for the infinite-horizon Lorentz Gas first by Bleher\cite{bleher1992} and then formalized by Szasz\cite{szasz2007}; in this case a non-normal CLT was proved with 
\begin{equation}
\lim_{t\to\infty} \frac{x_t-x_0}{\sqrt{Dt\log{t}}} 
\end{equation}
the correct stationary random process. This implies diffusion is anomalous, with $\langle x_t^2\rangle \sim t\log{t}$.

\section{Machta-Zwanzig approximation}\label{section-MZ-approx}

In strongly chaotic periodic billiards with a space configuration such that the escape time from a unit cell is longer than its ergodic time, i.e. the time mixing takes place within a unit cell, the motion of particles in the periodic lattice is well approximated as a (symmetric) random walk. Indeed, if a typical trajectory is \emph{trapped} within a unit cell for a time sufficiently long so that mixing takes place, then the decay of correlations imply the particle will lose memory of its initial state; therefore, its exit direction (into a neighboring unit cell) will be effectively random. We note that this excludes unit cells with infinite-horizon geometries. Expressed formally, given an initial phase space point $\bm X_0 = (\bm q_0, \bm{\hat{v}}_0, \bm I_0) \in \mathcal{D}_0\otimes S^1\otimes \mathbb{Z}^d$, when the mentioned conditions hold the unit cell index (winding number) dynamics $\bm I_n$ follows a Markov process, i.e.,  
\begin{equation}
\mbox{Pr}(\bm I_{n+1} \,|\, \bm I_{n}, \bm I_{n-1}, \ldots, \bm I_{0},\ldots) = \mbox{Pr}(\bm I_{n+1} \,|\, \bm I_{n}). 
\end{equation}
Under this approximation is possible to give an analytical expression for the diffusion coefficient as we discuss below. 

The idea to obtain the billiard diffusion coefficient in this way was first considered by Machta and Zwanzig for the particular case of the Lorentz Gas in the high density regime\cite{machta1983}. In this seccion we review their argument for a general one-dimensional billiard chain assumed to be strongly mixing and to possess narrow exits compared to the dimensions of the cavity. In an ergodic planar billiard, the average residence time in a unit cell $\mathcal{D}_0$ is given by \cite{chernov1997}
\begin{equation}\label{santalo}
\tau = \frac{\pi |\mathcal{D}_0|}{|\overline{\partial\mathcal{D}_0}|},
\end{equation}
where $|\mathcal{D}_0|$ is the fundamental domain area and $|\overline{\partial\mathcal{D}_0}|$ is the exit openings total length [for instance, consider the unit cell given by figure \ref{fig-sinai-buni} (b) where the dashed lines correspond to $\partial \mathcal{D}_0$]. On the other hand, it is well known that for a random walk in an isotropic\footnote{This means that the probability of a particle to exit through the right and left leads are the same and equal to 1/2 independent of its prior direction.} unidimensional lattice with period $a$ the diffusion coefficient is $ D = a^2/\tau$. Therefore, we obtain the Machta-Zwanzig diffusion coefficient,
\begin{equation}\label{machta-zwanzig}
D_{MZ} = \frac{Z^2 |\overline{\partial\mathcal{D}_0}|}{\pi |\mathcal{D}_0|}
\end{equation}
with $Z$ the unit cell length. We note that \eref{santalo}, usually called Santalo's formula, is an exact result so the only assumption to obtain \eref{machta-zwanzig} was the random walk approximation. As we have argued, this condition holds when the particles are trapped for a long time in a unit cell beacuse this means they collide many times with the obstacles; hence, since the billiard is assumed strongly chaotic this implies a fast correlation decay and effective markovian dynamics for the winding number $\bm I_n$ leading to the random walk approximation. Consequently, assuming strong chaotic dynamics in the unit cell, the validity of \eref{machta-zwanzig} rests only on the mean free path length $\tau_c$ since the average number of collisions of a trajectory before exiting a cavity is given by $\tau/\tau_c$. An analytical expression for $\tau_c$ is given by \eref{santalo} replacing $|\overline{\partial\mathcal{D}_0}|$ with $|\partial\mathcal{D}_0|$, the obstacles boundary length in the unit cell. Then, we have that the Machta-Zwanzig approximation for the diffusion coefficient \eref{machta-zwanzig} holds in the limit $\tau_c \ll \tau$, that is, when
\begin{equation}\label{MZ-condition}
|\overline{\partial\mathcal{D}_0}| \ll |\partial\mathcal{D}_0|.
\end{equation}

\section{Cosine billiard}\label{section-cosine-billiard}

In this section we define the periodic cosine billiard which will be used in the next chapters to numerically test some of our classical and quantum (wave) results; therefore we sometimes also refer to it as the cosine waveguide. The periodic cosine billiard is composed of a one-dimensional chain of hard-wall cavities with cosine shaped boundaries. We employ this system because our numerical method for the quantum scattering problem --which is very efficient in the semi-classical limit--  requires a smooth waveguide with connected boundaries, thus the Lorentz Gas or Bunimovich Stadium chain cannot be employed. Even though the quantum numerical method is highly efficient for our purposes, the cosine billiard has the disadvantage to be much harder (numerically expensive) to solve classically that the aforementioned systems. In addition, there are no analytical proofs of hyperbolicity and ergodicity for the cosine billiard, so we will have to investigate them numerically. In spite of these contretemps, the cosine billiard has been used before in the quantum chaos literature\cite{lunaacosta1996,ketzmerick2000,lunaacosta2002,mendoza2008} because of the possibility of easily changing its dynamic from a mixed phase space to (apparent) full chaos with the tuning of the cosine amplitude.
\begin{figure}[h]
 \centering
 \includegraphics[width=1\columnwidth]{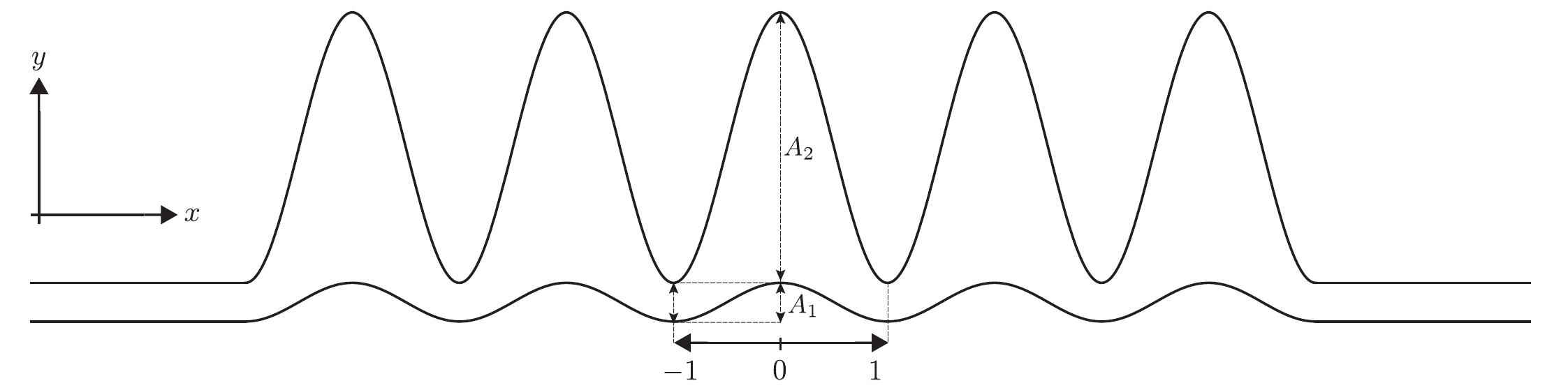}
 \caption[Cosine billiard]{Cosine billiard chain with five unit cells connected to two plane leads. The unit cell boundaries are defined in \eref{cosine-billiard-def}--\eref{cosine-billiard-def-2} as a function of the amplitudes $A_1$ and $A_2$ shown in the figure.}
 \label{fig-cosine-billiard} 
\end{figure}

Let $(x,y)$ be the coordinates in configuration  space. We recall we have set the particle's speed to $v=1$ unless otherwise stated; for general velocity $v$ the diffusion coefficient simple scales as $D\to D v$. We define the unit cell as the region enclosed by $h_1(x)<y<h_2(x)$ for each $x\in [-1,1]$, where 
\begin{eqnarray}\label{cosine-billiard-def}
h_1(x) &=& \frac{A_1}{2}\left[1+\cos{(\pi x)}\right] \quad \mbox{and}\\
h_2(x) &=& A_1+ \frac{A_2}{2}\left[1+\cos{(\pi x)}\right], \label{cosine-billiard-def-2}
\end{eqnarray}
Hence, $h_1(x)$ and $h_2(x)$ define the lower and upper waveguide boundaries, respectively. With the unit cell defined in this way our cosine billiard always has finite horizon, i.e., it does not allow unbounded collision-free trajectories for any values of $A_1>0$ and $A_2>0$. It is trivial to see this is guaranteed by the fact that $h_1(0)=h_2(\pm 1) = A_1$. As explained in section \ref{section-lorentz-gas}, besides strong mixing, the finite-horizon property is fundamental to obtain normal diffusive dynamics. 

Note that the unit cell mirror symmetry $x\to-x$ is not relevant for the classical transport properties of the billiard but makes the numerical solution of the quantum scattering problem faster (the transmission and reflexion matrices $\bm t$ and $\bm r$ are the same in both direction). However, this induces an anti-unitary symmetry in the quantum Hamiltonian which plays a role in the statistical and transport properties of the waveguide as we discuss in chapter \ref{chapter-nb}.

The character of the cosine billiard classical dynamic can be mixed or predominantly chaotic, depending on the parameters $A_1$ and $A_2$. In order to assess the dynamical properties of this system we use basically two tools, namely the unit cell Poincar\'e sections 
\begin{eqnarray}
\mathcal{S}_x &=& \{ (y,v_y) \, :\; \forall n\in\mathbb{N},\;  |x|= 1 + nZ \} \label{sx-section} \\
\mathcal{S}_y &=& \{ (s,v_s) \, :\;  \forall x(s)\in[-1,1] ,\; y=h_2[x(s)] \}
\end{eqnarray}
where $Z=2$ is the waveguide period and $(s,v_s)$ denote the Birkhoff coordinates, and the instantaneous velocity autocorrelation function 
\begin{equation}
C_v(t)=\langle v_x(t) v_x(0)\rangle.
\end{equation}
We note that $\mathcal{S}_x$ is the stroboscopic Poincaré section on the unit cell connecting border and $\mathcal{S}_y$ is the surface of section on the unit cell upper wall parametrized with Birkhoff coordinates. Transport properties of the cosine billiard depend mostly on its dynamic over $\mathcal{S}_x$, because when this map is sufficiently chaotic almost all stable quasi-periodic trajectories on $\mathcal{S}_y$ are trapped within the unit cell and therefore do not contribute to transport. 

\begin{figure}[h]
  \centering
(a)\\
 \includegraphics{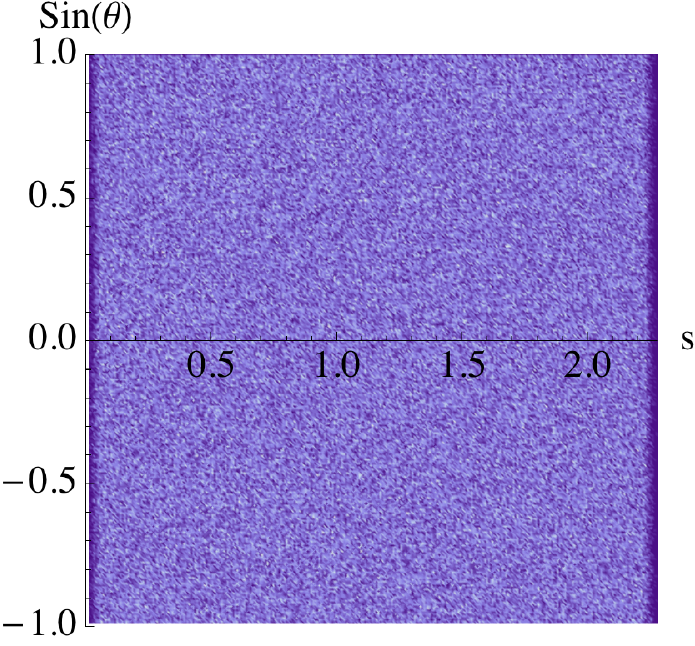}\hspace{1cm}
 \includegraphics{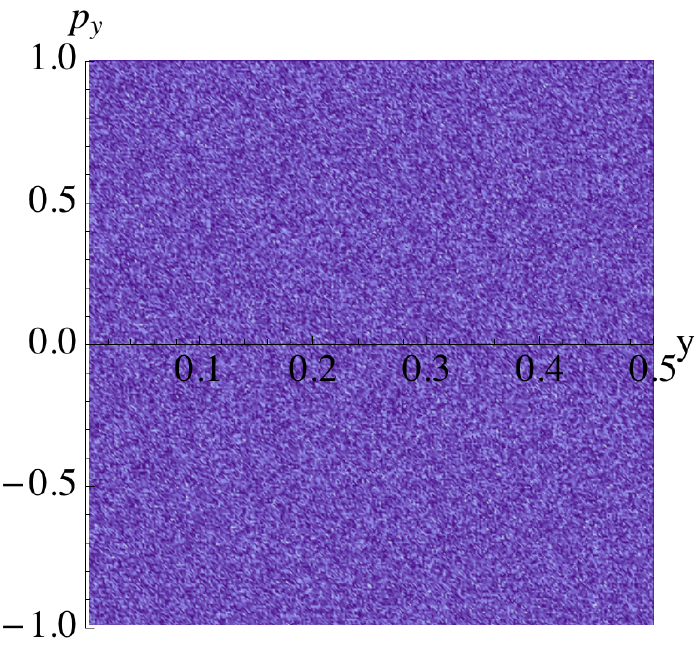}\\
 \vspace*{1cm}
 (b)\\
 \includegraphics{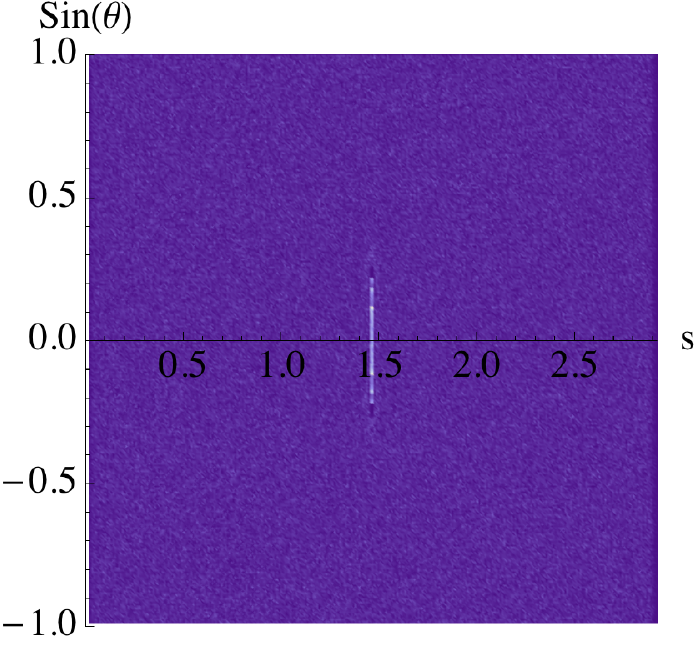}\hspace{1cm}
 \includegraphics{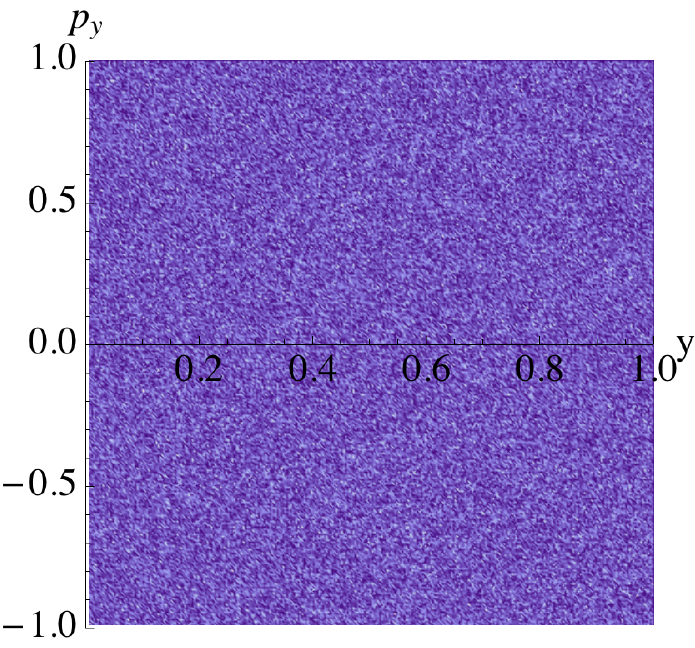}
 \caption[Poincare sections density plots of the cosine billiard]{Density plots of the Poincare sections $S_x$ (right column) and $S_y$ (left column) of the cosine billiard. Figures (a) correspond to a configuration with parameters $A_1=0.5$, $A_2=3.0$ where we observed strong chaotic dynamics; in this case there are no visible tori. In (b) we show a configuration with $A_1=1.0$, $A_2=0.45$, where the dynamics is also strongly diffusive but a small tori is present in $S_x$ related to families of integrable stable orbits trapped in the unit cell; this is not relevant for transport since the stroboscopic map is strongly chaotic as the density plot in $S_y$ reveals.}
 \label{fig-poincare-sections} 
\end{figure}

\begin{figure}[h]
 \centering
 \includegraphics[width=0.8\columnwidth]{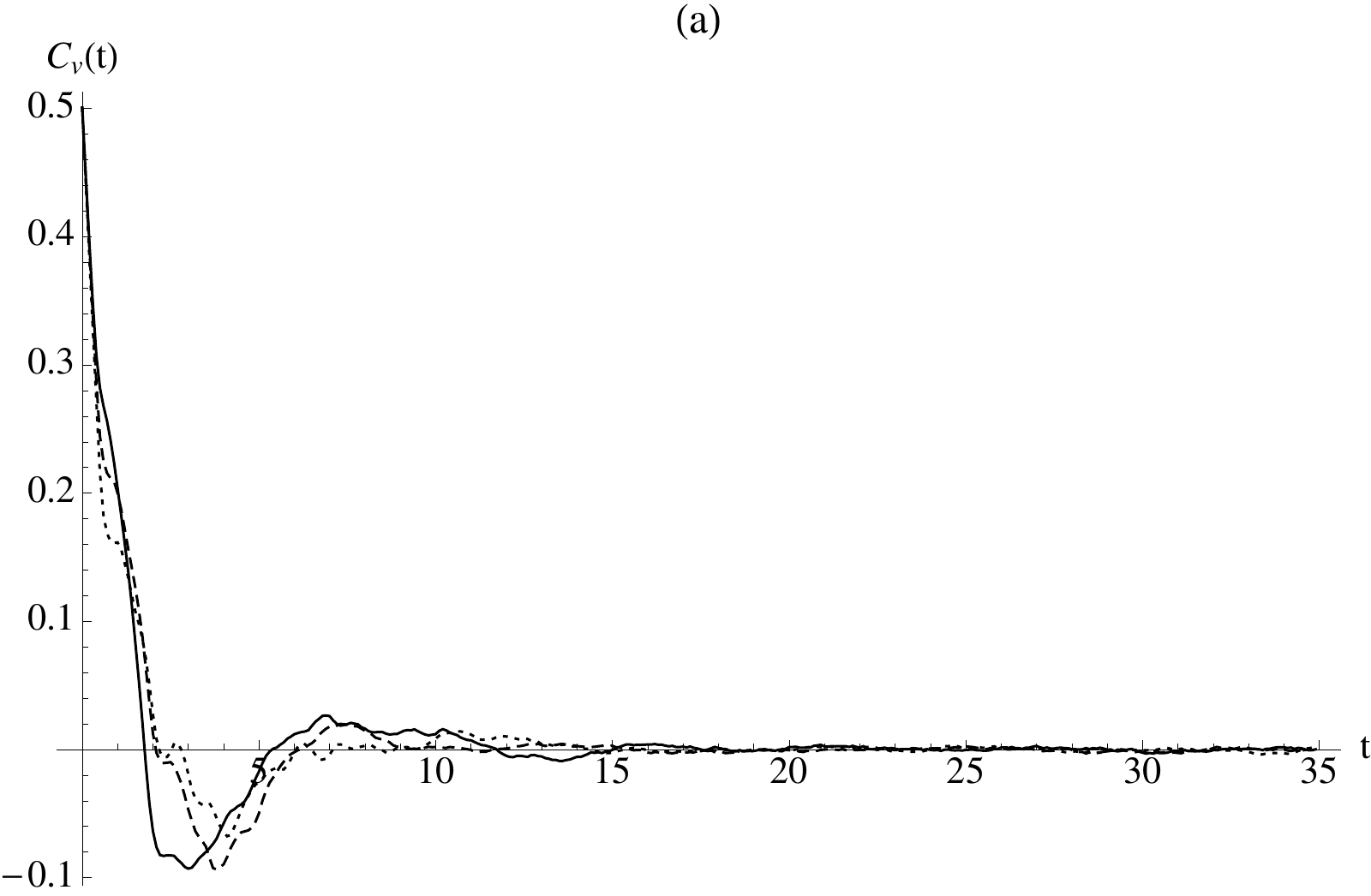}\\
 \vspace*{1cm}
 \includegraphics[width=0.8\columnwidth]{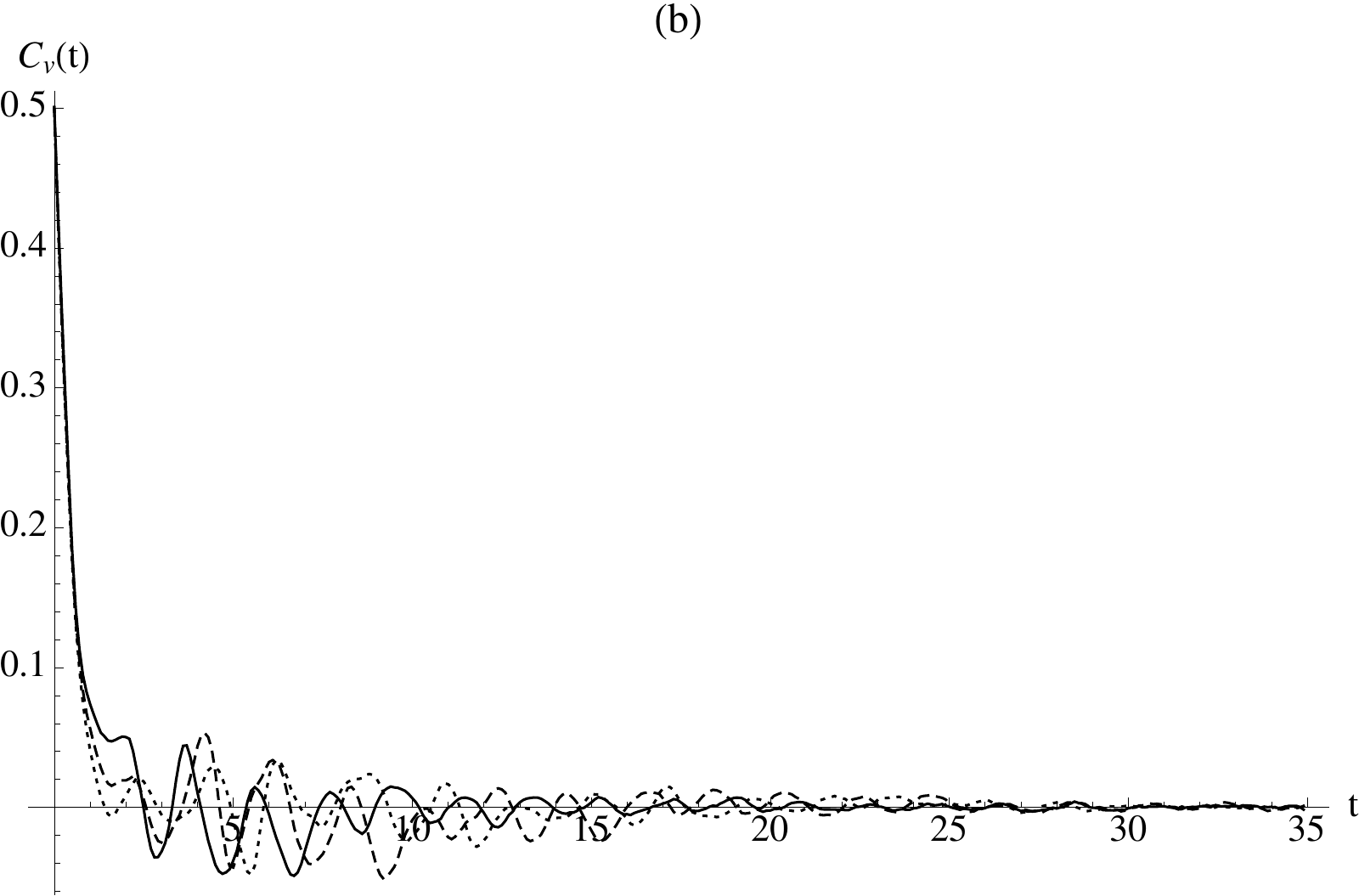}
 \caption[Cosine billiard velocity autocorrelation function]{Velocity autocorrelation funcion $C_v(t)=\langle v_x(t)v_x(0)\rangle$ for an ensemble of $10^5$ particles. In figure (a) we plot this function for the three configurations with $A_1=1.0$, $A_2=0.6,\,0.45,\,0.3$ (full, dashed and dotted lines). In figure (b) we show the same function for $A_1=0.5$, $A_2=2.5,\,3.5,\,4.5$ (full, dashed and dotted lines). The decay to noise level is quite fast in all cases; in (a) at around $t\sim20$ (which is approximately 25 collisions) correlations have decayed to noise level whereas in (b) the same happens for $t\sim30$ (which is approximately 9 collisions).}
 \label{fig-vx-correlation} 
\end{figure}

\begin{figure}
 \centering
 \includegraphics[width=.95\columnwidth]{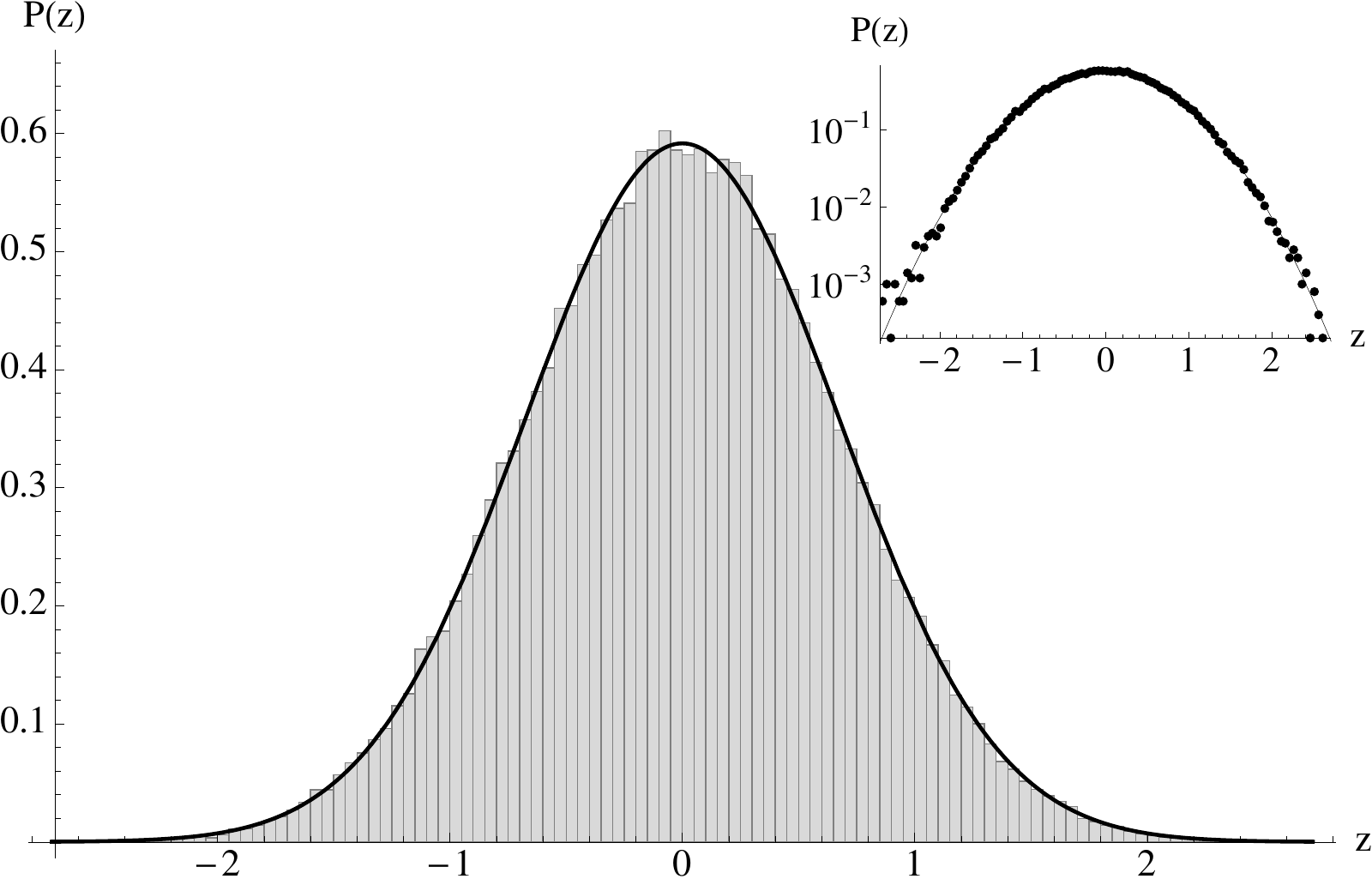}
 \caption[Cosine billiard normalized displacement histogram]{Histogram of the normalized displacement $\tilde{x}_t$ for a cosine waveguide with $A_1=1.0$, $A_2=0.3$ at time $t=50000$ using an ensemble of $10^5$ initial conditions. The best Gaussian fit is plotted over it. The inset show the same histogram (dots) in log-scale where convergence to a Gaussian (full line) is more evident to be achieved even deep in the tails. For the $A_1=0.5$ configurations of the cosine billiard we are considering a similarly good converge of $\tilde{x}_t$ is observed.} 
 \label{fig-dx-histogram} 
\end{figure}

\begin{figure}
 \centering
\includegraphics[width=0.95\columnwidth]{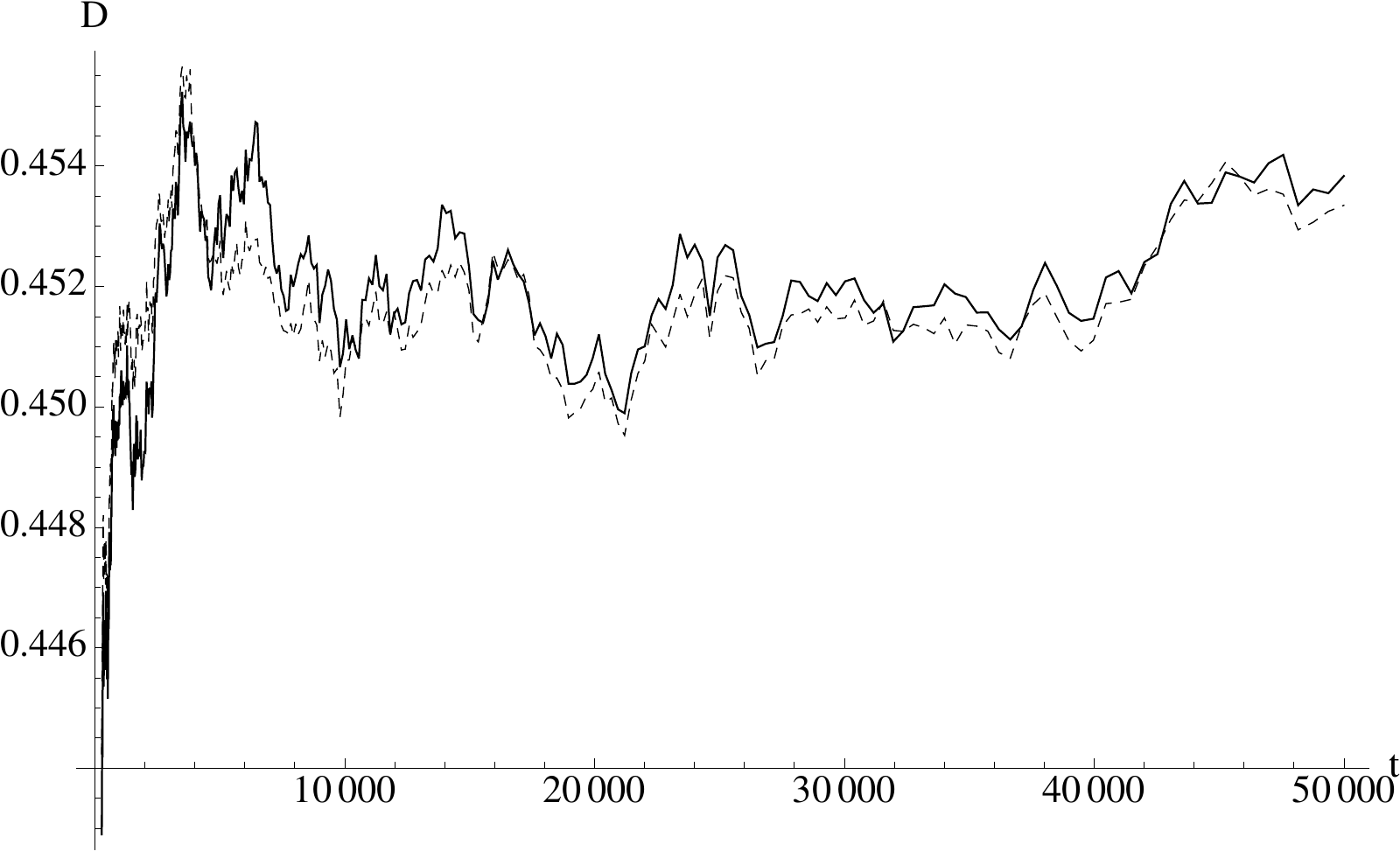}
 \caption[Diffusion coefficient for the cosine billiard]{Finite time approximations of the diffusion coefficient $D(t) = \langle x_t^2 \rangle / t$ for the $A_1=1.0$, $A_2=0.3$ cosine waveguide calculated from the variance (second moment) of the sample's displacements (black line) and from the first moment using $\frac{\pi}{2} \langle |\tilde{x}_t| \rangle^2 = \langle \tilde{x}_t^2 \rangle$, as expected for Gaussians, (dashed line) for an ensemble of $10^5$ initial conditions. To obtain the diffusion coefficient $D$ we average $D(t)$ for times after the transient relaxation regime. The diffusion coefficients obtained in this way from the first and second moments of $x_t$ are the same up to standard error.}
 \label{fig-moments-gauss}  
\end{figure}

We will consider two ranges of parameters. The first is given by $A_1=1$ and $A_2\in[0.3,\, 0.6]$ where the system is predominantly chaotic but a few small tori are observed in $\mathcal{S}_y$ for some values of $A_2$ [see figure \ref{fig-poincare-sections}]. Still, in all cases tested in this range, the connected ergodic component in $\mathcal{S}_x$ makes up 95\% or more of the section area and $C_v(t)$ decays exponentially [see figure \ref{fig-vx-correlation}]. This implies normal diffusive dynamics as can be seen by looking at the scaled asymptotic displacement
\begin{equation}
\tilde{x}= \lim_{t\rightarrow\infty} \frac{x(t)-x(0)}{\sqrt{t}} ,
\end{equation}
which is distributed as a Gaussian random variable with zero mean and variance $D=\langle \tilde{x}^2\rangle$ [see figure \ref{fig-dx-histogram}]. Since the velocity of autocorrelation decay is exponential, the convergence of $\tilde{x}$ distribution to its asymptotic Gaussian limit is quite fast. An alternative method to support our conclusion that $\tilde{x}/\sqrt{D_1}$ converges to a standard Normal distribution and that at least its first two moments also converge, is given by calculating $M_1=  \langle |\tilde{x}_t| \rangle$. It is easy to see that if this is the case then 
\begin{equation}\label{moments-relation}
\frac{\pi}{2}M_1^2  = \langle \tilde{x}_t^2 \rangle = D
\end{equation}
[see figure \ref{fig-moments-gauss}]. Note that, for instance, an anomalous diffusive system with ergodic phase-space but infinite horizon trajectories may exhibit convergence in distribution to a Gaussian for an appropriate normalized displacement but would fail to satisfy relation \eref{moments-relation} between its first and second moments for any finite time approximation of $\tilde{x}$ \cite{armstead2002}. 

As we have seen for the parameter range  $A_1=1$ and $A_2\in[0.3,\, 0.6]$, although there are small tori observable in the section $\mathcal{S}_y$, the billiard dynamic is strongly chaotic in $\mathcal{S}_x$ and effectively diffusive as we have seen from the fast decay of correlations $C_v(t)$ and strong convergence (at least up to the second moment) of the scaled displacement $\tilde{x}$. In chapter \ref{chapter-nb} we will use this configuration to test a semi-classical result regarding the average number of propagating Bloch modes in a diffusive waveguide and show excelent agreement with the analytical result. However, this range of parameters has the disadvantage of not fulfilling condition \eref{MZ-condition} for the Machta-Zwanzig approximation. In fact, it is easy to see that $|\partial \mathcal{D}_0| = 3A_1 + A_2$ and $|\overline{\partial \mathcal{D}_0}|  = A_1$, thus in the best case $|\partial \mathcal{D}_0| = 3.6 |\overline{\partial \mathcal{D}_0}|$. 

The second parameters range we consider is given by $A_1=0.5$ and $A_2\in[2.5,\,4.5]$, which is closer to the Machta-Zwanzig limit since in this case $|\partial \mathcal{D}_0|/ |\overline{\partial \mathcal{D}_0}|$ moves between 8 and 12. With this configuration the system satisfies all the properties previously discussed, i.e., \nopagebreak it shows fast decay of correlations and exhibits a diffusive dynamic [see figures \ref{fig-poincare-sections} and \ref{fig-vx-correlation}].

\chapter{Random matrix theory}\label{chapter-rmt}

The theory of random matrices was first formally developed in the nineteen-sixties mainly by Wigner, Dyson, Mehta and Gaudin, motivated by understanding the energy spectra statistics of heavy nuclei. Later in the same decade, these techniques were applied to small metal particles to study their microwave absorption properties. A brief account of the history of these developments as well as a complete treatment of the theory's current status can be found in the book by Metha \cite{book_mehta}.

Random matrix theory (RMT) experienced a revival of interest in recent decades and has found many applications in disciplines outside physics, for instance in statistics, finance, engineering and number theory. In physics, its domain of use shifted from nuclear physics to quantum chaotic systems since it was noted by Bohigas, Giannoni and Schmit\cite{bohigas1984} that the Wigner-Dyson ensembles of hermitian matrices applied generically to describe statistical properties of closed chaotic systems (see also \cite{book_haake} and \cite{book_gutzwiller}). This discovery was followed by the work of Altshuler and Shklovskii\cite{altshuler1986} on the universal conductance fluctuations of disordered systems, which led to the development of a random matrix theory of quantum transport. A nice review on this subject is given by Beenakker in \cite{beenakker1997}. 

In this chapter, we present some known properties and results of RMT which are relevant for our work. First, in section \ref{section-connectio-chaotic}, we discuss the connection between a quantum chaotic systems and random matrices. Then, in the following sections, we present the Wigner-Dyson and Circular ensembles which describe the Hamiltonians and Scattering matrices of chaotic systems, respectively. Finally, in section \ref{section-rmt-periodic-chain}, we define the RMT periodic chain model which we will use in the next chapters.

\section{Connection to chaotic systems}\label{section-connectio-chaotic}

Random matrix theory deals with the statistical properties of large matrices with randomly distributed elements. The basic assumptions for RMT to be a good description of a given physical system are that its evolution dynamics maintains phase coherence, it is chaotic and it is \emph{big} so that the number of energy levels (or degrees of freedom) is sufficiently large for a statistical description to make sense. Hence, for low-dimensional chaotic systems such as billiards, the connection between their quantum statistical properties and RMT is achieved in the semi-classical limit. 

Let us consider a quantum system whose energy spectrum is $E_1 \leq E_2 \leq \cdots \leq E_N$ for a fixed integer $N$ which is large but finite. One of the first spectral statistics to be studied semiclassically was the energy level nearest-neighbor spacing $P(S)$, which is the pdf of the \emph{unfolded}\cite{book_haake} spectrum nearest-neighbor spacings, namely
\begin{equation}
S_i = \frac{E_i - E_{i+1}}{\Delta E},
\end{equation}
where $\Delta E=1/\bar{\rho}_E$ is the mean level spacing with $\bar{\rho}_E$ the mean local density of states [see equation \eref{mean-density-of-states}]. Generic classically integrable systems with more than two degrees of freedom have a Poissonian level spacing distribution $P(S)=e^{-S}$ and display no correlations, hence energy levels tend to cluster together without any repulsion\cite{berry1977}. This implies that level crossings are not avoided when a parameter in the Hamiltonian is changed. On the other hand, classically chaotic (non-integrable) systems display different levels of repulsion depending on their symmetries; there are three universality classes with repulsion given by
\begin{equation}\label{lev-rep-intro}
P(S)\sim S^\beta \quad \mbox{for $S\to 0$},
\end{equation}
with $\beta=1, 2$ or 4. For systems possessing an anti-unitary symmetry (such as time reversal) $\beta=1$, for systems without such symmetries $\beta=2$ and for anti-unitary symmetric systems with broken spin-rotation symmetry $\beta=4$. It was noted\cite{bohigas1984} that the full distribution $P(S)$ observed in chaotic systems corresponded for all $S$ to the nearest-neighbor spacings observed in  the eigenvalues of large random hermitian matrices taken from the appropriate gaussian ensembles, called the Wigner-Dyson ensembles. Thus, it was conjectured by Bohigas, Giannoni and Schmit (BGS) that a system with chaotic phase-space dynamics is expected to show universal quantum level statistics in the semiclassical limit, consistent with the predictions of these ensembles, with the only remaining relevant physical parameter being the mean level spacing $\Delta E$. 

The main tool to understand this connection between Wigner-Dyson RMT and chaotic systems has been the Gutzwiller trace formula\cite{book_gutzwiller}, which was recently used in the first semiclassical proof of the BGS conjecture, given by Hakee \textit{et al.} in \cite{haake2004, haake2007}. Gutzwiller trace formula gives a semiclassical expression for the density of energy states as a sum over the periodic orbits of a chaotic system, $\rho(E) = \sum_\gamma A_\gamma e^{i S_\gamma/\hbar}$, where $S_\gamma$ and $A_\gamma$ are the action and stability amplitude of the periodic orbit $\gamma$. After unfolding the spectra of a chaotic system, all statistical properties of $\rho(E)$ are reproduced by the analogous function in an ensemble of random Hamiltonians with the appropriate symmetries. In disordered systems, the \emph{physical ensemble} represented by RMT is given by the set of allowed disorder realizations. In contrast, in chaotic systems where disorder does not play a role the physical ensemble being caricatured by RMT can be taken as a uniform set of energy realizations in the interval $[E-dE/2,E+dE/2]$, with $\Delta E \ll dE \ll E$ such that $\rho(E)$ is approximately constant in this interval. This means that $dE$ is classically small but large enough to contains many quantum levels. We call this the \emph{semiclassical ensemble}. 

On the other hand, the RMT of quantum transport is based on Dyson circular ensembles of scattering matrices, which can be defined by a principle of maximum entropy on a set of unitary matrices with the appropriate symmetries imposed (unitary, orthogonal or symplectic). First, Bl\"umel and Smilansky\cite{blumel1990} showed semiclassically that the distribution of scattering phase shifts of a finite-horizon chaotic cavity was well described by the circular ensembles and then, Jalabert, Pichard and Beenakker\cite{jalabert1994} showed that this RMT agreed with the semiclassical conductance in a quantum dot calculated by Baranger, Jalabert and Stone\cite{baranger1993}. This theory explains two of the most salient features of quantum transport of chaotic cavities, namely the weak localization effect and universal conductance fluctuations. The transport properties in a \emph{disordered wire} geometry was first studied by Dorokhov\cite{dorokhov1982} and Mello, Pereyra and Kumar\cite{mello1988}. They developed a differential equation for the evolution of the transmission eigenvalues pdf as a function of the wire length known as the Dorokhov-Mello-Pereyra-Kumar (DMPK) equation, which assumed a wire composed of small weakly scattering segments taken from the circular ensembles. This equation accounted for the ohmic and localized regimes observed in a disordered wire. Another method used in quantum transport is the non-linear $\sigma$ model, which allows us to compute the $n$-point correlation functions of the wire eigenvalues; this method has been proved equivalent\cite{brouwer1996} to the DMPK equation but provides less information since only give access to correlation functions. A different approach was later pursued by Iida, Weidenm\"uller and Zuk\cite{iida1990-2, iida1990} whom computed the conductance average and variance of a (discrete) chain of random quantum dots taken from the circular ensembles. They studied the relevance of the coupling of the quantum dots chain to the external leads and showed that, assuming orthogonal Wigner-Dyson statistics for the quantum dots hamiltonians, was sufficient to obtain universal conductance fluctuations. 

Later, Simons and Altshuler\cite{altshuler1993, simons1993} considered the statistical properties of the spectrum $E_i(X)$ when a chaotic system is subject to an external perturbation $X$, for instance in the form of a Aharonov-Bohm magnetic flux or a background potential. In this case, it is found that the statistical properties of $E_i(X)$ are universal after appropriate scaling dependent on two physical parameters: the mean level spacing $\Delta E$ and the generalized conductance $C(0) = \langle \partial_X E_i^2\rangle / (\Delta E)^2$. This allows us to examine the statistical properties of the energy bands in periodic systems\cite{mucciolo1994}, where $X$ is the Bloch phase which acts as an Aharonov-Bohm flux over the unit cell defined on a torus topology (periodic boundary conditions). In section \ref{section-nb-universal-correlator} we will employ Simons and Altshuler universal correlation function to obtain the average number of propagating Bloch modes in a periodic chaotic system.

\section{Wigner-Dyson ensembles}

\subsection{Definitions}

There are three different Wigner-Dyson ensembles of hermitian matrices, namely the Gaussian Orthogonal Ensemble (GOE), Gaussian Unitary Ensemble (GUE) and Gaussian Symplectic Ensemble (GSE). They receive their names because they contain random gaussian elements and are invariant with respect to orthogonal, unitary and symplectic transformation, respectively. A detailed treatment of this subject can be found in \cite{book_mehta}; in the present section we summarize the relevant results.

A generic physical system without time-reversal invariance possess a Hamiltonian not restricted by any symmetry but being hermitian. Given an hermitian matrix $\bm H = \bm H^\dagger$, the transformed matrix
\begin{equation}\label{automorf-gue}
\bm H' = \bm U^{-1} \bm H \bm U
\end{equation}
is also hermitian for any unitary matrix $\bm U\in{\rm U}(N)$. Hence, the GUE ensemble is defined over the space of hermitian matrices with a pdf $P(\bm H)$ satisfying 
\begin{equation}\label{gue-pdf-inv}
P(\bm H')\, d\bm H' = P(\bm H)\, d\bm H 
\end{equation}
for any unitary matrix $\bm U$ with $\bm H'$ given by \eref{automorf-gue} and where the volume element is given by
\begin{equation}\label{gue-vol-element}
d\bm{H} = \prod_{k\leq j} d H^{(\rm r)}_{kj}\, \prod_{k< j} d H^{(\rm i)}_{kj}
\end{equation}
with $H^{(\rm r)}_{kj}$ and $H^{(\rm i)}_{kj}$, the real and imaginary parts of $H_{kj} = (\bm H)_{kj}$, statistically independent variables. It follows from this definition that most general form of the GUE pdf is
\begin{equation}\label{gue-pdf}
P(\bm H) = c\, \exp\left( -a\,\mbox{Tr}\,\bm H^2 + b\,\mbox{Tr}\,\bm H   \right)
\end{equation}
with $a>0$, $b$ and $c$ real. The constant $b$ is usually taken as zero since it accounts to a simple change of the spectrum $E_i$ mean value.

On the other hand, if a (spinless) system possess time-reversal symmetry, its eigenfunctions are real, hence there exists a non-diagonal basis where the Hamiltonian is a real matrix. Thus, in this case the Hamiltonian can be taken as a real symmetric matrix. The GOE ensemble is defined over the set of these matrices requiring equivalent invariance properties than for the GUE case, namely that \eref{gue-pdf-inv} holds with $\bm H' = \bm W^{-1} \bm H \bm W$ for any orthogonal matrix $\bm W\in {\rm O}(N)$. The GOE volume element is similar to \eref{gue-vol-element} without the imaginary parts product. From this conditions follows that the GOE pdf have the same form as \eref{gue-pdf}. 

The third universality class corresponds to systems with time-reversal symmetry but without spin-rotation symmetry. It can be shown that in this case the system Hamiltonian can be chosen as a hermitian self-dual\footnote{This means that the Hamiltonian is a quaternion matrix and there exists a quaternion basis where it is quaternion real.} matrix $\bm H = \bm H^R$. The set of matrices with these symmetries is invariant to symplectic\footnote{Here we refer to the symplectic group ${\rm Sp}(N)$ which is the group of invertible quaternionic matrices preserving the standard hermitian norm.} transformations, hence the GSE is defined imposing this requirement in a similar fashion to what we have done for the GUE and CUE. 

Originally, Wigner and Dyson studied an ensemble of hermitian matrices $\bm H$ with $P(\bm H) = c\,\exp( -a\,\mbox{Tr}\,V(\bm H) )$ with $V(\cdot)$ a potential function. In case $V(\bm H) \sim \bm H^2$ we recover the Gaussian ensembles; however, in the limit of large matrices $N\to\infty$ the spectral correlations become independent of $V(\cdot)$ sufficiently far away from the spectrum edges. This is the heart of the universality of spectral correlations underlying BGS conjecture.

\subsection{Eigenvalue and levels spacing distributions}

Once we have obtained the pdf for the Gaussian ensembles, the next question is to determine the eigenvalues $\{ E_n\}$ pdf. This is accomplished by finding the Jacobian $J_\beta$ that allows us to write the volume element $d\bm H$ as a function of $E_n$ and the eigenvectors matrix.  Thus, we have that
\begin{equation}
d\bm H = J_\beta\, d\mu_\beta(\bm U)\, \prod_{i=1}^N dE_i ,
\end{equation}
where $\beta=1,2,4$ for the unitary, orthogonal and symplectic ensembles, $\bm U$ is in the relevant symmetry group and $d\mu_\beta(\bm U)$ is the Haar measure\footnote{Given a topological group $G$, the left (right) Haar measure is the unique measure over subsets of $G$ that is invariant to left (right) multiplication. For compact groups (such as ${\rm U}(N)$, ${\rm O}(N)$ and ${\rm Sp}(N)$) the right and left Haar measures are the same.} of this group. The Jacobian is given by\cite{book_haake}
\begin{equation}
J_\beta(\{E_n\}) = \prod_{i<j}^N |E_i - E_j|^\beta ,
\end{equation}
with $\beta=1,2,4$ defined previously. Since $\mbox{Tr}\,\bm H^2=\sum_{i=1}^N E_i^2$, $P(\bm H)$ given by \eref{gue-pdf} does not depend on the eigenvectors of $\bm H$. Therefore, the eigenvalue distribution takes the form
\begin{equation}
P(\{E_n\}) = c\,\prod_{i<j}^N |E_i - E_j|^\beta\, \exp\left( -a \sum_{i=1}^N E_i^2 \right), 
\end{equation}
which can be recasted in the form of a Boltzmann-Gibbs distribution as
\begin{equation}
P(\{E_n\}) = c\, \exp\left[ -\beta\left( \sum_{i<j}^N u(E_i,E_j) + a \sum_{i=1}^N E_i^2   \right)   \right]
\end{equation}
with a two-point potential energy $u(E,E')=-\ln|E-E'|$ and $\beta$ playing the role of inverse temperature. Hence, the energy eigenvalue dynamics when a system parameter is changed is expected to be analogous to a unidimensional Coulomb gas since the logarithmic repulsion $u(E,E')$ is equivalent to the Coulomb interaction of two line charges. Although the eigenvalue dynamics turns out to be integrable, their statistical equilibrium properties are reproduced by RMT.

The marginal probability density function
\begin{equation}
\bar{\rho}(E) = \int dE_2\ldots dE_N\, P(E,E_2,\ldots,E_N)
\end{equation}
is also of particular interest. Setting the normalization $\langle H_{ij} H_{ji}\rangle = 1/4N$ (fixing the constant $a$ in \eref{gue-pdf}) in order to avoid divergences in the limit $N\to\infty$, one obtains the well-known Wigner semicircle law
\begin{equation}
\bar{\rho}(E) = \left\{
\begin{array}{ll}
\displaystyle (2/\pi)\, \sqrt{1-E^2}  & \mbox{for $|E|<1$} \\
0  & \mbox{for $|E|>1$}  
\end{array}\right.
\end{equation}
valid for the unitary, orthogonal and symplectic ensembles. 

Finally, using $P(\{E_n\})$ we can compute the nearest-neighbor spacing pdf. For the most trackable case of dimension $N=2$ is possible to obtain a simple closed form solution, namely
\begin{equation}\label{level-repulsion}
P(S) = \left\{
\begin{array}{ll}
(\pi S/2) e^{-\pi S^2/4}  				&\;  \mbox{orthogonal} \\
(32 S^2/\pi^2) e^{-4 S^2/\pi}  			&\;  \mbox{unitary}   \\
(2^{18} S^4/3^6\pi^3) e^{-64 S^2/9\pi}	&\;  \mbox{symplectic}.
\end{array}
\right.
\end{equation}
The distribution $P(S)$ in the limit $N\to\infty$ is quite close to \eref{level-repulsion} so for most practical purposes this is an adequate approximation. In particular, it gives the correct level repulsion \eref{lev-rep-intro} for the three universality classes.

\section{Circular ensembles}\label{section-circular-ensembles}

\subsection{Definitions}

The circular ensembles are probability spaces defined over unitary matrices satisfying a maximum entropy principle, i.e. their probability measures are uniform in the appropriate parametrization sets. In the same fashion as with the Wigner-Dyson Gaussian ensembles there are three classes of circular ensembles, namely the Circular Unitary Ensemble (CUE), Circular Orthogonal Ensemble (COE) and Circular Symplectic Ensemble (CSE). These ensembles were originally developed by Dyson\cite{dyson1962a} as a more manageable alternative to the Gaussian ensembles but later a microscopic justification was given by Brouwer\cite{brouwer1995} relating them to the Gaussian ensembles; given a quantum dot whose Hamiltonian is taken from GUE, GOE or GSE, its scattering matrix belongs to CUE, COE or CSE, respectively.

The scattering matrix of a system without time-reversal symmetry is restricted only to be unitary. Since any unitary matrix $\bm S\in {\rm U}(N)$ can be written as $\bm S=\bm{UV}$, with $\bm U$ and $\bm V$ also unitary, we define a neighborhood $d\bm S$ in the set of unitary matrices as
\begin{equation}\label{cue-ds}
\bm S + d\bm S = \bm U(\bm 1 + i\, d\bm M)\bm V,
\end{equation}
where $d\bm M$ is an infinitesimal hermitian matrix with elements $dM_{ij} = dM_{ij}^{(\rm r)} + idM_{ij}^{(\rm i)}$ and we note that here $N$ is the total number of scattering channels in the system (for instance, in a two-leads waveguide with $M_1$ propagating channels on lead 1 and $M_2$ on lead 2, $N=M_1+M_2$). The real and imaginary parts of $d\bm M$ are allowed to vary independently over small intervals $d\mu_{ij}^{(\rm r)}$, $d\mu_{ij}^{(\rm i)}$, so the volume element measure is
\begin{equation}
\mu_{\rm U} (d\bm S) = \prod_{i\leq j}^N d\mu_{ij}^{(\rm r)} \, \prod_{i<j}^N d\mu_{ij}^{(\rm i)},
\end{equation}
independent of the choice of $\bm U$ and $\bm V$. Since we require a uniform probability density function $P_{\rm U}(\bm S)$ over the unitary group $U(N)$, 
\begin{equation}
P_{\rm U}(\bm S)d\bm S = \frac{1}{V_{{\rm U}(N)}} \mu_{\rm U}(d\bm S)
\end{equation}
with the normalization constant $V_{{\rm U}(N)}$ given by the measure of the complete group. Dyson defined the CUE ensemble by property \eref{cue-ds}, namely requiring an ensemble pdf invariant to every automorphism $\bm S \to \bm{U'SV'}$ with $\bm U', \bm V' \in{\rm U}(N)$, which implies that $\mu_{\rm U}$ is the Haar measure of the unitary group.

On the other hand, for a system possessing time-reversal invariance and spin-rotation symmetry, the scattering matrix must be a unitary symmetric matrix $\bm S$, which can always be written as 
\begin{equation}\label{coe-decomp}
\bm S = \bm U^T\bm U
\end{equation}
with $\bm U\in {\rm U}(N)$. Thus, in this case the infinitesimal neighborhood is defined by
\begin{equation}\label{coe-ds}
\bm S + d\bm S = \bm U^T(\bm 1 + i\,d\bm M)\bm U,
\end{equation}
where $d\bm M$ is a real symmetric matrix with elements $dM_{ij}$, $i\leq j$, varying independently in an small interval $d\mu_{ij}$. Then, the measure of this differential is
\begin{equation}\label{mu_rmt_orth}
\mu_{\rm O} (d\bm S) = \prod_{i\leq j}^N d\mu_{ij}.
\end{equation}
Hence, we define the COE probability density $P_{\rm O}(\bm S)$ to be uniform over the set of unitary symmetric matrices, this is
\begin{equation}
P_{\rm O}(\bm S) d\bm S = \frac{1}{V_{{\rm Us}(N)}} \mu_{\rm O}(d\bm S),
\end{equation}
where $V_{{\rm Us}(N)}$ is the normalization constant. Dyson proved that the COE ensemble is uniquely defined by property \eref{coe-ds}, this is by demanding an ensemble pdf invariant to every automorphism $\bm S \to \bm V^T \bm S \bm V$ with $\bm V\in{\rm U}(N)$. Note that the set of unitary symmetric matrices do not form a group, thus, in particular do not have an associated Haar measure. However, using decomposition \eref{coe-decomp} it is possible to obtain $\mu_{\rm O}$ as an induced measure from $\mu_{\rm U}$. 

Finally, for time-reversal invariant systems but with broken spin-rotation symmetry, the scattering matrix is a unitary self-dual quaternion matrix. Any such matrix can be decomposed as
\begin{equation}\label{cse-decomp}
\bm S = \bm U^R \bm U
\end{equation}
with $\bm U\in{\rm U}(N)$ and $\bm U^R$ its dual. In this case Dyson proved that the CSE ensemble is defined uniquely by the property of being invariant to every automorphism $\bm S \to \bm W^R \bm S\bm W$ where $\bm W$ is a unitary quaternion matrix. The CSE measure can be obtained in a similar way to $\mu_{\rm O}$, namely as the induced measure from $\mu_{\rm U}$ and decomposition \eref{cse-decomp}.

\subsection{Transmission eigenvalues distribution}

Since the circular ensembles pdfs are uniform, it is clear that when transforming to its diagonal form representation the only non-trivial term is the Jacobian $J$ related to the eigenvalue variables. Given a scattering matrix $\bm S$ from one of the circular ensembles, we can write $d\bm S = J\, \mu_{E}(d\bm U) \prod_{i} d\phi_i$, with $\phi_i\in[0,2\pi)$ the phase-shift (i.e. the eigenvalue phases), $\bm U$ its eigenvectors matrix (which is orthogonal, unitary or symplectic, depending on the ensemble considered) and $\mu_E$ the respective unique ensemble measure. The Jacobian,
\begin{equation}\label{circ-jac}
J = \prod_{l<j}^N |e^{i\phi_l} - e^{i\phi_j}|^\beta,
\end{equation}
was also computed by Dyson and is basically the joint eigenvalue pdf, $P(\{\phi_l\}) = c\, J$, with $c$ a normalization constant. The repulsion exponent $\beta$ in \eref{circ-jac} is the same as for the Gaussian case, namely $\beta=1,2,4$ for the Circular Unitary, Orthogonal and Symplectic Ensembles, respectively.

As we have seen in chapter \ref{chapter-waveguides}, the $\bm S$ matrix eigenstates mix incident and outgoing waves [see equation \eref{s-matrix-def}] and its eigenvalues are not directly related to transport properties. This is the reason why we have introduced the polar decomposition in order to relate the scattering matrix to the transmission eigenvalues $T_i$ [see section \eref{section-polar-decomp}].  Therefore, from a transport perspective, it is more natural to consider the ensemble density as a function of these variables; in other words, the Jacobian associated to this representation must be computed. The corresponding measure in this representation was obtained by Baranger and Mello\cite{mello1994} and is given by
\begin{equation}\label{P(S)-T}
P(\bm S)  = \prod_{i<j}^N |T_i - T_j|^\beta\, \prod_{k}^N T_k^{1 - \beta/2}\, \prod_{i} dT_i\, \prod_{n}^4 \mu(d \bm U_n)
\end{equation}
with $\bm U_n$ the four unitary matrices in \eref{s-matrix-polar} which in addition are symmetric or self-dual for COE or CSE, respectively.  

\subsection{Quantum dot and DMPK equation}\label{section-dmpk}

As we have mentioned in section \ref{section-connectio-chaotic}, the Circular Ensembles where first employed in the context of quantum transport in order to model disorder cavities (quantum dots) and disordered wires. In the case of quantum dots, the scattering matrix is directly taken from the appropriate ensemble depending on the desired symmetries of the system. Then, all transport information is encoded in $P(\bm S)$ given by \eref{P(S)-T}, in particular the conductance $g=\sum_i T_i$. Since the conductance can also be written as $g= \sum_{n,m}|t_{nm}|^2$, an important quantity is given by  $\langle |t_{nm}|^2 \rangle$ for the Circular ensembles. In fact, from the construction of the Circular ensembles is easy to show that the squared average of the scattering matrix elements\footnote{Here we assume a quantum dot connected to two identical leads with $N$ propagating modes. The general case with differents leads is also easy to obtain\cite{beenakker1997}.} is given by
\begin{equation}\label{S-matrix-circular-average}
\langle |S_{nm}|^2 \rangle = \left\{
\begin{array}{ll}
\displaystyle \frac{1}{2N} & \mbox{for CUE}   \\
& \\
\displaystyle\frac{1+\delta_{nm}}{1+2N}    &   \mbox{for COE} .  
\end{array}
\right.
\end{equation}
In the CUE case, transmission and reflexion probabilities are all equal, independent of the incoming and outgoing channel; this is the quantum analog to the Machta-Zwanzig assumption of an effective random walk dynamics [see section \ref{section-MZ-approx}]. On the other hand, \eref{S-matrix-circular-average} shows explicitly that in the COE ensemble scattering from mode $n$ back to mode $n$ is twice as probable as from mode $n$ to a different channel, which is a consequence of time-reversal symmetry. 

From \eref{S-matrix-circular-average}, we obtain the average conductance in a quantum dot with two identical contacts with $N$ channels,
\begin{equation}
\langle g \rangle = \frac{N}{2}\left[1 + \frac{1}{4}\left(1 - \frac{2}{\beta}\right) \right].
\end{equation}
This means that
\begin{equation}\label{qd-loc}
\langle g \rangle_{\rm COE} = \langle g \rangle_{\rm CUE} - \frac{1}{4},
\end{equation}
implying that the conductance in time-reversal symmetric systems is lower than when this symmetry is broken. This effect, called \emph{weak localization}, is explained semiclassically by the enhancement of reflection probability due to constructive interference of time-reversed trajectories \cite{book_datta}. Weak localization is a purely quantum effect; classical transmission and reflection probabilities off a quantum dot are equal [see section \ref{section-MZ-approx}].

Another well known universal feature observed in quantum dots are the \emph{universal conductance fluctuations}: for two equal contacts with $N$  channels the conductance variance is
\begin{equation}\label{qd-var}
\mbox{Var}[g] = \frac{1}{8}\beta^{-1},
\end{equation}
independent of $N$ and hence of $\langle g\rangle$. 

On the other hand, the standard model to describe a disordered wire is the DMPK equation. This is a partial differential equation for the transmission eigenvalues pdf as a function of the wire length. Its derivation starts from a finite wire segment with transmission spectrum $T_n$ which is composed with a new segment of infinitesimal length. Then, the compounded wire transmission matrix is obtained using [see equation \eref{t-r-comp}]
\begin{equation}
\bm t_2 = \bm t_1 (\bm 1 - \bm r_0 \bm r_1)^{-1}\bm t_0 ,
\end{equation}
where the index $i=0,1,2$ denote the $\bm S$ matrix components of the original, infinitesimal and resultant wires, respectively. It is assumed that the conductor is weakly disordered (with an order parameter given by $l/L_0\ll 1$, where $L_0$ and $l$ are the infinitesimal segment length and mean free path) so that we can write the resulting transmission eigenvalue perturbatively as $T_n + \delta T_n$. In order to obtain $\delta T_n$, a second order perturbative expansion of $\bm t_2\bm t_2^\dagger$ is performed. Then, the transmission and reflection matrices are written in polar form [see equation \eref{s-matrix-polar}] with the matrices $\bm U, \bm V, \bm U', \bm V'$ taken from the desired circular ensemble. This allows us to compute the first two moments of $\delta T_n$, which imply a Fokker-Planck equation for $T_n$ with the wire length playing the role of time parameter. For a detailed derivation of the DMPK equation see the book by Mello and Kumar \cite{book_mello}. 

The DMPK equation shows that in a disordered wire there are three regimes as the systems length $L$ is increased, namely a ballistic regime for $L/l\ll 1$, then a metallic regime for $1\ll L/l \ll N$ and finally a localized regime for $N\ll L/l$. In the metallic regime, the average conductance decays linearly with $L$ displaying ohmic scaling $\langle g \rangle = N/(s+1)$ with $s=L/l$. In addition, in this regime the conductance exhibits a phenomenology similar to the quantum dot, showing weak localization and universal conductance fluctuations with
\begin{align}
\delta g &= -\frac{1}{3} \\
\mbox{Var}[g] &= \frac{2}{15}\beta^{-1},
\end{align}
where $\delta g = \langle g \rangle_{\rm COE} - \langle g \rangle_{\rm CUE} $. Note, however, that this result differs slightly from the quantum dot [equations \eref{qd-loc} and \eref{qd-var}], which is a consequence of correlations in the DMPK equation solution being not geometric as for the quantum dot in \eref{P(S)-T}. 

The \emph{localized regime} is characterized by the exponential decay of the conductance $\langle g \rangle = \exp{(-L/2\xi)}$ with $\xi \sim  N \beta l$, which is also known as Anderson localization \cite{anderson1979}. This effect is again a consequence of wave interference but in this case not restricted to time-reversed trajectory pairs; multi-scattering in a long wire can produce strong inference among different ray paths causing the absence of transmission. In this regime, the conductance pdf is log-normal with $\mbox{Var}[\log{g}] = - 2 \langle \log{g}\rangle = 4L/\xi$.

\subsection{Periodic chain of quantum dots}\label{section-rmt-periodic-chain}

The periodic chain model we use in the next chapters is constructed by joining together a series of identical quantum dots defined by a scattering matrix taken from the CUE or COE ensembles. The scattering matrix for arbitrary chain length can be obtained using composition rules \eref{t-r-comp}, thus each quantum dot (circular) ensemble induces a periodic chain ensemble. Note that this ensemble of scattering matrices is different from the Circular ensembles because, as we have seen in section \ref{section-oseledet}, in the long chain limit transmission eigenvalues $T_i$ associated to non-propagating Bloch modes converge exponentially to zero whereas $T_i$ linked to propagating Bloch modes are $\mathcal{O}(1)$. This implies a distribution $P(\bm S)$ different to \eref{P(S)-T} and, in fact, dependent on the chain length. In chapter \ref{chapter-conductance-props}, we will study the conductance pdf of this model as a function of its length [see figure \ref{fig-p_g}] where some of the particular features of the periodic chain ensemble become evident.

\chapter{Number of propagating Bloch modes}\label{chapter-nb}
\chaptermark{Number of Bloch Modes}

As we have shown in \ref{section-landauer-conductance}, the number of propagating Bloch modes $N_B$ plays an important role for the transport properties of a long periodic chain. The asymptotic semiclassical behavior of $\langle N_B \rangle$ depends on the phase-space dynamics of the unit cell, growing linearly with the wavenumber $k$ in systems with a non-null measure of ballistic trajectories and going like $\sim \sqrt{D k}$ in diffusive systems (with $D$ the associated classical diffusion coefficient) as was shown by Faure\cite{faure2002}. We review this result in section \ref{nb-semiclassical-limit}, where we also present numerical evidence of its validity in a physically realistic waveguide. In addition, we show an alternative derivation of $\langle N_B\rangle$ using Simons and Altshuler's universal parametric autocorrelation function\cite{altshuler1993, simons1993}. Finally, in section \ref{section-nb-rmt} we obtain an exact expression for $\langle N_B\rangle$ in a RMT periodic chain using Faure's result and the Machta-Zwanzig approximation for the diffusion coefficient.

\section{\texorpdfstring{$N_B$}{NB} in the semiclassical limit $\hbar\to\infty$}\label{nb-semiclassical-limit}
\sectionmark{$N_B$ in the semiclassical limit}
\subsection{Faure's result}\label{section-faure-result}

We start reviewing the main result obtained by Faure in \cite{faure2002} for $\langle N_B\rangle$ in general Hamiltonian quasi-one-dimensional periodic systems and then obtain the particular result for a billiard. Let us consider the stationary Schrodinger equation in coordinates $(x,y)$,
\begin{equation}\label{schrod-eq}
\left( -\frac{\hbar^2\nabla^2}{2m} + V(x,y)\right)\psi(x,y) = E\,\psi(x,y) ,
\end{equation}
with $V(x,y)$ the potential assumed to be $Z$-periodic in $x$, i.e. $V(x+Z,y)=V(x,y)$, and constrictive in $y$, this is,
\begin{equation}
V(x,y\to\pm\infty) = \infty.
\end{equation}
As we have recalled in section \ref{waveguide-bloch-spec}, since the Hamiltonian is periodic we known from Bloch theorem that the solutions of \eref{schrod-eq} are given by $E_n(\theta)$, $\psi_{n,\,\theta}(x,y)$ with $n\in\mathbb{N}$ and $\theta\in[0,\,2\pi)$ such that $\psi_{n,\,\theta}(x+Z,y) = e^{i\theta }\psi_{n,\,\theta}(x,y)$. That is, the energy spectrum forms continuos bands parametrized by an integer index $n$ and the Bloch parameter $\theta$, and the system eigenfunctions are $Z$-periodic up to a phase $\theta=qZ$ with $q$ the Bloch quasi-momentum. Therefore, we argued that wave propagation is ballistic with velocity given by
\begin{equation*}\tag{\ref{vh_def2}}
v_{n}(\theta) = \frac{Z}{\hbar}\left.\frac{dE_n}{d\theta'}\right|_{\theta'=\theta},
\end{equation*}
which is proportional to the band slope. Since 
\begin{equation}
N= \int_{a}^{b} dx\, \delta(C-f(x)) \Theta[f'(x)] f'(x)
\end{equation}
is the number of points such that $f(x)=C$ with positive derivative in the interval $(a,b)$, Faure expressed the number of Bloch modes with energy $E$ propagating with positive velocity as
\begin{equation}
N_B(E)= \sum_n \int_{-\pi}^{\pi} d\theta\, \delta(E-E_{n}(\theta)) \Theta[v_{n}(\theta')] \left.\frac{dE_{n}}{d\theta}\right|_{\theta'=\theta},
\label{Nb_def2}
\end{equation}
where  $\delta(\cdot)$ is Dirac's delta function and $\Theta(\cdot)$ is Heaviside step function. Considering the distribution of quantum velocities $P_\hbar(v,E)$, defined as \cite{asch1998}
\begin{equation}
 \label{Ph_def}
P_\hbar(v,E) =h^2 \sum_n \int \frac{d\theta}{2\pi}\, \delta(v-v_{n}(\theta))\delta(E- E_{n}(\theta)) 
\end{equation}
with $v_{n}(\theta)$ defined in (\ref{vh_def2}), one obtains
\begin{equation}
 \label{exact}
 N_B(E)  = \frac{1}{h L} \int_0^\infty v P_\hbar(v,E) \,dv. 
\end{equation}
The normalization of $P_\hbar(v,E)$ over velocities $v$  is 
\begin{equation}
\int dv P_\hbar(v,E)=h^2 \int \frac{d\theta}{2\pi}\rho(E,\theta) ,
\label{normalizacion}
\end{equation}
with 
\begin{equation}\label{level-density-function}
\rho(E,\theta)= \sum_n \delta(E- E_{n}(\theta)) 
\end{equation}
the density of states with fixed Bloch parameter $\theta$. We note that (\ref{exact}) is an exact result, a particular case of the Kac-Rice formula \cite{wilkinson1999}, valid for quasi-one-dimensional periodic systems.

The number of propagating Bloch modes \eref{exact} is a rapidly fluctuating quantity as a function of $E$ at scales near the mean level spacing $\Delta E=1/\bar{\rho}_E$, where 
\begin{equation}\label{mean-density-of-states}
\bar{\rho}_E =\int \frac{d\theta}{2\pi} \langle \rho(E,\theta)\rangle_E
\end{equation}
is the mean density of states and the average $\langle\cdot\rangle_E$ is the semiclassical ensemble average defined in section \ref{section-connectio-chaotic}. The fluctuations of $N_B(E)$ are eliminated by averaging it over this ensemble. Hence, we now consider the average number of propagating Bloch modes,
\begin{equation}
\label{promedio}
 \langle N_B(E)\rangle  = \frac{1}{h Z} \int_0^\infty v \langle P_\hbar(v,E) \rangle \,dv.
\end{equation}
In what follows, we will drop the average label $E$ unless it is necessary. As follows from (\ref{normalizacion}), $\langle P_\hbar(v,E) \rangle$ is normalized to $h^2\bar{\rho}_E=\nu_E$ where $\nu_E$ is the Liouville measure of the constant energy surface $\Sigma_E = \{ (x,p_x,y,p_y) \,:\, 0\leq x\leq Z,\; H(x,p_x,y,p_y)=E \} $ (Weyl law). The averaged velocity distribution in the rhs of \eref{promedio} can be analyzed semiclassically to obtain $\langle N_B(E)\rangle$, as we show bellow.

We now turn to obtain the leading order term of the semiclassical expansion of $\langle N_B(E)\rangle$. Asch and  Knauf \cite{asch1998} proved the semiclassical convergence of $P_\hbar(v,E)$ towards the classical asymptotic mean velocities distribution $P_a(v,E)$, 
\begin{equation}
 \label{asch-knauf}
 P_\hbar(v,E) = P_a(v,E) \qquad \mbox{when}\quad
\hbar \rightarrow 0,
\end{equation}
that holds for test functions independent of $\hbar$ (i.e., over intervals of order $\delta v \sim\mathcal{O}(\hbar^0)$, $\delta E \sim\mathcal{O}(\hbar^0)$), a coarse graining consistent with the average taken in (\ref{promedio}). In the right-hand side of (\ref{asch-knauf}), $P_a(v,E)$ is the probability density of the asymptotic mean velocity
\begin{equation}
 \label{classical-asymptotic-velocity-dist}
 v_a=\lim_{t\rightarrow\infty} \frac{x(t)}{t},
\end{equation}
with $x(t)$ the longitudinal position at time $t$ of a particle in the waveguide whose initial condition was taken randomly with a uniform probability distribution on the surface $\Sigma_{E}$. When the classical dynamics is not purely ergodic, so there are some tori associated to ballistic trajectories in the stroboscopic Poincare section $\mathcal{S}_x$, the leading order term in the semiclassical limit of $N_B$ can be obtained just using equivalence (\ref{asch-knauf}) in the integrand of (\ref{promedio}). In this case, one obtains
\begin{equation}
\langle N_B(E)\rangle = \frac{1}{h} \mu_{\rm bal}(E) + o(\hbar^{-1}),
\end{equation}
where 
\begin{equation}\label{mu-bal}
\mu_{\rm bal}(E) = \frac{1}{Z} \int_0^\infty v \langle P_\hbar(v,E) \rangle \,dv
\end{equation}
can be identified \cite{faure2002} as the measure of ballistic trajectories in $\mathcal{S}_x$ [see equation \eref{sx-section}]. On the other hand, in case the dynamics is completely ergodic, $P_a(v,E)dv$ is a punctual measure at $v=0$ because, as implied by the CLT \eref{clt}, 
\begin{equation}
v_a=\lim_{t\to\infty}\frac{1}{\sqrt{t}} \frac{x(t)}{\sqrt{t}} = \lim_{t\to\infty} \frac{\chi}{\sqrt{t}}
\end{equation}
with $\chi$ a Gaussian random variable with null mean. Hence, $v_a=0$ almost surely. In order to obtain the leading semiclassical contribution in $\langle N_B(E)\rangle$, it is necessary to quantify more precisely how the quantum level velocity distribution
approaches the classical asymptotic velocity distribution as $\hbar \to 0$, that is, we need a refinement of (\ref{asch-knauf}).

As discussed in section \ref{corr-decay-and-diff}, when the classical dynamics is fully ergodic and the mixing rate is rapid enough, we expect that (almost) any initial ensemble of particles in the system will relax by diffusion, with a Gaussian density profile spreading according to $\langle x(t)^2 \rangle = D_E t $ for large times, with $D_E$ the diffusion coefficient for $\Sigma_{E}$. Hence, the mean velocity distribution at time $t$ is given by
\begin{equation}
 \label{P_t}
 P_t(v,E) = \frac{\nu_E}{\sqrt{2\pi}\bar{\sigma}_t} \exp{\left(-\frac{v^2}{2\bar{\sigma}_t^2}\right)}
\end{equation}
where $\bar{\sigma}_t^2 = D_E/t$. In the limit $t \rightarrow\infty$, $P_t(v,E)$ converges to a punctual measure at $v=0$. 
It has been shown \cite{eckhardt1995}, using the Gutzwiller trace formula and also with RMT arguments, that in systems with hyperbolic classical dynamics the variance 
of a quantum operator in the semiclassical limit is equal to the variance of the associated classical observable at time $t=t_H/g$, with $t_H =h/\triangle E=\nu_E/h$ the Heisenberg time of the unit cell and $g$ a factor depending on the anti-unitary symmetries of the system. For the billiard we study in section \ref{section-cosine-billiard} the factor is $g=2$ due to the transversal mirror  symmetry of the unit cell \cite{berry1986} [see the discussion in section \ref{section-nb-universal-correlator}]. Then, we have a relation between the variance $\sigma_\hbar^2=\langle v_{n}(\theta)^2\rangle$ of the quantum level velocity and the variance of the classical asymptotic mean velocity $\bar{\sigma}_t$, namely
\begin{equation}
 \label{eckhardt_equiv}
 \sigma_\hbar^2 = \frac{g D_E}{t_H} =\bar{\sigma}_{t_H/g}^2 \quad \mbox{for}\quad \hbar\rightarrow 0. 
\end{equation}
This relation between level velocity variance and the classical diffusion coefficient $D_E$ was studied in detail for the kicked rotor \cite{lakshminarayan1999} where the diffusion coefficient displays a nontrivial behavior as a parameter is varied. In a different context, \cite{barra2001} explores the relation between classical diffusion coefficient and quantum evolution properties.

Inspired by (\ref{eckhardt_equiv}), Faure conjectured a higher order semiclassical equivalence between 
the quantum velocity distribution and the classical mean velocity distribution at time $t=t_H/g$,
\begin{equation}
 \label{faure_conjecture}
   P_\hbar(v,E) = P_{t=t_H/g}(v,E) \quad \mbox{when}\quad  \hbar \rightarrow 0 \, ,
\end{equation}
over widths of order $\delta E\sim\mathcal{O}(\hbar^0)$, $\,\delta v\sim\mathcal{O}(\sqrt{\hbar})$. This conjecture is based on result (\ref{eckhardt_equiv}) and the Quantum Ergodicity Theory \cite{backer1998}. 
It can also be seen as a consequence of the universality in the parametric variation of energy levels discovered by Simons and Altshuler \cite{altshuler1993}, as we will discuss in section \ref{section-nb-universal-correlator}. Using (\ref{faure_conjecture}) in the integral of (\ref{promedio}) we have, for a purely diffusive waveguide, 
\begin{equation}
 \label{nb_k_diff}
 \langle N_B(E) \rangle= \frac{1}{\sqrt{h}}\sqrt{\frac{g\,\nu_E D_E}{2\pi Z^2}} + o(\sqrt{1/h}).
\end{equation}
Equation (\ref{nb_k_diff}) was derived in \cite{faure2002} --without the anti-unitary symmetry factor $g$-- and checked numerically for the kicked Harper model.

\subsection{\texorpdfstring{$\langle N_B\rangle$}{<NB>} in a periodic diffusive billiard}\label{section-nb-billiards}

Billiards where presented in chapter \ref{chapter-billiards}; they are a particular class of Hamiltonian systems defined by free point particles which collide elastically with the billiard (waveguide) walls. Therefore, they have the simplifying property of equivalent phase-space dynamics on every energy shell except for a time scale change. For two-dimensional billiards, $\nu_E=2\pi A_c$ depends only on $A_c$, the area of the unit cell. Hence, the semiclassical limit $\hbar \to 0$ is equivalent to $k \to \infty$ due to the relation $\hbar^2k^2=2 E$ [see \eref{helmholtz}]. In $\Sigma_E$, the asymptotic mean speed $v_a$ takes values in the interval $[0,\sqrt{2E}]$. Considering the change of variables $v=\sqrt{2E}u$ we have from \eref{promedio} and \eref{asch-knauf},
\begin{equation}
\langle N_B(E)\rangle = \frac{\sqrt{2E}}{h Z} \int_0^1 u P_a(u) \,du 
\end{equation}
with $P_a(u)$ the classical asymptotic mean velocity distribution in $\Sigma_{1/2}$ where the particles move with constant unit speed $v=1$, after setting their mass $m=1$. 
Since $\sqrt{2E}=\hbar k$,
\begin{equation}
 \label{nb_quant}
\langle N_B(k) \rangle = \frac{k}{2\pi Z} \int_0^1 v P_a(v) \,dv.
\end{equation}
Thus, for ballistic waveguides, i.e. when $P_a(v)$ is not a degenerate punctual measure at $v=0$, the average number of propagating Bloch modes goes asymptotically for $k\to\infty$ as $\langle N_B(k)\rangle \sim \mu_{bal} k/2\pi  + o(k)$ where $\mu_{bal}$ is defined in \eref{mu-bal}. 

For diffusive billiard chains, the diffusion coefficient satisfies $D_E=|v|D_1$ with $|v|= \sqrt{2E}$ and $D_1$ the diffusion coefficient in the energy shell $\Sigma_{1/2}$. Then, using $|v|=\hbar k$ and $\nu_E=2\pi A_c$ in (\ref{nb_k_diff}), we obtain that the average number of propagating modes in a diffusive waveguide is
\begin{equation}
 \label{nb_kk_diff}
 \langle N_B(k) \rangle= \frac{\sqrt{g\,A_c D_1}}{\sqrt{2\pi} Z} \sqrt{k} + o(\sqrt{k}) \quad \mbox{for}\quad k\to\infty.
\end{equation}

\subsection{Connection to the universal parametric correlation function}\label{section-nb-universal-correlator}

In section \ref{section-faure-result} we closely followed Faure's derivation of the semiclassical expression for $\langle N_B\rangle$ given by (\ref{nb_k_diff}). Here, we provide an alternative argument allowing us to better understand the connection of this result with \cite{altshuler1993,dittrich1997,dittrich1998}. We define the probability distribution of quantum level velocities used in section \ref{section-faure-result} in natural units $w$,
\begin{equation}
p(w,E)=\frac{1}{2\pi}\sum_n \int_{-\pi}^{\pi} d\theta\, \delta[ E-E_{n}(\theta)]\, \delta\hspace{-3pt}\left(w-\left.\frac{dE_n}{d\theta'}\right|_{\theta'=\theta}\right),
\end{equation}
so that 
\begin{equation}
     N_B(E) = 2\pi \int_0^\infty w p(w,E)\,dw.
     \label{26}
\end{equation}
Then, as was noted in \cite{zirnbauer1992}, the distribution of level velocities satisfies $p(w,E)=\lim_{\phi \to 0}\phi K(w\phi,\phi)$ with $K(\Omega,\phi)$
the autocorrelation function
\begin{equation}
K(\Omega,\phi)=\langle \rho(E+\Omega,\theta+\phi)\rho(E,\theta)\rangle_{E,\theta}-\bar{\rho}_E^2
\end{equation}
of the energy density $\rho(E,\theta)$ defined in \eref{level-density-function}. We recall that the average is over $\theta$ and over a classically small energy interval around $E$ (the semiclassical ensemble). According to Simons and Altshuler \cite{altshuler1993}, 
\begin{equation}
k(\omega,\chi)=K(\Omega,\phi)/\bar{\rho}_E^2
\end{equation}
is a universal function with the scaling 
\begin{eqnarray}
\omega &=&\bar{\rho}_E\,\Omega,\\
\chi &=& \sqrt{C(0)}\phi \quad\mbox{and}\\
C(0) &=& \bar{\rho}_E^2\left\langle \left(\partial_\theta E_{n}(\theta)\right)^2\right\rangle.\label{c(0)}
\end{eqnarray}
They showed this result for disordered systems and verified numerically its validity for chaotic systems in the semiclassical limit. It follows from their analytical expression for $k(\omega,\chi)$ that the probability distribution of level velocities $p(w,E)$ is a Gaussian with variance $C(0)/\bar{\rho}_E^2$. Thus, from (\ref{26}), we obtain  
\begin{equation}\label{nb-c(0)}
N_B(E)=\sqrt{2\pi C(0)},
\end{equation}
where $C(0)$, defined in \eref{c(0)}, is system dependent and must be computed, for instance, semiclassically. In our case, $C(0)$ is simply related to the average of the squared quantum velocity, 
\begin{equation}\label{c(0)-periodic-system}
C(0)=\frac{\bar{\rho}_E^2\hbar^2}{L^2}\left\langle v_{n}(\theta)^2\right\rangle
\end{equation}
that is given in (\ref{eckhardt_equiv}). Then, from \eref{nb-c(0)} and \eref{c(0)-periodic-system}, we obtain the dominant term in Faure's expression \eref{nb_k_diff} for $\langle N_B(E)\rangle$. Hence, equation (\ref{nb_k_diff}) can be seen as a consequence of the parametric level correlations universalitiy. The validity of this universal correlation function for periodic systems was studied by Dittrich et al. \cite{dittrich1998} in the semiclassical limit. They computed $K(\Omega,\phi)$ by Fourier transforming their semiclassical expression for the form factor identifying a $C(0)$ that agrees with the previous result.

As a last remark, we note that the universal correlation function $k(\omega,\chi)$ is different for GOE and GUE systems. For periodic systems, where $\chi$ plays the role of the Bloch parameter, time reversal symmetry $T$ is broken for all $\chi$ besides three exceptional points \cite{dittrich1997}. Thus, when no further anti-unitary symmetries exist, $k(\omega,\chi)$ moves between GOE and GUE at different parts of the band. For the periodic cosine billiard we consider in our numerical calculations, the unit cell is invariant under the operator $S_x$ that transforms $x \to -x$ (mirror reflection along the transversal direction). So, even though the unit-cell Hamiltonian do not commute neither with $S_x$ nor $T$, it does with the operator $T S_x$ which is anti-unitary. Then, as $\chi$ is varied, the system always belongs to GOE \cite{berry1986}. The presence of this anti-unitary symmetry explains the factor $g=2$ in \eref{nb_k_diff} and \eref{nb_kk_diff}, which enhance $\langle N_B\rangle$ by a factor $\sqrt{2}$.

\subsection{Numerical results in a periodic cosine waveguide}\label{section-nb-cos-billiard}

We now study numerically the semiclassical behavior of $\langle N_B(k)\rangle$ in the cosine billiard chain. In order to compute the transmission and reflection matrices we employ the admittance multi-modal method described in \cite{pagneux2010}. For each value of $k$, the field is described using the local waveguide transverse modes truncated to a finite number of channels, as described in chapter \ref{chapter-waveguides}. The number of propagating modes $N_B(k)$ is obtained by counting the solutions of the unit cell transfer matrix eigenvalue problem
\begin{equation*}\tag{\ref{m-bloch-problem}}
\bm{M}_1 \psi = \lambda \psi,
\end{equation*}
with $|\lambda| = 1$ and positive group velocity. As we have discussed in this chapter, $N_B(k)$ fluctuates very fast as a function of $k$ at scales $\Delta k = 2\pi/(kA_c)$ \footnote{$\Delta k$ follows from the Weyl law $N=k^2 A_c/(4\pi)$ }. We consider an average of the form 
\begin{equation}
 \label{nb_av}
 \langle N_B(k) \rangle_r \equiv \int N_B(k') f_r(k-k')\, dk' ,
\end{equation}
where $f_r(k)$ is a positive function normalized to one with compact support in an interval of length $r \Delta k$ with $r\gg 1$. The dependence of $\langle N_B(k) \rangle_r$ in the width of $f_r(k)$ is discussed at the end of this section but for what follows we drop explicit reference to $r$. 

\begin{figure}
 \centering
\includegraphics[width=.95\columnwidth]{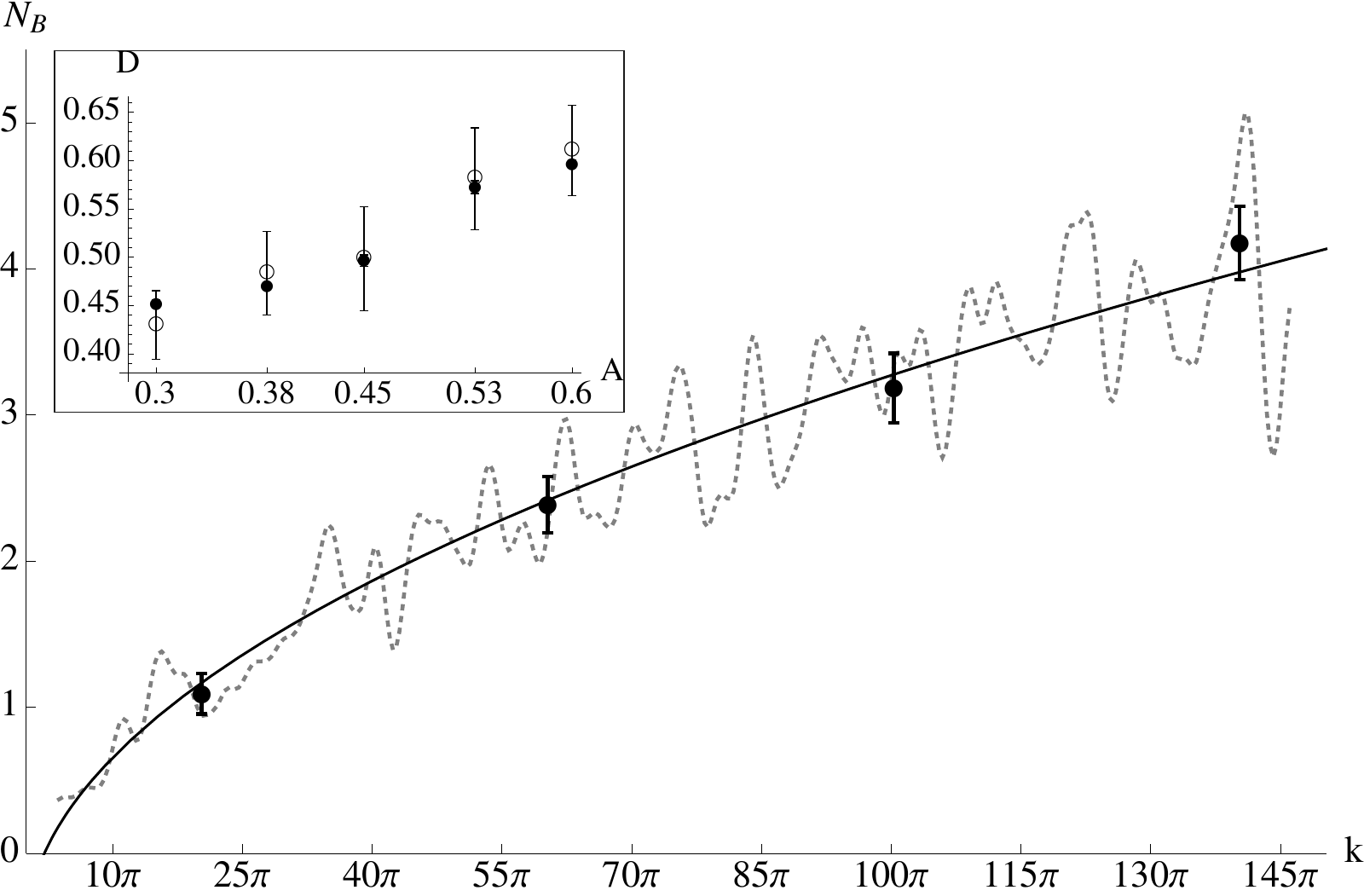}
 \caption[$\langle N_B(k)\rangle$ in a diffusive periodic cosine billiard]{The dashed line is the Gaussian smoothed moving average (with standard deviation $\sim \pi$) of the number of propagating Bloch modes $N_B(k)$ in the cosine billiard chain with $A_1=1$ and $A_2=0.3$ for a uniform $k$ sampling with increments $\delta k =0.37147\pi$. The expected semiclassical result for $\langle N_B(k)\rangle$ (full line) shows good agreement with the moving average. The four black points (with their associated standard error) were computed for the same geometry using a uniform sampling of 153 wavenumbers in an interval $300$ mean level spacings wide [see figures \ref{fig-bb_fat_points_convergence} and  \ref{fig-nb-k-fat-points}], which constitutes better statistics than the moving average (a sampling at least 15 times more dense). In the inset, the diffusion coefficient calculated classically (black dots) and the one obtained from the best fit following (\ref{nb_kk_diff}) for the quantum computation of $N_B(k)$ (circles) for five values of $A_2$ is shown. In all cases, a good agreement with the expected semiclassical behaviour given by (\ref{nb_kk_diff}) is observed.}
\label{fig-nb_k_0.30} 
\end{figure}

In figure \ref{fig-nb_k_0.30}, we show $\langle N_B(k)\rangle$ for $k\in [\pi, 150\pi]$, for a cosine billiard with $A_1=1$ and $A_2=0.3$ [see section \ref{section-cosine-billiard} for the cosine billiard definition]. As we have discussed in section \ref{section-cosine-billiard}, for this billiard configuration the phase-space does not exhibit noticeable tori and its dynamic is strongly diffusive. In the $k$ interval explored, the number of open transverse modes in plane leads ranges from 1 up to 150. For all wavenumbers, 100 evanescent modes where used in the numerical calculation. The dashed line shows a Gaussian smoothed moving average of $N_B(k)$ with variance $\pi^2$ for a uniform sampling of wavenumbers with spacing $\delta k \sim 0.37\pi$, which is bigger than the mean level spacing in the whole $k$ range. The full line is the expected semiclassical result (\ref{nb_kk_diff}) adjusted with an additive constant. This constant appears because in the system under study $\langle N_B(k^*) \rangle= 0$ for some $k^*>0$ and we are not sufficiently deep in the semiclassical regime for its relative value to be negligible. $N_B(k)$ fluctuates very rapidly with $k$, as expected, but the agreement between the moving average and the semiclassically predicted curve is quite good in the whole wavenumber interval explored.  We observe similar good results for all other configurations of the unit cell discussed in section \ref{section-cosine-billiard}, even when there are some small stable tori in the $\mathcal{S}_y$ Poincare section. This is not surprising because the classical dynamics in those systems is also strongly mixing, and for waves to resolve such small structures wavenumbers of order $k\gtrsim 2000 \pi$ would be needed, a fairly large value compared to the range considered. In the inset of Fig.  \ref{fig-nb_k_0.30}, we show the diffusion coefficient calculated classically (from Montecarlo simulations) and also the one obtained from the best fit following (\ref{nb_kk_diff}) for the quantum computation of $N_B(k)$ for five values of $A_2$. 

\begin{figure}
 \centering 
 \includegraphics[width=1\columnwidth]{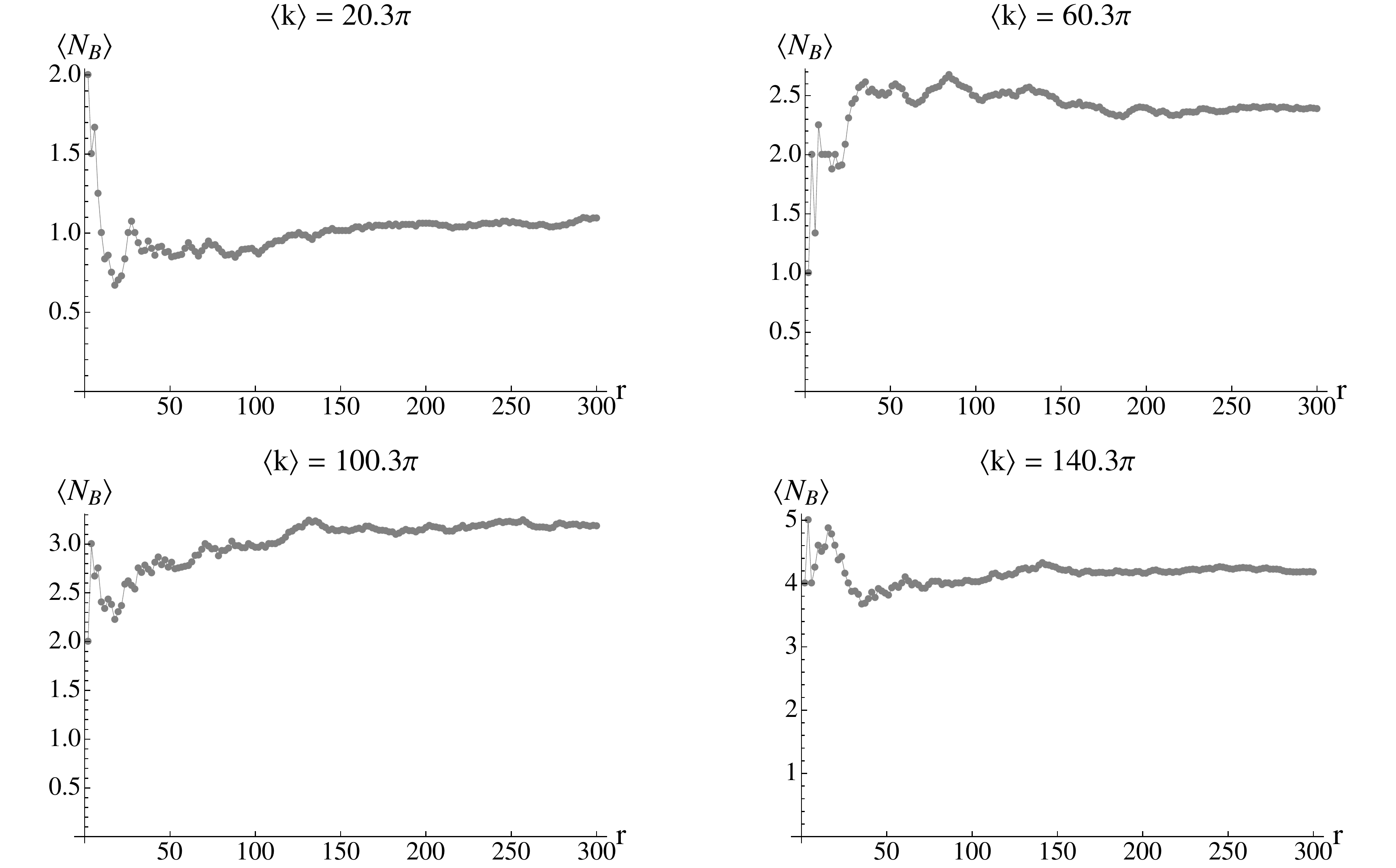}
 \caption[Convergence of the semiclassical ensemble for $\langle N_B\rangle$]{$\langle N_B(k)\rangle_r$ as a function of $r$ for four energies in cosine billiard with $A_1=1$ and $A_2=0.3$. A uniform sampling of 153 wavenumbers distributed in a 300 mean-level-spacing wide interval was used. In all cases after $r \sim 200 $ the average fluctuations are notably reduced and the result for $\langle N_B(k)\rangle$ is in agreement with the expected semiclassical result [see figure \ref{fig-nb_k_0.30}] indicating that this is a good smoothing window to eliminate the spectrum fluctuations.}
 \label{fig-bb_fat_points_convergence} 
\end{figure}

As we mentioned in the definition of the average number of propagating Bloch modes (\ref{nb_av}), the width of the interval $r\Delta k$ over which we take the average is arbitrary and it is only assumed to be much larger than the mean level spacing $\Delta k$ and classically small. In order to find out how relevant this choice is, we calculate $\langle N_B(k)\rangle_r$ for four different values of $k$ as a function of $r\in [2, 300]$ using a uniform $k$ sampling of 153 points around the four wavenumbers. In figure \ref{fig-bb_fat_points_convergence} we see that, after a transient phase,  at $r\sim 200$ the value of the average reaches an approximately stable region in the four cases.  The four black dots in figure \ref{fig-nb_k_0.30} show $\langle N_B(k)\rangle_r$ for $r=300$ and the $k$ values of figure \ref{fig-bb_fat_points_convergence}, displaying better agreement to the theoretical curve since this average takes more semiclassical realizations.

We end this section remarking how deep into the semiclassical limit the presented results are. In our calculations $\lambda/h_0$ --the wavelength in units of the waveguide width-- is as small as $0.012$.  We note that the calculations in \cite{dittrich1997}, which are also compared with semiclassical results for periodic billiards, were computed with the parameter $\lambda/h_0 \sim 0.12$.

\begin{figure}[h]
 \centering
 \includegraphics[width=1\columnwidth]{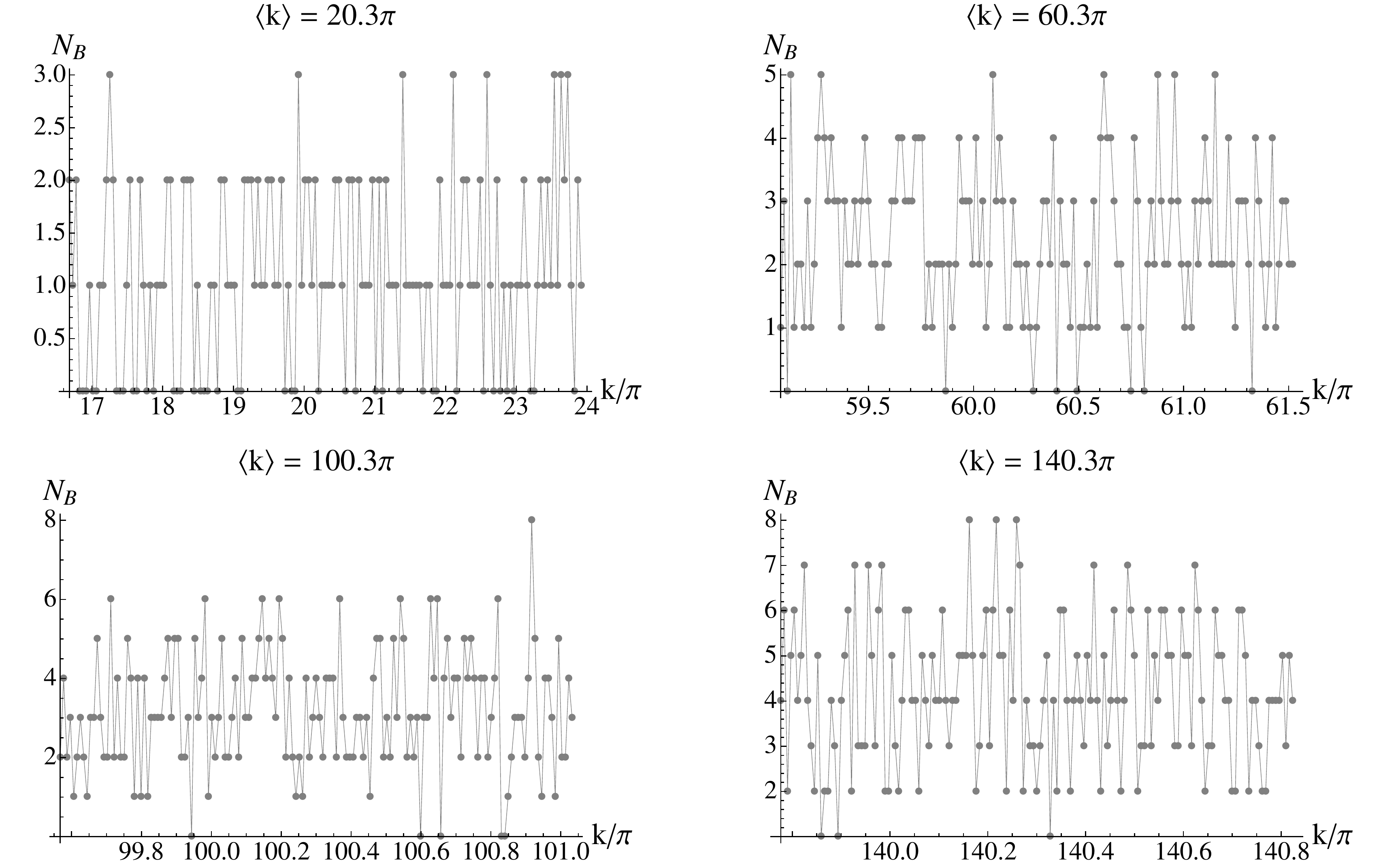}
 \caption[$N_B(k)$ in a diffusive periodic cosine billiard]{Number of propagating Bloch modes $N_B$ as a function of $k$ for the cosine billiard with $A_1=1$ and $A_2=0.3$ for the four \emph{fat points} shown in figure \ref{fig-nb_k_0.30}.  Each fat point is the average of the 153 uniformly distributed wavenumbers $k$ shown in the four plots.}
 \label{fig-nb-k-fat-points} 
\end{figure}

\section{Random matrix periodic chain model}\label{section-nb-rmt}

\subsection{\texorpdfstring{$\langle N_B\rangle$}{<NB>} for a random matrix periodic chain}\label{subsection-nb-rmt}

A periodic waveguide random matrix model was introduced in chapter \ref{chapter-rmt}. It consisted of a periodic chain of strongly scattering cavities whose $\bm S$ matrix is taken from the COE or CUE ensembles, depending on the desired symmetry of the system. In this section, we obtain an expression for $\langle N_B\rangle$ in such systems by means of Faure's result \eref{nb_kk_diff} and the Machta-Zwanzig diffusion coefficient approximation \eref{machta-zwanzig}. 

In chapter \ref{chapter-rmt}, we discussed how the circular ensembles, COE and CUE, characterize the statistical properties observed in a quantum dot scattering matrix. We defined a quantum dot as a coherent strongly scatterering cavity, with chaotic classical dynamics and such that its mean residence time $\tau$ is much larger than its ergodic time. From these properties follow that the $\bm S$ matrix is insensitive to the microscopic details of the quantum dot [see equation \eref{S-matrix-circular-average}] and hence implies universality in the sense of RMT. Not by coincidence, as we argued in section \ref{section-dmpk}, the conditions under which the Machta-Zwanzig diffusion coefficient approximation \eref{machta-zwanzig} holds are the same and thence it gives the universal diffusion coefficient for (the ray dynamics) in a chain of quantum dots, which depends only on the unit cell perimeter and openings length. 

\begin{figure}
\includegraphics[width=1\columnwidth]{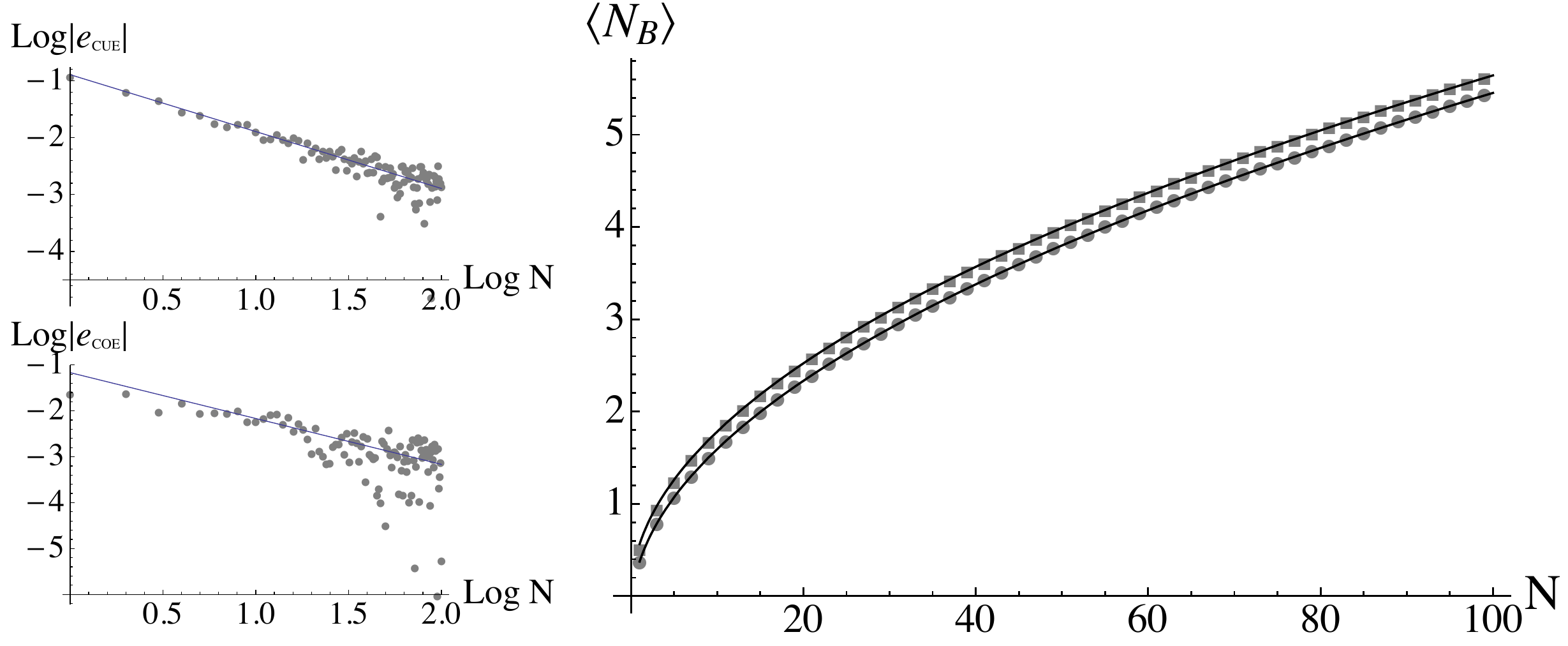}
 \caption[Theoretical and numerical result for $N_B(N)$ in RMT periodic chain models]{(Right) Number of propagating Bloch modes $\langle N_B\rangle$ calculated numerically for the COE (circles) and CUE (squares) periodic chain model; the black lines show the theoretical result given by \eref{nb-rmt}. (Left) The circles show the $\log_{10}$ of the relative error $e=\left| 1 - \langle N_B\rangle^{\rm (numeric)}/\langle N_B\rangle^{\rm (theoric)}\right| $ as a function of $\log_{10}(N)$ for both ensembles and the line, which have a slope $-1$, indicates that $\alpha=0.5$ [see equation \eref{nb-rmt}].}
 \label{FIG-NB_N-RMT} 
\end{figure}

Let the quantum dot periodic chain be defined as in section \ref{section-rmt-periodic-chain}, with cavities of area $A_c$ and opening of width $h_0$. Then, its Machta-Zwanzig diffusion coefficient \eref{machta-zwanzig} is given by
\begin{equation}\label{MZ-quantum-dot}
D_{\rm MZ} = \frac{2 Z^2 h_0}{\pi A_c}.
\end{equation}
We know that for a periodic chain of chaotic cavities in the semiclassical limit, the average number of propagating Bloch modes $\langle N_B\rangle$ is given by \eref{nb_kk_diff}. Hence, inserting \eref{MZ-quantum-dot} into this expression we obtain
\begin{equation}
\langle N_B \rangle = \sqrt{\frac{h_0 k}{\pi^2}}.
\end{equation}
Recalling that the number of open modes in the leads is given by 
\begin{equation*}\tag{\ref{N-open-modes}}
N = \lfloor h_0k/\pi \rfloor, 
\end{equation*}
we finally have the average number of propagating Bloch modes in a periodic quantum dot chain, 
\begin{equation}\label{nb-rmt-1}
\langle N_B \rangle = \sqrt{\frac{N}{\pi}} + o(\sqrt{N}) \quad \mbox{for $N\to\infty$}.
\end{equation}
In figure \ref{FIG-NB_N-RMT}, we compare this result with montecarlo calculations for the COE and CUE models showing perfect agreement in the CUE case and a constant additive correction $-0.2$ for the COE case. The latter can be attributed to a weak-localization effect, as we will argue in chapter \ref{chapter-conductance-props}. We also studied the next order correction to \eref{nb-rmt-1}, which turned out to vanish as $\sim N^{-1/2}$ for large $N$ for COE and CUE as is also shown in figure \ref{FIG-NB_N-RMT}. Therefore, we have that
\begin{equation}\label{nb-rmt}
\langle N_B \rangle = 
\left\{
\begin{array}{ll}
\displaystyle  \sqrt{\frac{N}{\pi}} - 0.2 + \mathcal{O}(N^{-1/2}) &\quad \mbox{for COE}  \\
\\
\displaystyle  \sqrt{\frac{N}{\pi}} +  \mathcal{O}(N^{-1/2})  &\quad \mbox{for CUE} . 
\end{array}
\right.
\end{equation}

\subsection{Characterization of $N_B$ distribution}\label{section-p-nb-dist}

As we have seen in the previous sections, once the semiclassical ensemble or RMT model is introduced, $N_B$ can be considered as a random variable taking integer values for each realization of the waveguide. We now study numerically some aspects of its probability density function, $P_{N_B}(n)$, using the RMT periodic chain ensemble. 

\begin{figure}[t]
 \centering
 \includegraphics[width=0.49\columnwidth]{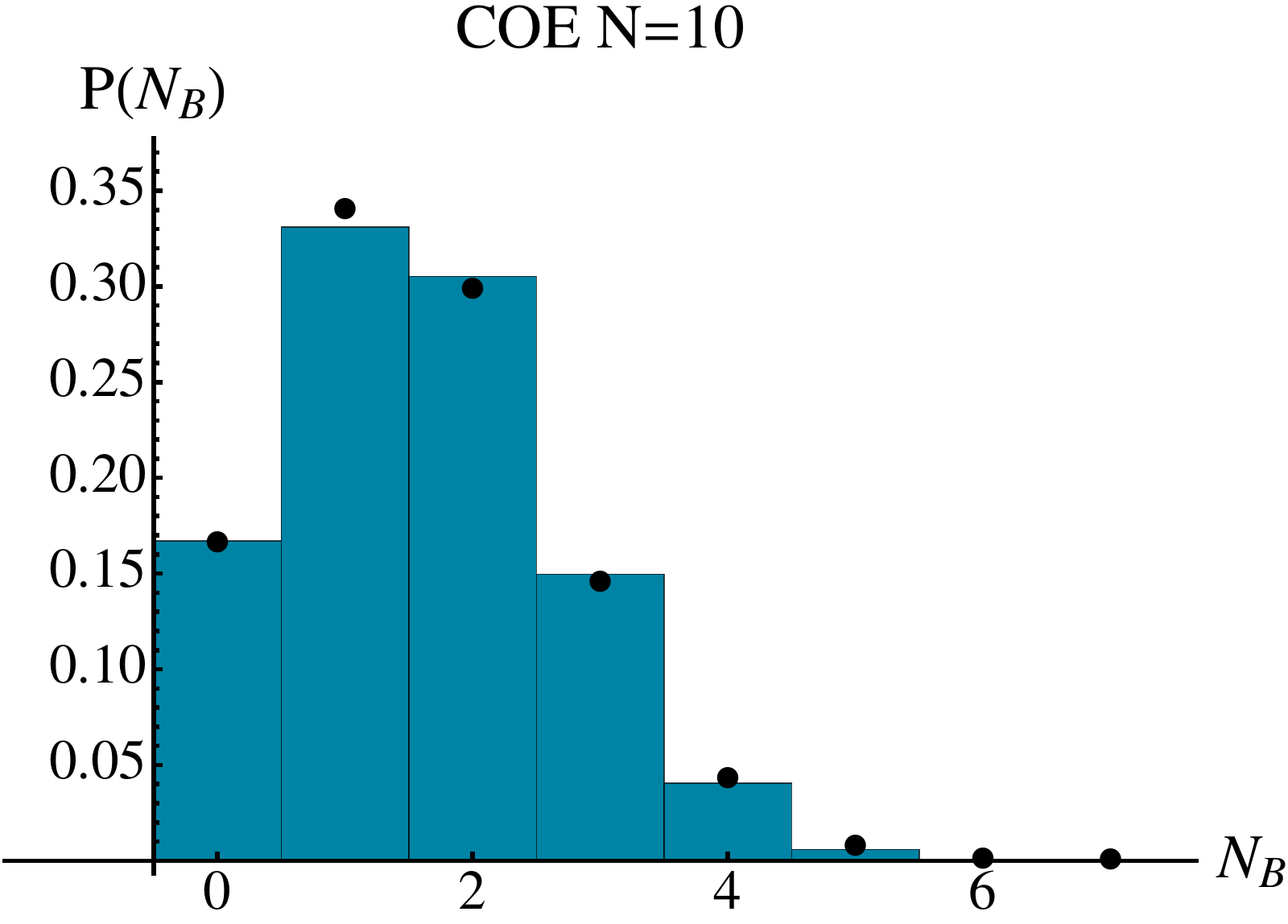}
 \includegraphics[width=0.49\columnwidth]{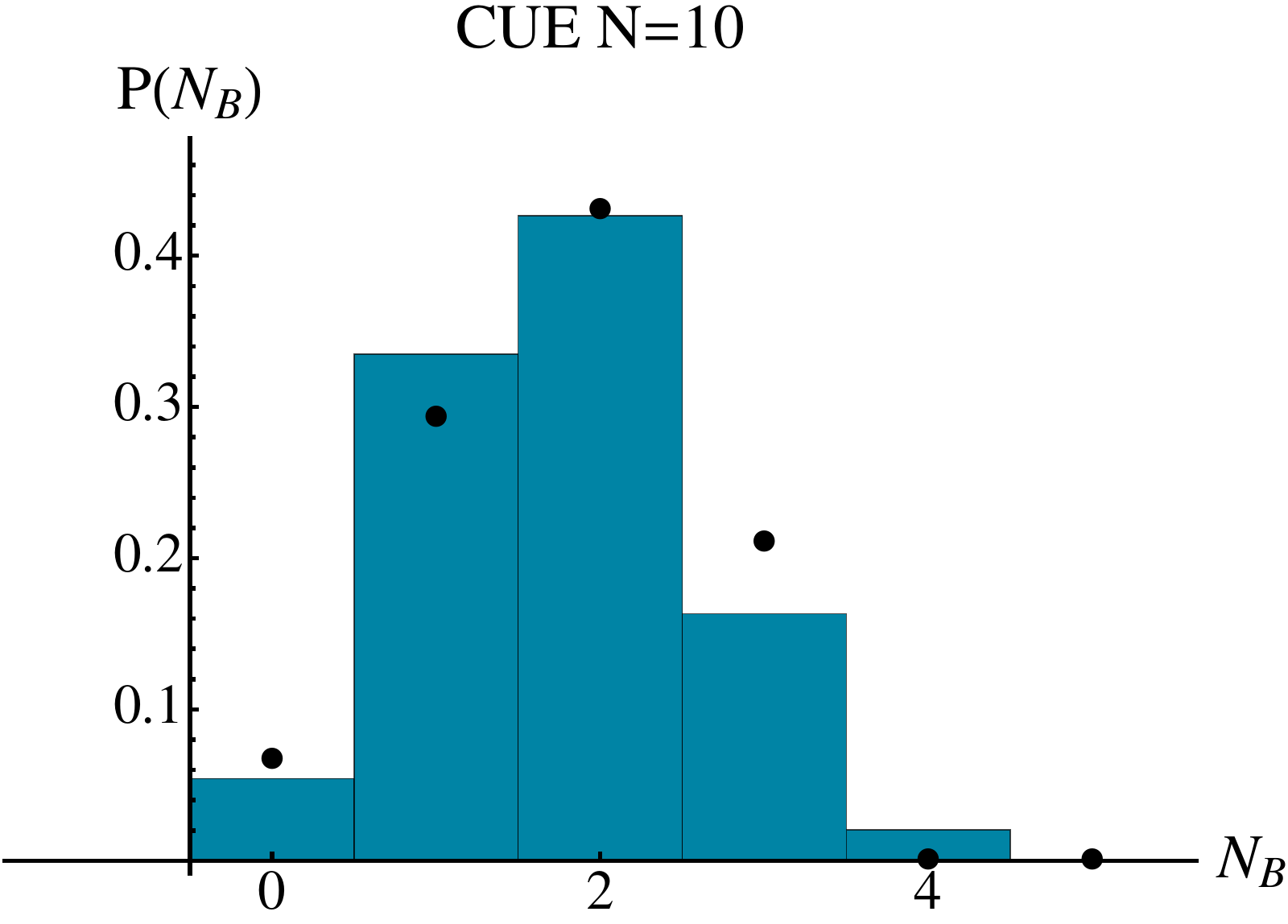}\\
 \vspace{0.5cm}
 \includegraphics[width=0.49\columnwidth]{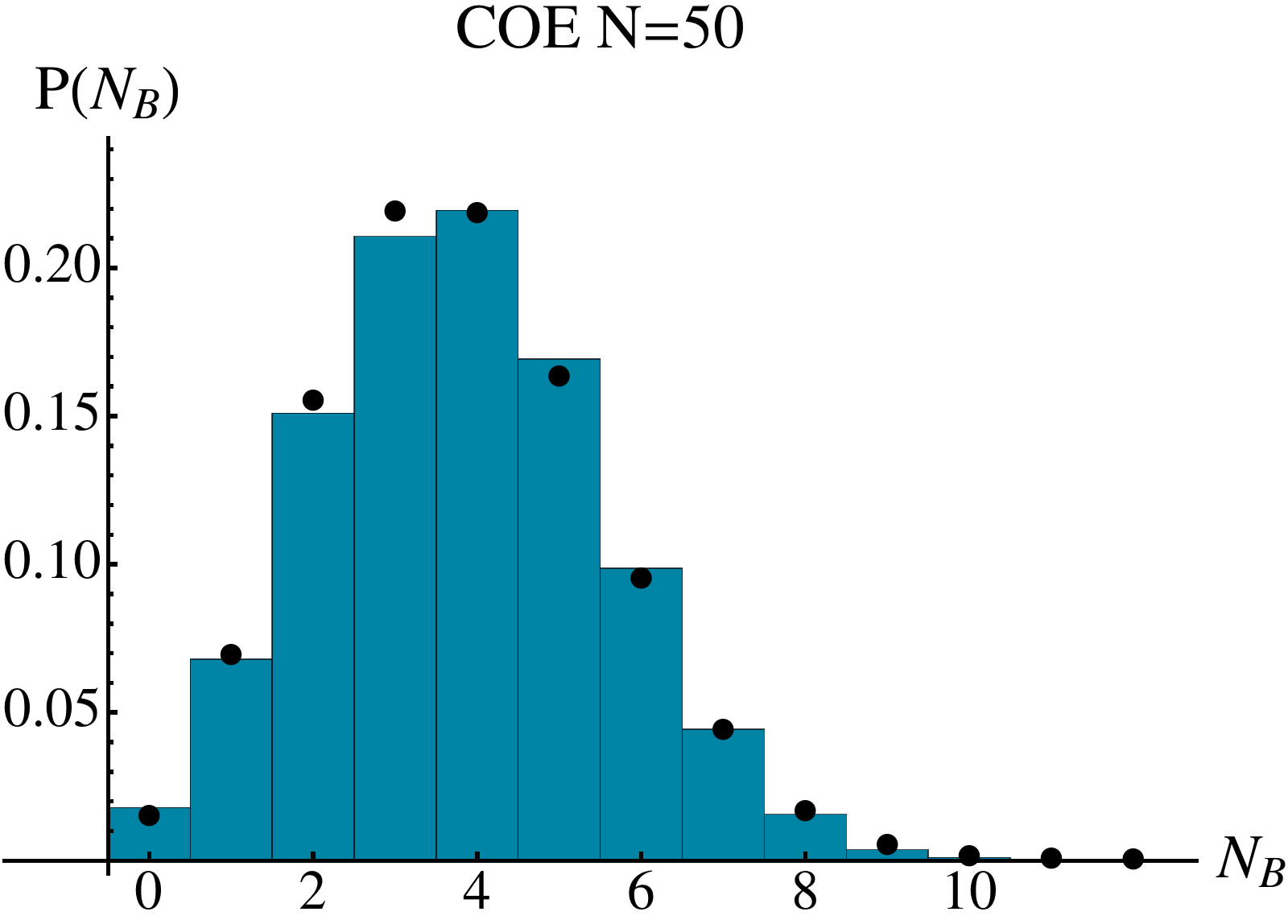}
 \includegraphics[width=0.49\columnwidth]{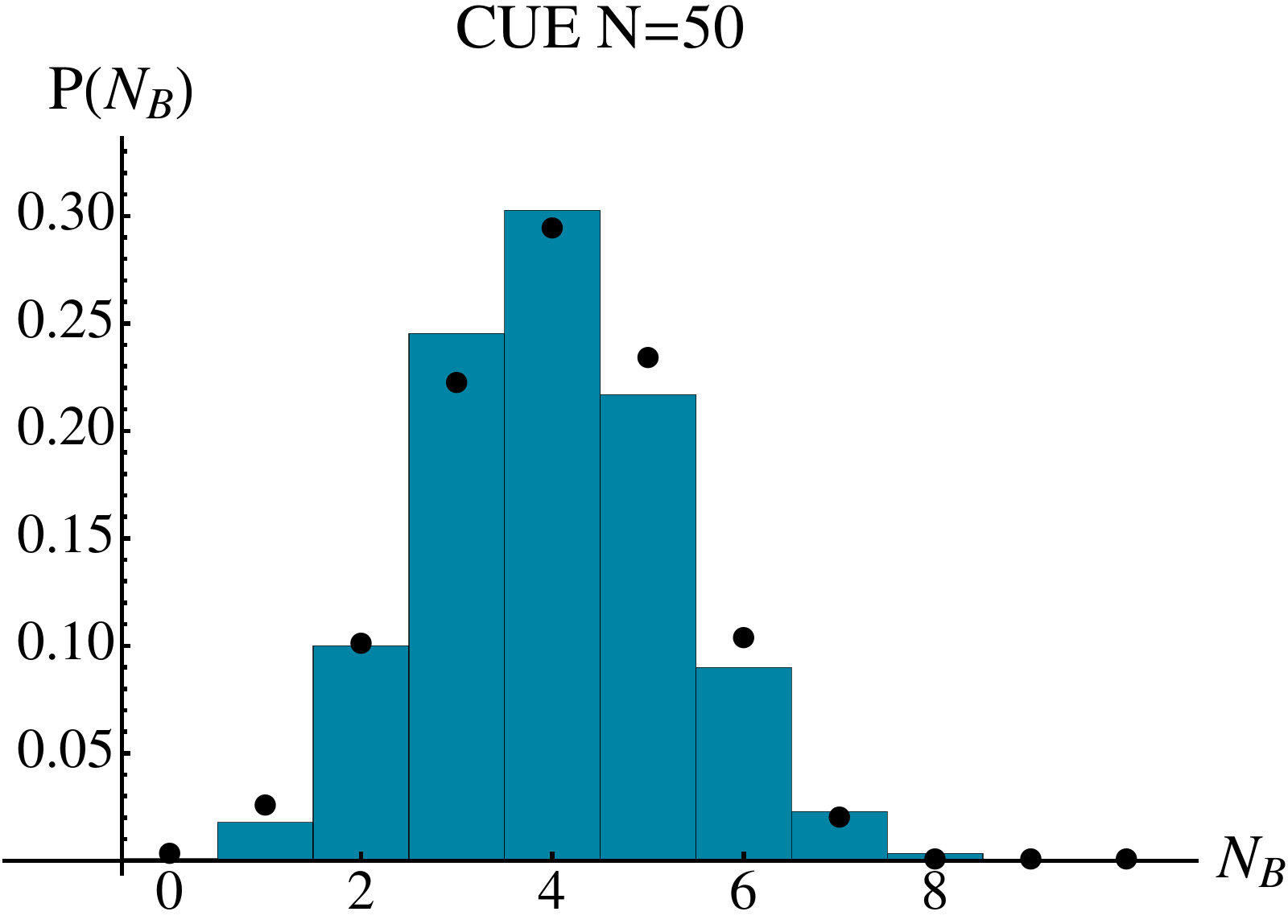}\\
  \vspace{0.5cm}
 \includegraphics[width=0.49\columnwidth]{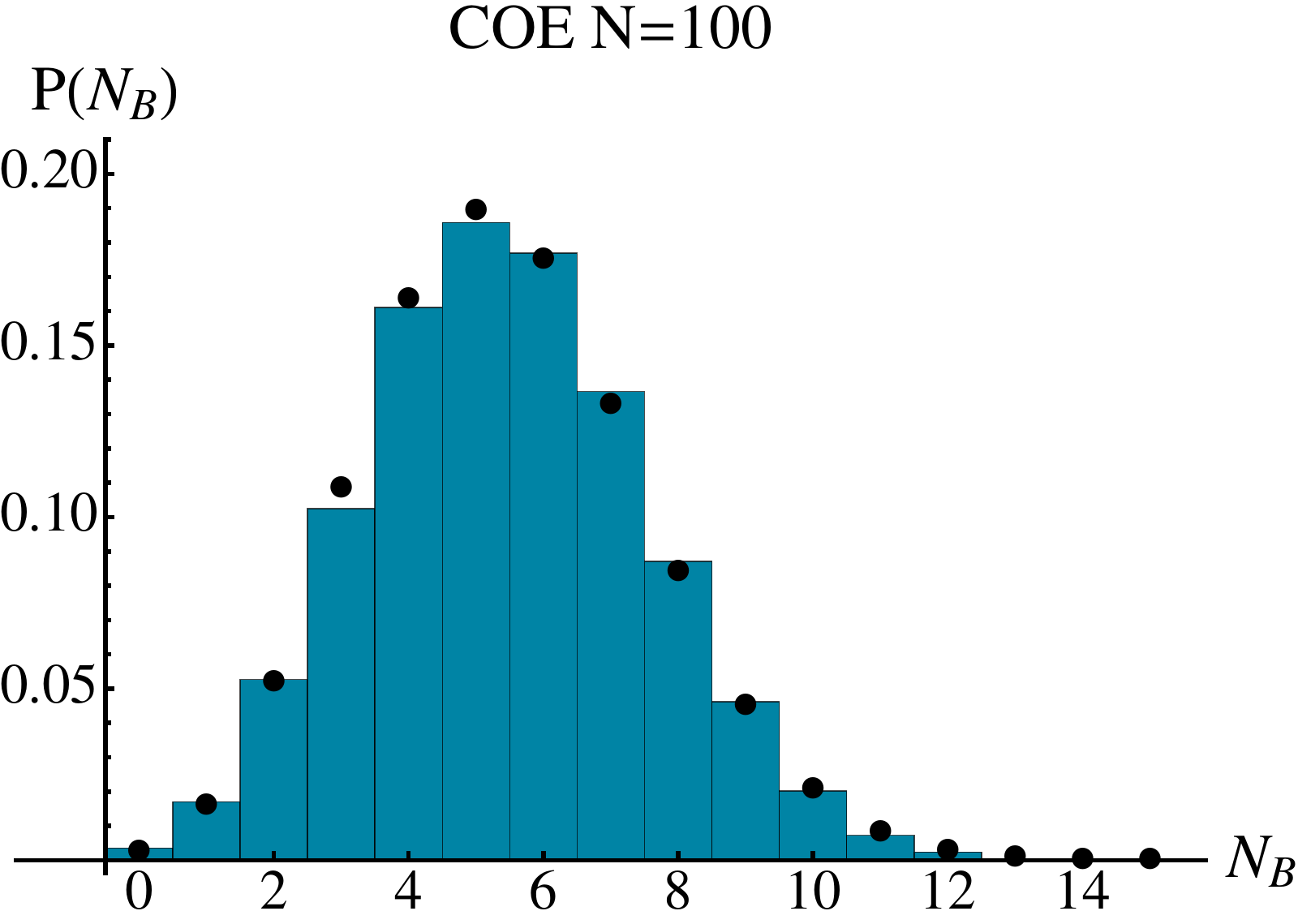}
 \includegraphics[width=0.49\columnwidth]{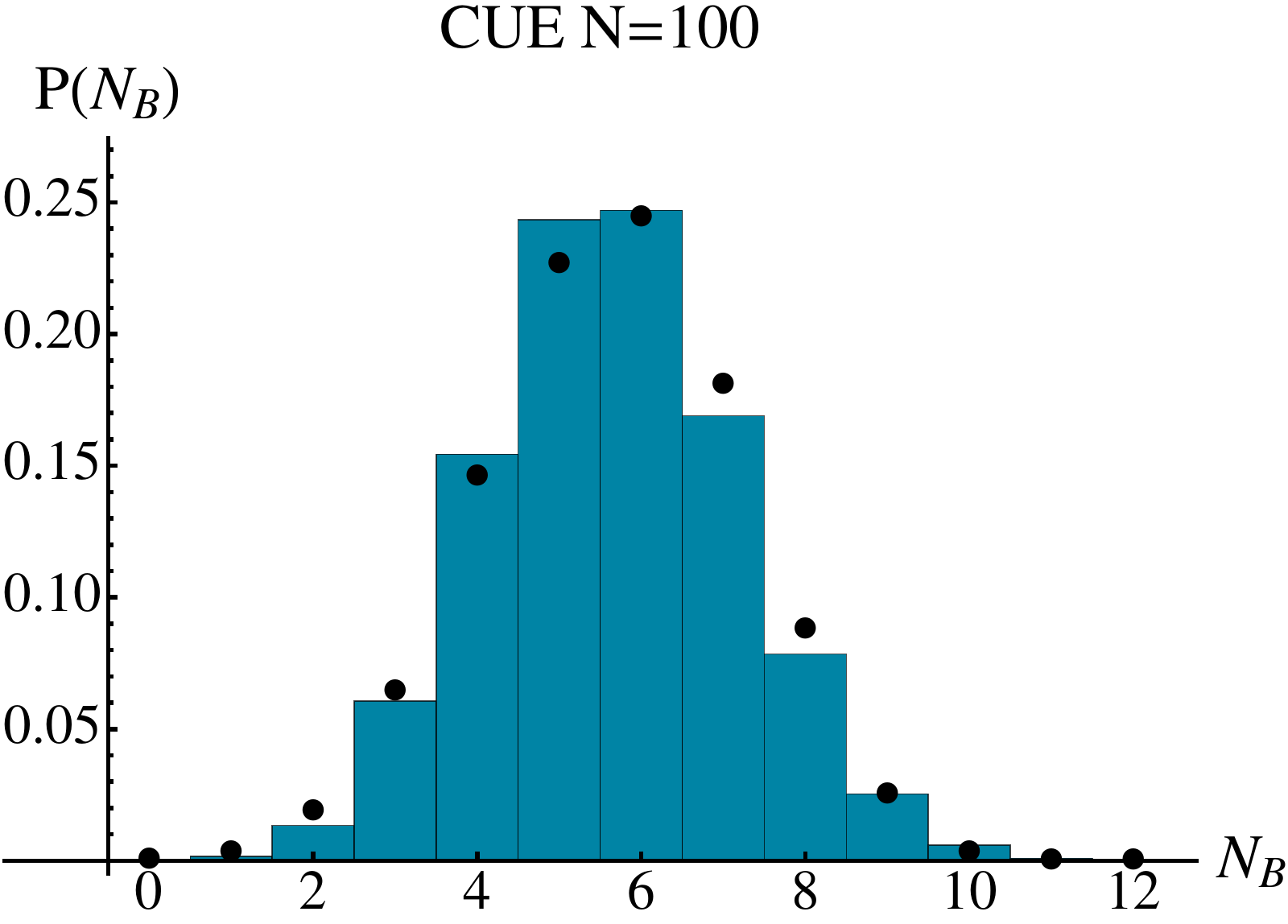}
 \caption[$N_B$ probability density functions for a RMT periodic chain]{$N_B$ Probability density function (bars) for a COE and CUE periodic waveguide with $N=10, 50, 100$. It is also shown the best fitted binomial distribution (dots) with mean $\langle N_B\rangle$ [see equation \eref{nb-rmt}] and variance given by \eref{var-nb}.}
 \label{fig-nb-dist} 
\end{figure}

In figure \ref{fig-nb-dist}, we show the pdf histograms for COE and CUE with $N=10,50,100$ and compare them with the best fitted binomial distribution imposing that $\langle N_B\rangle$ follows \eref{nb-rmt}; the agreement is good in both cases even for rather small $\bm t$ matrix dimension $N=10$. The binomial distribution is the discrete pdf of the number of successes in $n$ independent realizations of a dichotomic experiment with success probability $p$; its mean and variance are given by $np$ and $np(1-p)$, respectively. In our case, the analogy works taking $n=N$, $p=\langle N_B\rangle /N = 1/\sqrt{\pi N}$ and each success meaning the presence of a propagating Bloch mode in the spectrum. However, this is not a very good approximation because the \emph{experiments} are not independent. In fact, given there is a Bloch mode in the spectrum, the probability of a second one is not the same as for the first because of correlations. As we see below, this fact causes the variance of $N_B$ to be less than the expected from the binomial distribution, or equivalently, we have to set $n\neq N$ in the fitted binomial distributions shown in figure \ref{fig-nb-dist}.

\begin{figure}[t]
 \centering
 \includegraphics[width=.8\columnwidth]{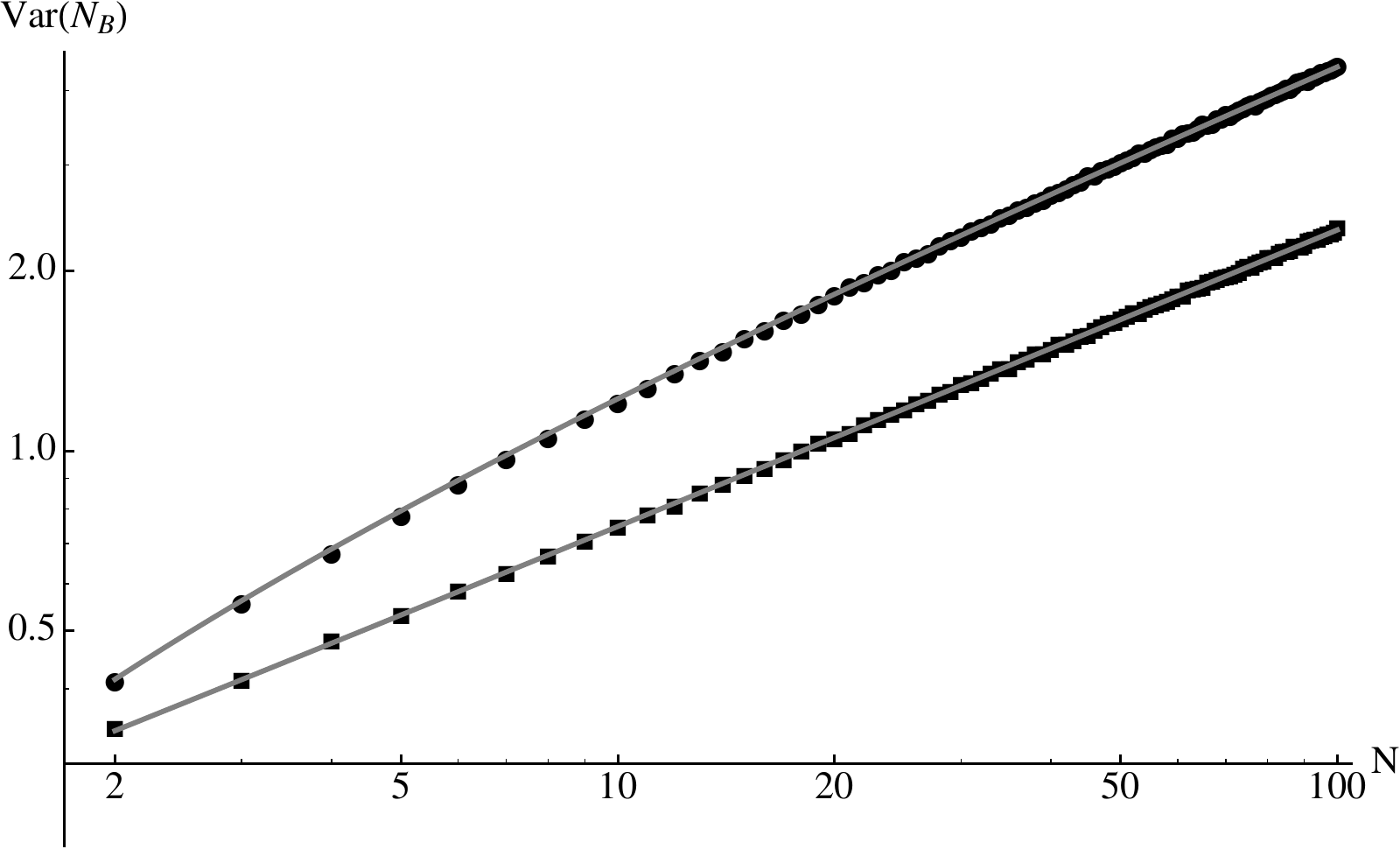}
 \caption[$\mbox{Var}(N_B)$ as a function of $N$ in a RMT periodic chain]{Variance of the number of propagating Bloch modes $N_B$ as a function of the number of open modes in the leads $N$, in a RMT periodic chain model with the COE (circles) and CUE (squares) ensembles. The gray lines show the fit given by \eref{var-nb}.}
 \label{fig-nb-var} 
\end{figure}

The variance of $N_B$ is shown in figure \ref{fig-nb-var} and is found to follow
\begin{equation}\label{var-nb}
\mbox{Var}[N_B] = \left\{
\begin{array}{ll}
\displaystyle 0.82\sqrt{\frac{N}{\pi}} - 0.24  &\quad \mbox{for COE} \\
&\\
\displaystyle 0.41\sqrt{\frac{N}{\pi}}   &\quad \mbox{for CUE}.
\end{array}\right.
\end{equation}
We note that this means $\langle N_B\rangle / \mbox{Var}[N_B] \to \alpha$ with $\alpha=0.82$ in the COE case and $\alpha=0.41$ in the CUE case, for $N\to\infty$. The fact that $\langle N_B\rangle / \mbox{Var}[N_B]$ converge to a constant value remind us of shot noise\cite{dejong1996} with Bloch modes playing the role of the quantized packet (electron, photon, etc). Since conductance $g\sim N_B$ as we discussed in section \ref{section-oseledet}, the conductance in a long chain will display this shot noise fluctuations [see section \ref{section-conductance-fluct}].

\begin{figure}[t]
 \centering
 \includegraphics[width=.8\columnwidth]{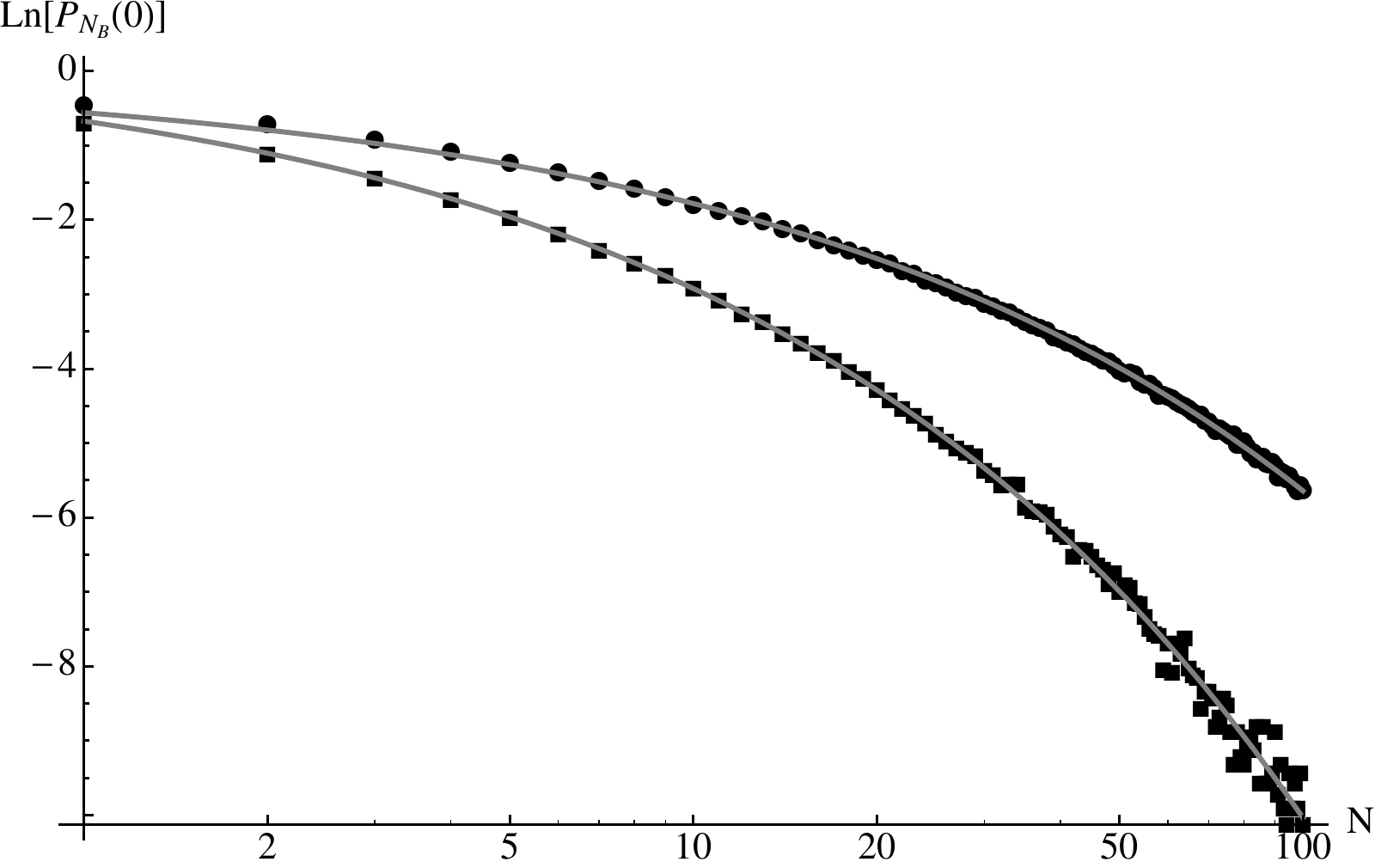}
 \caption[Probability of zero Bloch modes as a function of $N$]{Logarithm of the probability of having zero Bloch modes $P_{N_B}(0)$ as a function of $N$ in the RMT periodic chain model for COE (circles) and CUE (squares). The gray lines show the fit given by \eref{p0-nb}.}
 \label{fig-p0-nb} 
\end{figure}

To finish this section, we consider $P_{N_B}(0)$ as a function of $N$. This quantity gives us a measure of the average band gaps density in the ensemble. In figure \ref{fig-p0-nb} we show that $P_{N_B}(0)$ is well described by
\begin{equation}\label{p0-nb}
P_{N_B}(0) = \left\{
\begin{array}{ll}
\displaystyle\exp{\left(-\sqrt{\frac{N}{\pi}}\,\right)}  &\quad \mbox{for COE} \\
&\\
\displaystyle\exp{\left(0.37 - 1.8\sqrt{\frac{N}{\pi}}\, \right)}  &\quad \mbox{for CUE} .
\end{array}\right.
\end{equation}
Surprisingly, expression \eref{p0-nb} for the COE case agrees exactly with what would be expected from a Poisson distribution, in which the probability of observing zero events (in our case zero Bloch modes) is $e^{-\lambda}$ with $\lambda$ the distribution average. The Poisson distribution is a binomial distribution in the limit $n\to\infty$ and $p\to0$, which is consistent with our case where $n\sim N$ and $p\sim1/\sqrt{N}$. Of course, again the analogy is broken for the rest of the distribution because of the non-null correlations in Bloch spectra.

\chapter{Conductance in diffusive periodic waveguides}\label{chapter-conductance-props}
\chaptermark{Periodic diffusive waveguides}

In this chapter, we study the conductance of periodic quasi-one-dimensional waveguides with diffusive classical dynamics. The scattering approach to waveguides was presented in chapter \ref{chapter-waveguides}, where we defined the Landauer conductance which is the main focus in our analysis. On the other hand, we have discussed relevant classical properties of chaotic billiards in chapter \ref{chapter-billiards}, and introduced a random matrix model of periodic diffusive waveguides in chapter \ref{chapter-rmt}. We employ this model to obtain statistical properties that hold in general diffusive periodic waveguides. 

We consider a periodic waveguide with $L$ unit cells connected to two semi-infinite plane leads at its extremes as defined in section \ref{section-s-matrix}. The classical dynamics of unit cell is assumed diffusive, i.e. chaotic, with finite-horizon and strongly mixing.  We will take averages over the two ensembles defined in chapter \ref{chapter-rmt}; the first is given by the RMT periodic chain model which we denote $\langle\cdot\rangle$ (we use the same notation for the ensemble average in disordered systems but the context will distinguish which we are referring to), and the second by the semiclassical ensemble which we write $\langle\cdot\rangle_k$. 

Classically, for disordered or periodic chaotic systems, Ohm's law states that the conductance scales as $g\sim L^{d-2}$ where $d$ is the dimension, so for a wire ($d=1$), we have $g\sim 1/L$. In quantum systems, the conductance also scales in this way if wave coherence is broken at lengths of the order of the mean free path. Otherwise, in a wire with disorder of any strength, the average conductance decay exponentially $\langle g\rangle \sim e^{-2L/\hat\xi}$ for long wires, due to the Anderson Localization effect\cite{anderson1980}. This is also observed for $d=3$ in the presence of strong disorder but in two-dimensions the decay is not necessarily exponential. The conductance exponential decay is triggered for wires longer than the localization length $\hat\xi\equiv \lim_{L\rightarrow\infty} - 2(\partial\langle\log g(L) \rangle/\partial L)^{-1}$ and for shorter wires an Ohmnic regime $\langle g\rangle \sim 1/L$ is still present, which is known as the metallic regime and characterized by the presence of weak localization and universal conductance fluctuations \cite{lee1985}. 

In contrast, it is clear that in quantum periodic waveguides there is no localization (outside band gaps) because of Bloch ballistic modes. In section \ref{asymptotic-conductance-props}, we study the conductance of long periodic chains and establish in more detail their dependence on $N_B$ as proposed in section \ref{section-oseledet}. Then, in section \ref{approach-npg} we show that the non-propagating part of the conductance has an asymptotic long-chain limit that decays as $L^{-2}$. The main result of this chapter is the existence of an ohmic regime in periodic diffusive waveguides as discussed in section \ref{section-ohms-law}. Finally, we study the conductance fluctuations and distribution in the ohmic and Bloch-ballistic regimes.

\section{Long-chain conductance properties}\label{asymptotic-conductance-props}

\subsection{Number of Bloch modes and conductance average}

\begin{figure}[t]
 \centering
 \includegraphics[width=.8\columnwidth]{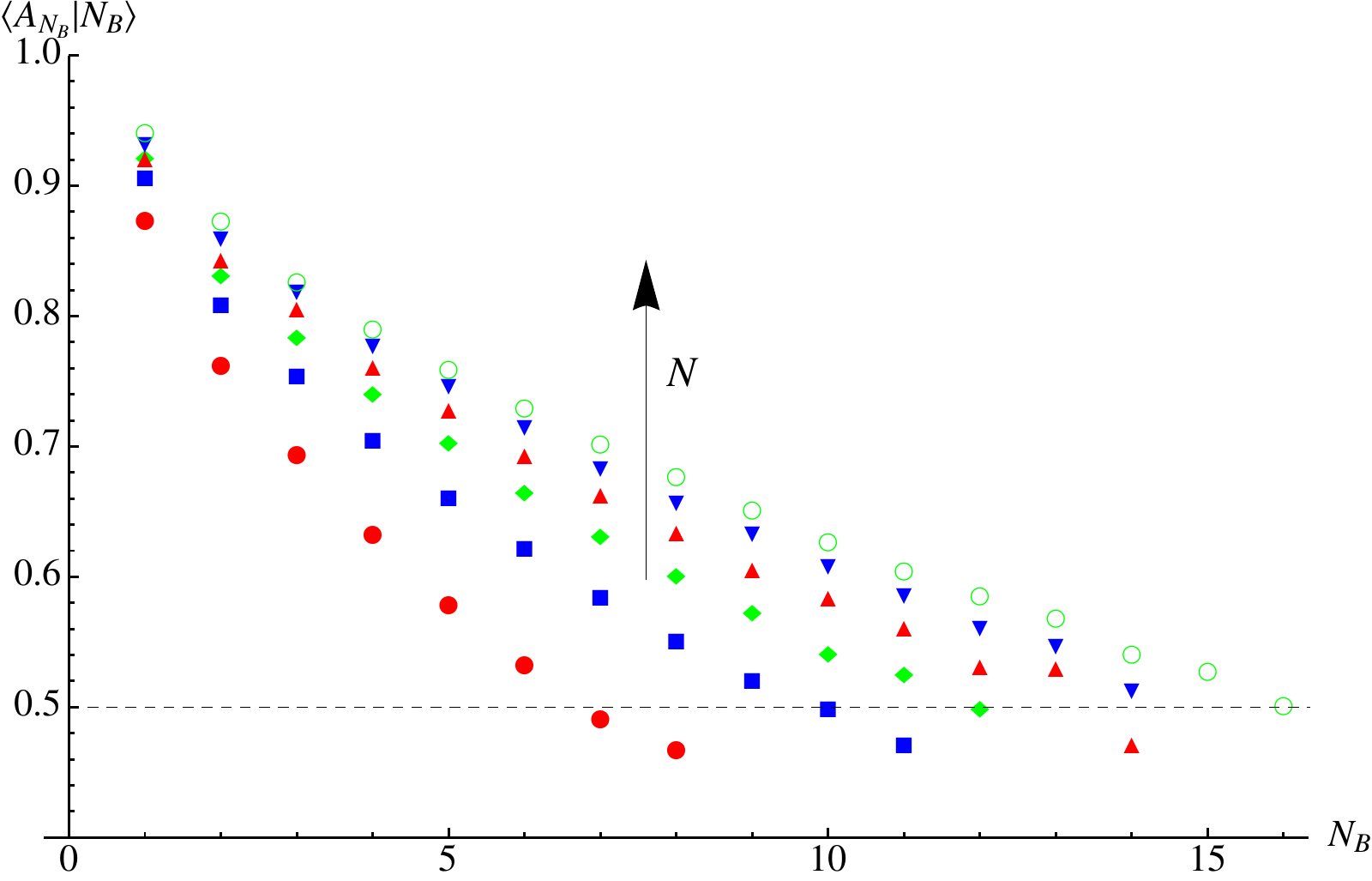}
 \caption[Conditional mean $\langle A_{N_B}  | N_B\rangle$ for the COE periodic chain]{Conditional mean $\langle A_{N_B}  | N_B\rangle$ for six COE samples of 100000 realizations for dimension $N=25,50,75,100,125,150$.}
 \label{fig-Anb} 
\end{figure}

As seen in section \ref{section-long-chain-conductance}, in a chaotic waveguide with periodic structure the \textit{typical} conductance, i.e. the conductance for a fixed energy taken from the semiclassical ensemble, displays exponential decay towards asymptotic quasi-periodic oscillations in the long chain limit [see equations (\ref{g_long_L}) and (\ref{t_periodic_decomp})]. From \eref{adim-g}, we can write the conductance as
\begin{equation}\label{g_typical}
g(L) = g_\infty(L) +  \sum_{|\lambda_i|\neq1} T_i(L) ,
\end{equation}
where
\begin{equation}\label{g-inf}
g_\infty(L)=\sum_{|\lambda_i|=1} T_i(L).
\end{equation}
Note that the last term in the rhs of \eref{g_typical} decays exponentially [see section \eref{section-oseledet}], so $g(L)=g_\infty(L)$ for $L\gg \ell$, with
\begin{equation*}\tag{\ref{T-decay-length}}
\ell = \left(\min_{|\lambda_i|>1}\{\log|\lambda_i|\}\right)^{-1}  
\end{equation*}
the largest evanescent mode decay length, defined in chapter \ref{chapter-waveguides}. The asymptotic long-chain conductance $g_\infty(L)$ is quasi-periodic in $L$ as can be inferred from (\ref{t_periodic_decomp}). Also, note that all $T_i$ and hence $g$ depend on the ensemble realization, this is, on the wavenumber $k$ or on the RMT realization. Then, taking $g_\infty(L)$ semiclassical average one obtains a constant value, 
\begin{equation}\label{g-asympt-average}
\langle g_\infty(L\to\infty)\rangle_k = \langle g_\infty\rangle_k \lesssim  \langle N_B\rangle_k ,
\end{equation}
as is plotted in figure \ref{g_L}. In order to make the inequality in \eref{g-asympt-average} more precise, we write the long-chain conductance as
\begin{equation}\label{g_inf}
g_\infty(L) = N_B\, A_{N_B}
\end{equation}
with 
\begin{equation}
A_{N_B} = \left\{
\begin{array}{rr}
\displaystyle \frac{1}{N_B} \sum_{i=1}^{N_B} T_i(L)  &\;\mbox{for $N_B\neq0$}  \\
  &\\
  1 & \mbox{for $N_B=0$}  
\end{array} \right.
\end{equation}
the average Bloch transmission, which is a random variable (of the semiclassical or RMT ensemble) dependent on $N_B$ (and $L$) and bounded between in $[0,1]$. Figure \ref{fig-Anb} plots how the conditional mean $\langle A_{N_B}  | N_B\rangle$ decreases as a function of $N_B$ from a value close to 1 for $N_B=1$ and saturates around $1/2$ for large $N_B$. Hence, we have
\begin{equation}\label{g_low_bound}
\langle g_\infty(L)\rangle = \sum_{n=1}^{N}  \langle A_{N_B} | N_B=n \rangle \,nP_{N_B}(n) > \frac{1}{2} \langle N_B\rangle ,
\end{equation}
where $P_{N_B}(n)$ is the probability of having $n$ Bloch modes [see section \ref{section-p-nb-dist}]. Therefore, from \eref{g_long_L} and \eref{g_low_bound} follows that
\begin{equation}\label{g-boundaries}
\langle N_B\rangle >\langle g(L)\rangle > \frac{1}{2}\langle N_B\rangle \quad\mbox{for $L\to\infty$}
\end{equation}
in the semiclassical limit. Numerically, it is observed using our RMT model that a good approximation for large $N$ is given by 
\begin{equation}\label{g_nb_exact}
\lim_{L\to\infty}\langle g(L)\rangle = \left\{
\begin{array}{cc}
0.75\, \langle N_B\rangle_{\rm COE} - 0.20   &\quad\mbox{for COE}\\
0.75\, \langle N_B\rangle_{\rm CUE} - 0.15   &\quad\mbox{for CUE}.
\end{array}\right.
\end{equation}
For the cosine billiard this result is also good and in fact should be universal since it stems from the RMT $T_i$ eigenvalues distribution. The difference in the conductance for COE and CUE can be seen as consequence of weak localization as discussed in section \ref{section-weak-loc-g}.

\subsection{Approach to the asymptotic regime}\label{approach-npg}

\begin{figure}[t]
 \centering
 \includegraphics[width=.8\columnwidth]{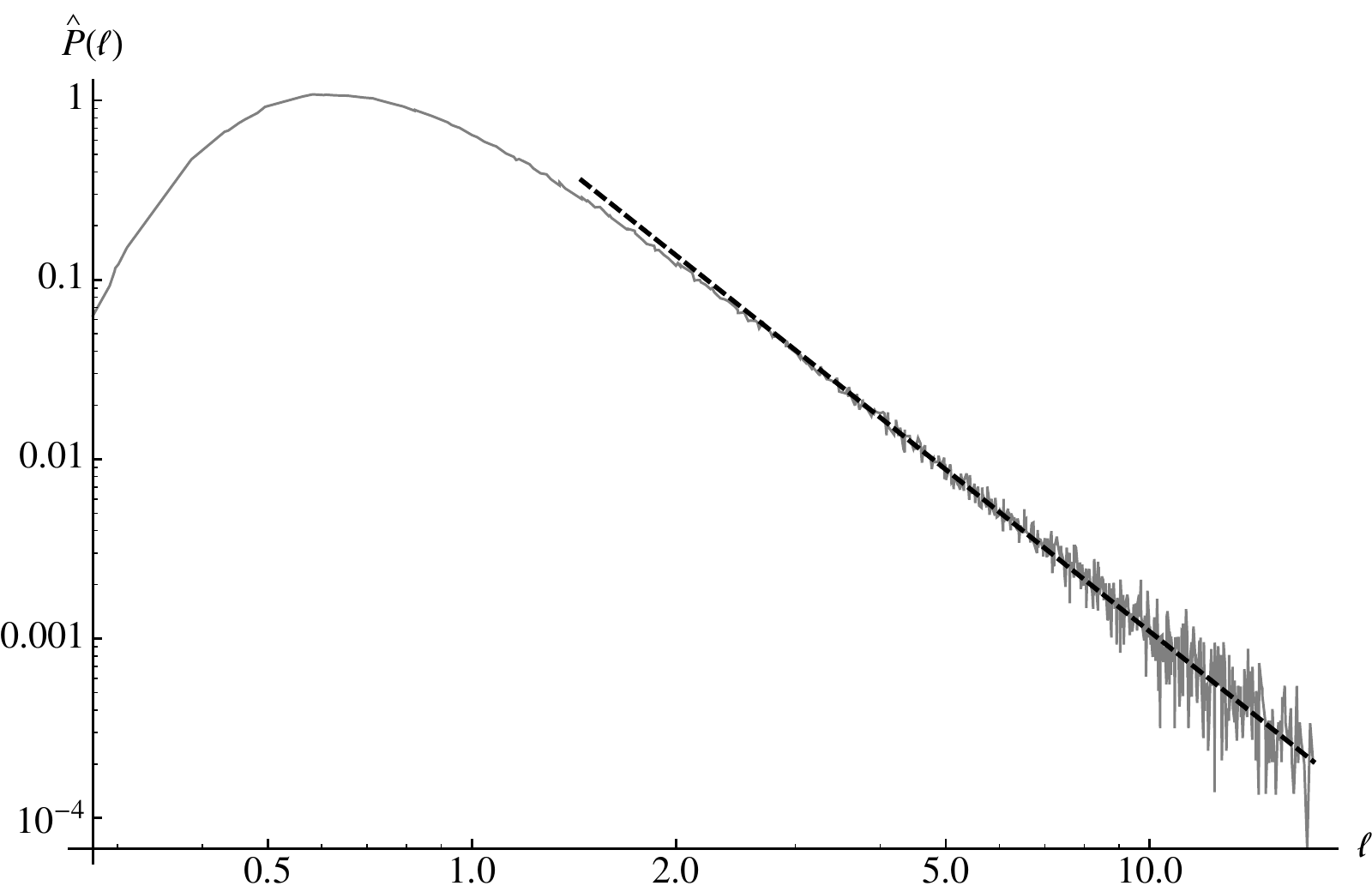}
 \caption[Scaled pdf $\hat{P}(\ell)$ of the longest decay length $\ell$]{Scaled pdf $\hat{P}(\ell)$ of the longest decay length $\ell$ for the COE ensemble (gray line), constructed by superposing the pdf $N^{3/4} P(\ell N^{3/4}, N)$ for $N = 25, 50, 75, 100, 125, 150$. The dashed black line decays as $\ell^{-3}$.}
 \label{figure-l-pdf} 
\end{figure}

\begin{figure}[t]
 \centering
 \includegraphics[width=.8\columnwidth]{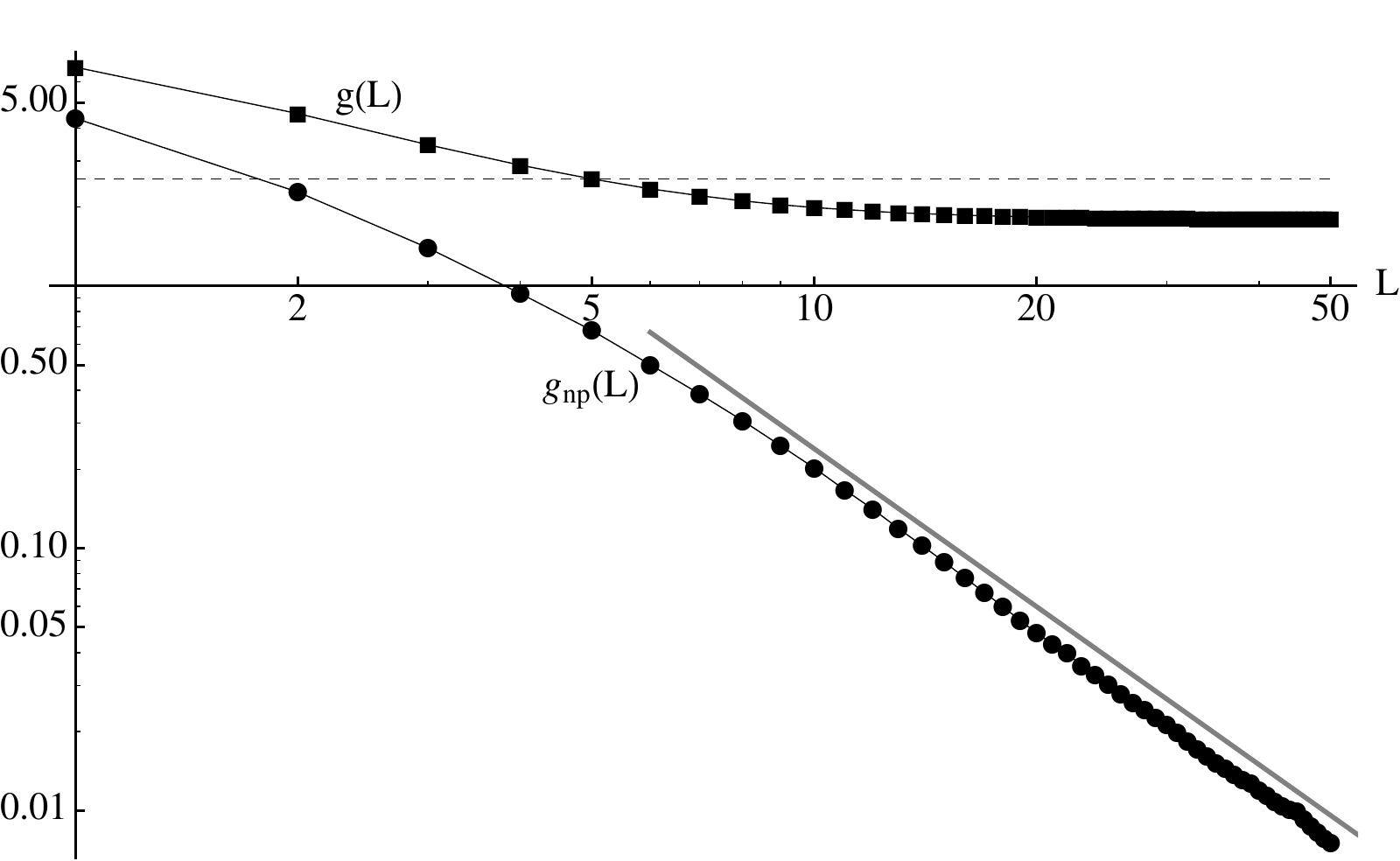}
 \caption[Average conductance as a function of $L$ in a periodic diffusive billiard chain]{Average conductance $\langle g(L)\rangle_k$ (squares) and its non-propagating part $\langle g_{\rm np}(L)\rangle_k$ (circles) for the cosine billiard chain with $A_1=0.5$, $A_2=4.5$ and $k=30.33\pi$. The average number of propagating modes $\langle N_B\rangle_k=2.6$ (dashed line) is an upper bound for $\langle g(L)\rangle_k$ as $L\to\infty$ [see equation \eref{g-boundaries}]. On the other hand, $\langle g_{\rm np}(L)\rangle_k$  displays an algebraic decay $\sim L^{-2}$ (gray line) for large $L$ as predicted by the pdf $\hat{P}(\ell)$ [see  equation \eref{non-prop-decay}].}
 \label{g_L} 
\end{figure}

The decay rate of the conductance $g(L)$ to its asymptotic regime depends on $\ell$ [see equations \eref{g_long_L} and \eref{T-decay-length}]. We begin this section focusing on the conductance non-propagating part
\begin{equation}
g_{\rm np}(L) = g(L)-g_\infty(L) =  \sum_{|\lambda_i|\neq1} T_i(L) ,
\end{equation}
which is dominated by the slowest to decay evanescent transmission amplitude,
\begin{equation}
T_m(L) = \frac{4}{2+\Lambda_m(L) + \Lambda_m(L)^{-1}}  = 4a_m(L)^{-1}e^{-2L/\ell}  \quad\mbox{for $L\gtrsim\ell$},
\end{equation}
where $a_m(L)^{-1}\sim\mathcal{O}(1)$ is quasi-periodic as we discussed in section \ref{section-oseledet}. Hence, taking the average we can drop $a_m(L)$ since the decay will be dominated by the exponential, this is, we have
\begin{equation}
\langle g_{\rm np}(L)\rangle = \langle 4e^{-2L/\ell}\rangle.  
\end{equation}
In order to calculate this average we will study the pdf of the decay length $\ell$. We can inspect numerically this distribution using RMT; we focus on COE since this is the ensemble representing the cosine billiard without a magnetic field. We consider the pdf 
\begin{equation}\label{L-pdf-def}
P(\ell,N) = \int_{\bm{U}\in U(2N)} d\mu_{\rm COE} (\bm{U}^T\bm{U})\, \delta\left[\ell -  \left( \max |\mathcal{L}(\bm{U^TU})|\right)\right] ,
\end{equation}
where $\mu_{\rm COE}$ is the invariant COE measure\footnote{The 2N-dimensional COE ensemble measure $\mu_{\rm COE}$ is the induced probability measure in the symmetric space $U(2N)/O(2N)$ by $U(2N)$ Haar measure.} and 
\begin{equation}
\mathcal{L}(\bm{S}) = \{(\log{|\lambda|})^{-1} \, : \, \lambda \in \mbox{eig}(\bm{M}(\bm{S})) \,\wedge\, |\lambda|>1\},
\end{equation}
is the decay lengths set associated to the spectrum of the transfer matrix $\bm{M}(\bm{S})$ obtained from a scattering matrix $\bm S$. We have found an asymptotic scaling, as shown in figure \ref{figure-l-pdf}, given by 
\begin{equation}\label{l*-pdf}
N^{3/4}\, P(\ell N^{3/4},\, N) \underset{N\rightarrow\infty}{\longrightarrow} \hat{P}(\ell) 
\end{equation}
with the scaled pdf $\hat{P}(\ell)$ displaying an algebraic tail, 
\begin{equation}\label{l-pdf-tail}
\hat{P}(\ell) \sim \ell^{-3} \quad\mbox{for $\ell\to\infty$}.
\end{equation}
Horvat and Prosen\cite{horvat2007} studied a pdf similar to (\ref{L-pdf-def}) but for the maximum modulus $\bm M$ eigenvalue and found and analogous asymptotic scaling with exponent $1/2$. 
Finally, using the scaled pdf \eref{l*-pdf} and its algebraic tail \eref{l-pdf-tail}, we have
\begin{equation}\label{non-prop-decay}
\langle g_{\rm np}(L) \rangle = \langle 4e^{-2L/\ell} \rangle  \sim L^{-2}.
\end{equation}
This result, which was obtained using the RMT periodic chain model, also holds for the semiclassical ensemble as expected and can be seen in figure \ref{g_L} for the cosine billiard chain.
\\

\begin{figure}[t]
 \centering
 \includegraphics[width=.8\columnwidth]{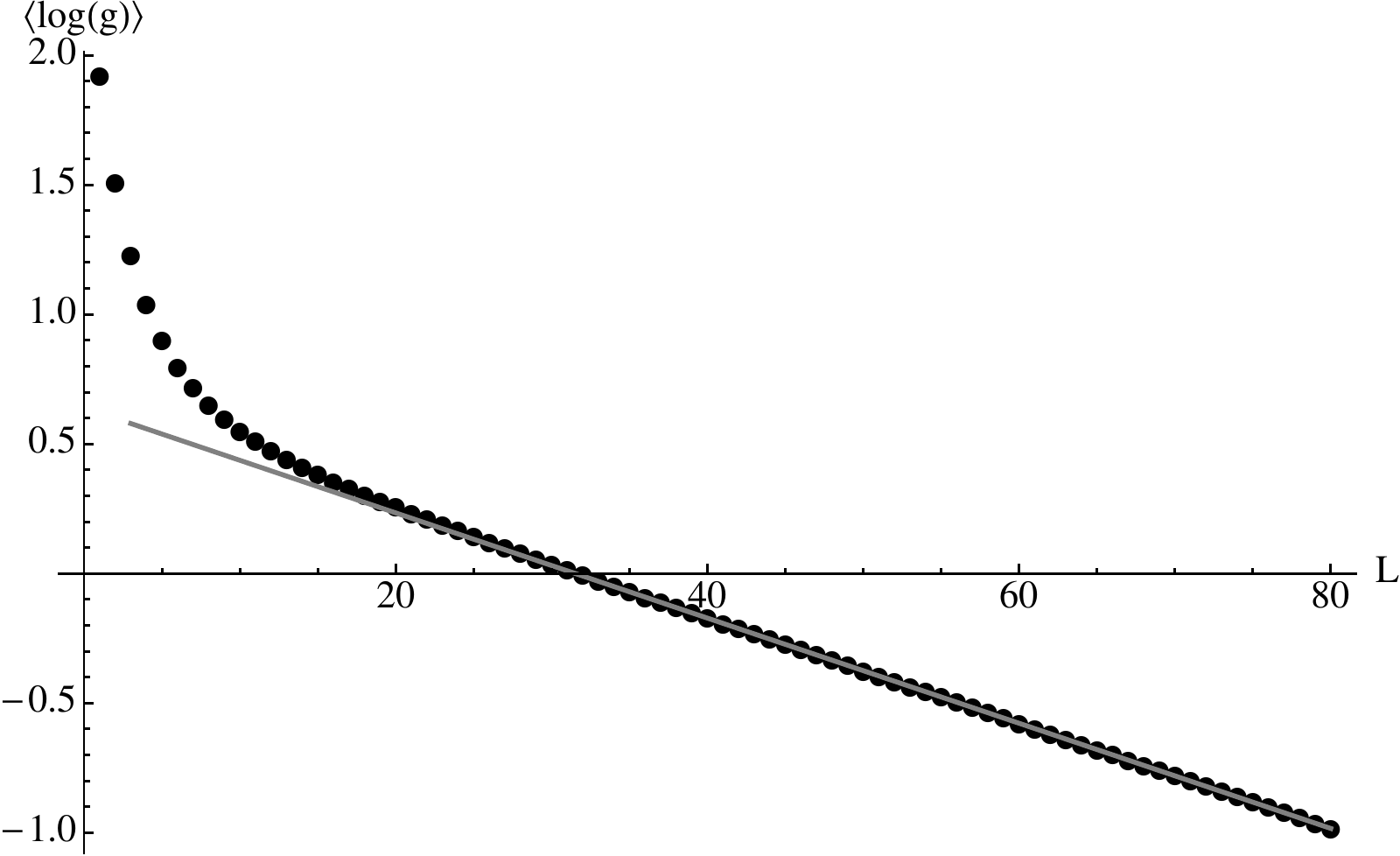}
 \caption[Average logarithmic conductance as a function of $L$]{Average logarithmic conductance $\langle \log{g(L)}\rangle_k$ (dots) for the cosine billiard chain with $A_1=0.5$, $A_2=4.5$ and $k=30.33\pi$ as a function of $L$. The linear decay $\sim -L/\xi$ predicted by \eref{log_g-2} is clearly observed (gray line). In this case $\xi=98.6$, $P_{N_B}(0)=0.06$ and $N=15$ hence \eref{inv-chi} implies the harmonic mean decay length $\langle \ell^{-1}\rangle_\infty^{-1} = 0.83$ (in unit cell length units). }
 \label{FIG-log_g} 
\end{figure}
In disordered systems, a standard characterization of the conductance decay to the asymptotic regime (which in that case is localization with $\langle g \rangle\to0$) is the linear decay of the average logarithmic conductance $\langle \log{g(L)}\rangle \sim -L$ for long wires. Interestingly, although in periodic chains the asymptotic conductance is not null because of the presence of propagating Bloch modes, in diffusive chains $\langle \log{g(L)}\rangle_k$ shows a similar behavior. In fact, we have for $L\gg \langle g_\infty \rangle_k ^{-1/2}$,
\begin{align}
\langle \log{g(L)}\rangle_k &= P_{N_B}(0)\,\langle \log{4e^{-2L/\ell}}\rangle_k  + (1-P_{N_B}(0))\, \langle\log{g_\infty(L)}\rangle_k  \label{log_g}    \\ 
& \approx -\langle 2\ell^{-1}\rangle_k\,P_{N_B}(0)\, L + \mbox{constant} \label{log_g-2}
\end{align}
assuming correlations between $g_\infty$ and $\ell$ are negligible. Therefore, even though in a periodic chain there is no exponential decay of the average conductance and its limit is non-null, 
\begin{equation}
\langle\log{g(L)}\rangle_k\sim -\frac{2L}{\xi} \quad\mbox{for large $L$}
\end{equation}
just like in a disordered wire. However, the constant $\xi^{-1}$, which corresponds to the localization length in disordered wires, in periodic chains can be seen as a measure of the band-gaps in the spectrum, since it is proportional to the probability of having zero Bloch modes, 
\begin{equation}\label{inv-chi}
\xi^{-1} = N^{-3/4}\, \langle \ell^{-1}\rangle_\infty\,P_{N_B}(0), 
\end{equation}
where $\langle \cdot\rangle_\infty$ is the average taken over the scaled pdf $\hat{P}(\ell)$. The probability $P_{N_B}(0)$ as a function of $N$ was studied in section \ref{section-p-nb-dist} for the COE and CUE ensembles. We have numerically estimated the scaled harmonic average decay length to be $\langle \ell^{-1}\rangle_\infty^{-1} = 0.84$ using the scaled pdf \eref{l*-pdf}. Alternatively, we can compute this number indirectly measuring $\xi^{-1}$ and using \eref{inv-chi}. In figure \ref{FIG-log_g}, we plot $\langle\log{g_\infty(L)}\rangle_k$ in the cosine billiard chain and use $\xi$ and $P_{N_B}(0)$ to calculate the scaled harmonic average decay length $\langle \ell^{-1}\rangle_\infty^{-1}$.

\begin{figure}[h!]
 \centering
 \includegraphics[width=0.8\columnwidth]{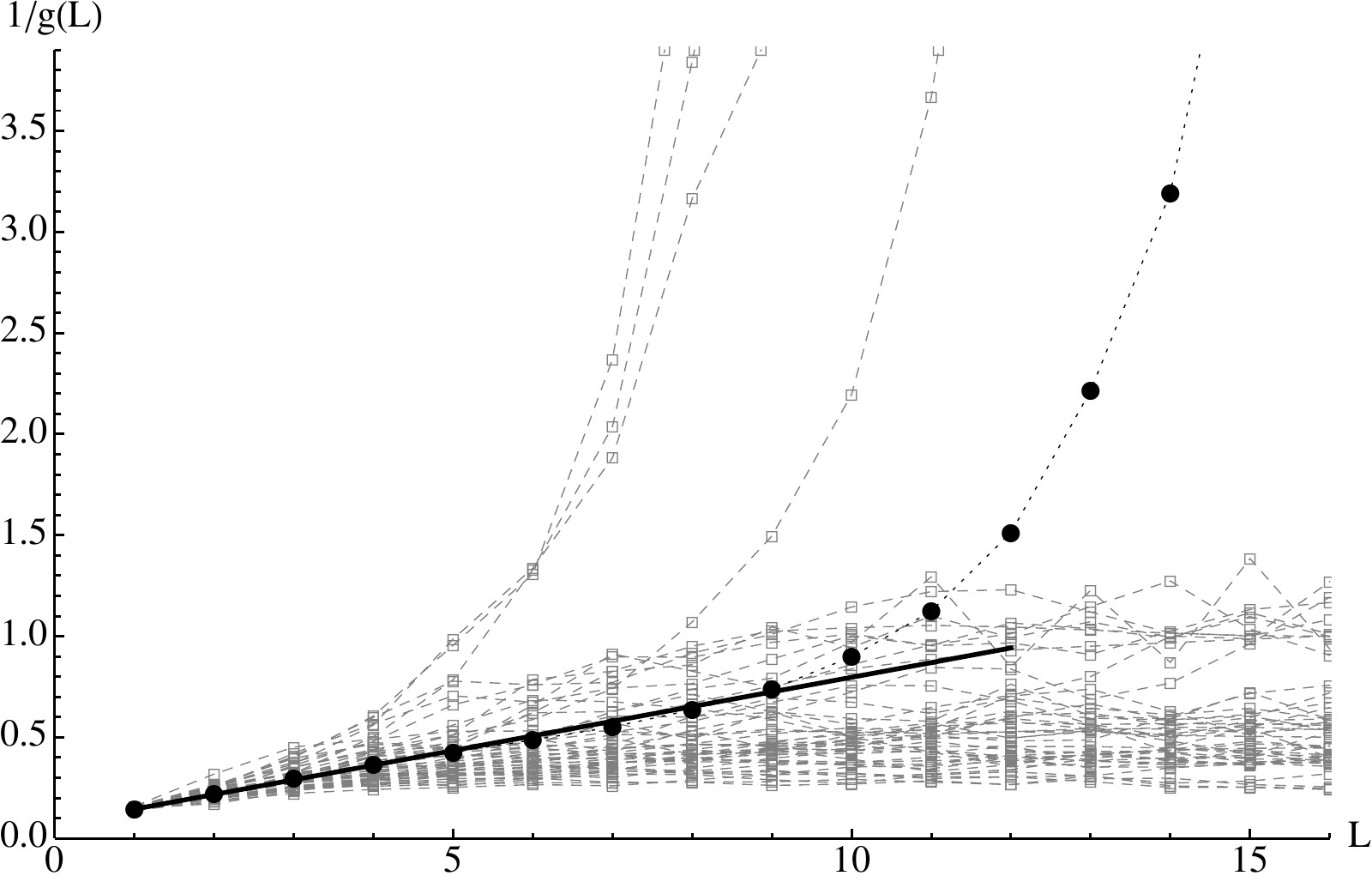}
 \caption[Ohmic regime in the periodic cosine billiard]{Average resistance $\langle 1/g(L)\rangle_k$ (black dots) for the cosine billiard with $A_1=0.5$, $A_2=4.5$ and $k=30.33\pi$. The full line shows the ohmic regime holding for $L<10$. Also, several realizations of $1/g(L)$ are displayed in gray: a few cases with $N_B=0$ grow exponentially and the rest reach an asymptotic oscillating value $1/g_\infty(L) \sim \mathcal{O}(1/N_B)$. The exponential grow of the average resistance $\langle R\rangle_k$ for $L>10$ is explained by the presence of the rare cases with $N_B=0$. }
 \label{fig-1_g} 
\end{figure}
\begin{figure}[h!]
 \centering
 \includegraphics[width=.95\columnwidth]{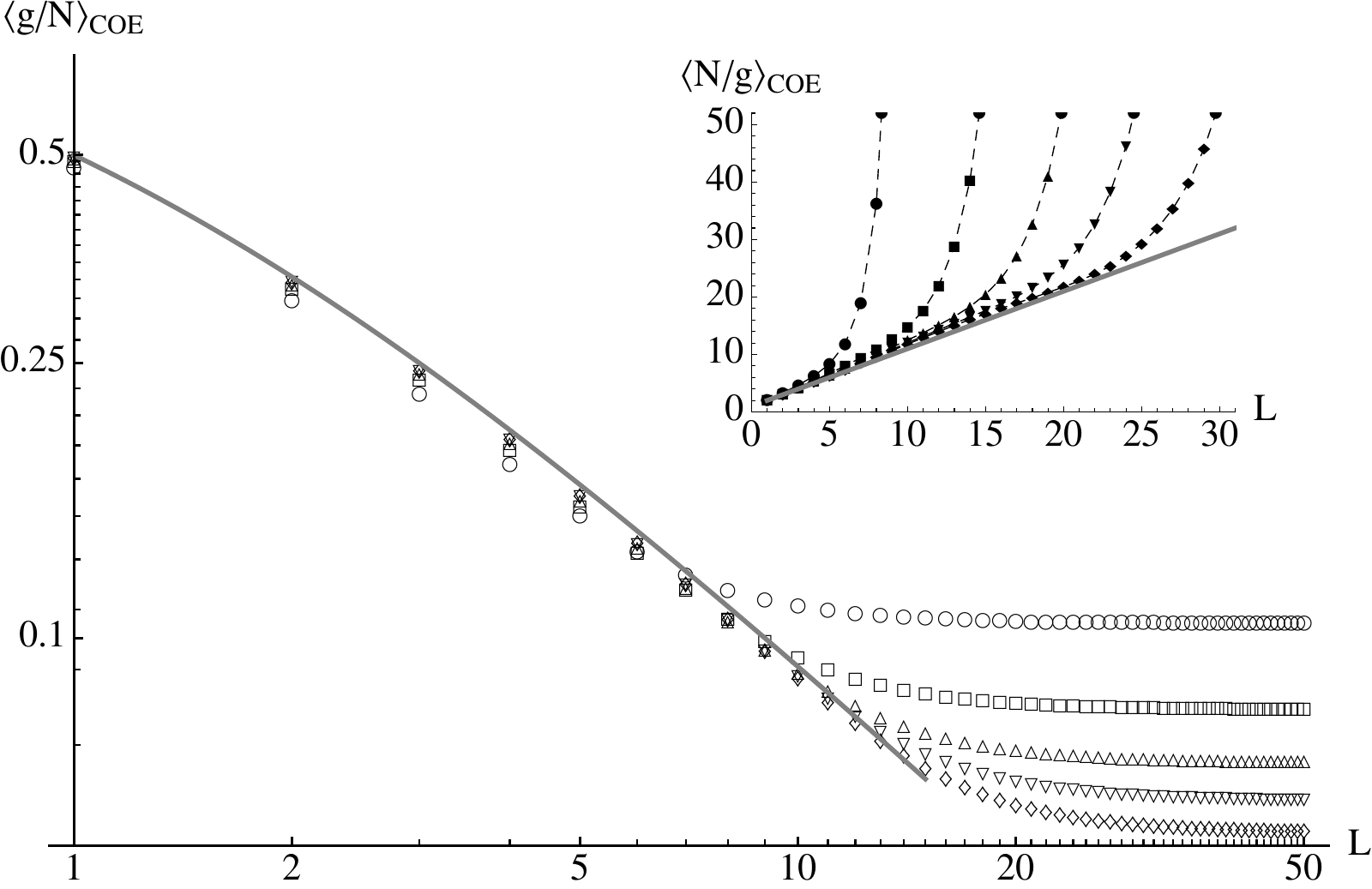}
 \caption[Ohmic regime in the COE periodic chain model]{Plot of the average normalized conductance $\langle g(L)/N\rangle$ and resistance $\langle N/g(L)\rangle$ (inset) for the COE periodic chain model with $N = 10, 20, 30, 40, 50$ (dots, squares, triangles up, triangles down, diamonds) as a function of $L$. The average conductance collapse to $\langle g(L)/N\rangle\sim (L+1)^{-1}$ and the average resistance to $\langle N/g(L)\rangle\sim L+1$ (full lines) for large $N$ during the ohmic regime which is observed in a range $1<L\lesssim\sqrt{N}$ [see discussion around \eref{lambda-conjecture}].}
 \label{fig-1_g-COE} 
\end{figure}
\newpage

\section{Ohmic regime}\label{section-ohms-law}

As we have seen, in periodic waveguides, even if its classical dynamics is chaotic, quantum transport is ballistic due to the existence of propagating Bloch modes. Interestingly, however, this asymptotic ballistic regime is preceded by an \emph{ohmic regime} for short chains, similar to what is observed in disordered system. As we have discussed in chapter \ref{chapter-introduction}, an ohmic conductor shows the conductance scaling $G=\sigma W/L$, where $W$ and $L$ are the sample width and length respectively, and $\sigma$ is its conductivity. In mesoscopic systems, the coherent propagation of waves causes this scaling to break up at large lengths; in disordered systems there is an ohmic regime for $L\ll\xi$ followed by localization and, as have been discussed in this work, in periodic systems propagation is ballistic in long-chains with $g(L\to\infty)\sim N_B$. In this section, we study the existence of the ohmic regime in diffusive periodic waveguides. We consider the resistance defined as
\begin{equation}
R = \frac{1}{g(L)},
\end{equation}
which is plotted in figure \ref{fig-1_g} for a cosine waveguide. Qualitatively, we found Ohm's law $\langle R\rangle = (L+1)/N$ holds for small $L$, i.e. before $\langle g(L)\rangle$ reaches its asymptotic value. This regime is followed by localization-like exponential grow of the resistance. The latter may seem surprising but is a consequence of the non-null probability of $g_\infty=0$ (or equivalently $P_{N_B}(0)\neq0$). In fact, although $P_{N_B}(0)\neq0$ decays to zero as $N\to\infty$, it dominates in the average $\langle 1/g(L) \rangle$ for long chains. Of course, for particular realizations of $1/g(L)$ the most probable is $1/g(L)=1/g_\infty(L)<\infty$ with $N_B\neq0$.

\begin{figure}[h!]
 \centering
 \includegraphics[width=0.8\columnwidth]{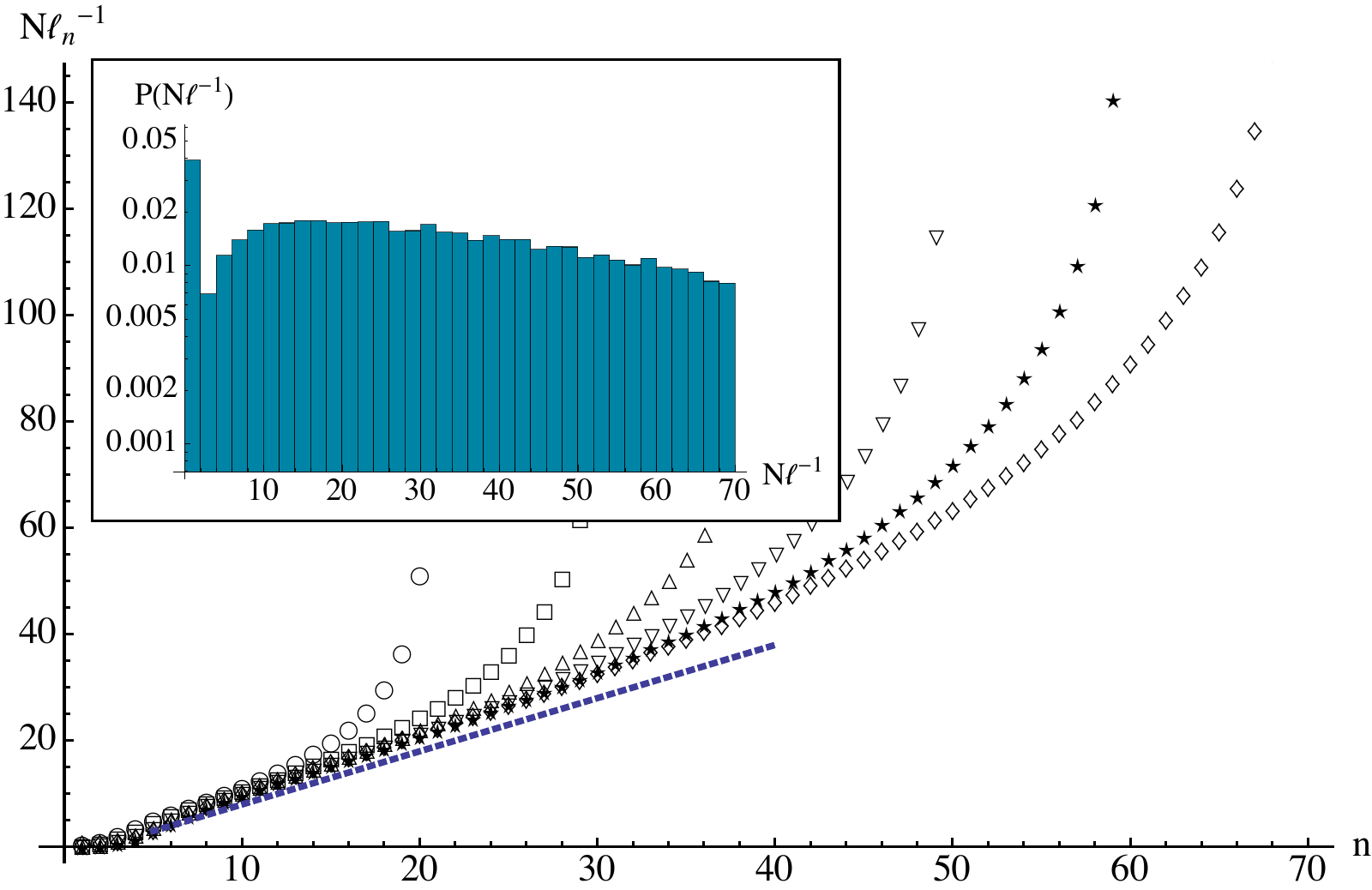}
 \caption[Average inverse decay lengths spectrum $\ell_n^{-1}$ for the COE periodic chain model]{Average inverse decay lengths spectrum $\ell_n^{-1} = \log{|\lambda_n|}$ for the COE periodic chain model. The plot shows $N\ell_n^{-1}$ for $N=10,20,30,40,50,60,70$ (circles, squares, up triangles, down triangles, stars, diamonds). The integer $n$ is the absolute-value-sorted $\lambda_n$ eigenvalue index. For this system $\langle N_B\rangle = 4.7$. As can be seen in the plot $N\ell_n^{-1}$ is close to zero for $n\lesssim \langle N_B \rangle$ and is followed by a range with linear growth $N\ell_n^{-1} \sim n$ (dashed line) which explains the existence of the ohmic regime in the periodic chain for $L< \langle N_B \rangle/N$. In the inset, part of the pdf $P(N\ell^{-1})$ for $N=70$ is plotted, showing a constant range for $10\lesssim N\ell^{-1}\lesssim 30$ which is equivalent to $P(|\lambda|)\sim|\lambda|^{-1}$ [see \eref{lambda-conjecture}]. }
 \label{fig-coe-lambdas} 
\end{figure}

On the other hand, the existence of the ohmic regime is a consequence of the full $\bm{M}$ matrix spectrum and is observed for small $L$, where several non-propagating $T_i$ still contribute to the sum in the rhs of (\ref{g_typical}). In order to understand this result we digress briefly to disordered systems, where the linear scaling of $R(L)$ observed before localization sets in can be explained by the DMPK localization lengths spectrum $\hat{\ell}_n$, $n=0,\ldots,N-1$. It is known\cite{dorokhov1982} that  $\hat{\ell}_n(N)^{-1} \sim n/N\hat{\ell}$, where $N \hat{\ell}$ denotes the disordered wire localization length. From the polar decomposition \eref{polar-rel}, the disordered wire conductance is given by 
\begin{equation}
g = \sum_{n=0}^{N-1}\frac{4}{2+e^{L/\hat{\ell}_n} + e^{-L/\hat{\ell}_n}}.
\end{equation}
Then, the ohmic scaling $1/g \sim L$ for $\hat{\ell}\ll L\ll\hat{\ell}N$ in disordered system can be seen as a consequence of $\hat{\ell}_n^{-1}$ lineal dependence on $n$. It turns out that the spectrum $|\lambda_n|$ of $\bm{M}_1=\bm{M}_1(\bm{S})$ with $\bm{S}$ taken from COE has a similar property, namely $\ell_n^{-1}=\log{|\lambda_n|} \approx n/N$ for $\langle N_B\rangle\leq n \ll N$ which explains the existence of an ohmic regime in diffusive periodic waveguides [see figure \ref{fig-coe-lambdas}]. This is a statistical characteristic of the transfer matrix spectrum, which can be recasted stating that 
\begin{equation}\label{lambda-conjecture}
p(|\lambda|) \sim \frac{1}{|\lambda|} \;,\quad \mbox{for}\quad \frac{\langle N_B\rangle}{N} < \log{|\lambda|} \ll 1
\end{equation}
with $p(|\lambda|)$ the marginal pdf of the spectrum absolute values $|\lambda_n|$. From this follows that $1/g\sim L$ for $L \lesssim N/\langle N_B\rangle  \sim \sqrt{N}$, which can also be understood from the two following arguments. 

First, one expects the ohmic regime to be observed only for chains whose diffusion time $t_D=L^2/D$ is smaller than the unit cell Heisenberg time $t_H = \nu_E /h$ at which Bloch states are resolved, where we recall that $D=D_1 v$, with $v$ the particle's speed, and for a billiard $\nu_E= kh A_c/v$. Then, we obtain that $t_D<t_H$ if and only if $L \lesssim \langle N_B\rangle \sim \sqrt{N}$ which is the expected result. Note that $\langle N_B\rangle = \alpha N/\langle N_B\rangle$ with $\alpha = A_c D_1/(h_0 Z^2) \sim \mathcal{O}(1)$ so there is no contradiction with our original estimate. The second argument comes from the matching of the diffusive regime conductance $\langle g(L)\rangle \approx g_0/(1+L)$, with $g_0=N$, to its asymptotic average value $\langle g_\infty\rangle \sim \langle N_B\rangle$ from where we obtain again that the ohmic regime is confined to $L\lesssim N/\langle N_B\rangle$. 

The ohmic regime of diffusive periodic systems is of course also observed in our RMT periodic chain model as plotted in figure \ref{fig-1_g-COE}, where we see that  
\begin{equation}\label{g-ohm-reg}
\langle g(L)/N\rangle = \langle N/g(L)\rangle^{-1}=(L+1)^{-1}.
\end{equation}
We note that this is one of the signatures of the metallic regime in disordered wires and a consequence of the $\mathcal{O}(1)$ fluctuations of the conductance. We come back to this in section \ref{section-g-dist}. Note that we have shown that Ohm's law in periodic diffusive waveguides is a property of the \textit{typical} resistance $R=1/g$ (without ensemble average) in contrast to its exponential regime which is observed in a set with measure going to zero as $N\rightarrow\infty$ [see figure \ref{fig-1_g}].

\begin{figure}[h]
 \centering
 \includegraphics[width=0.8\columnwidth]{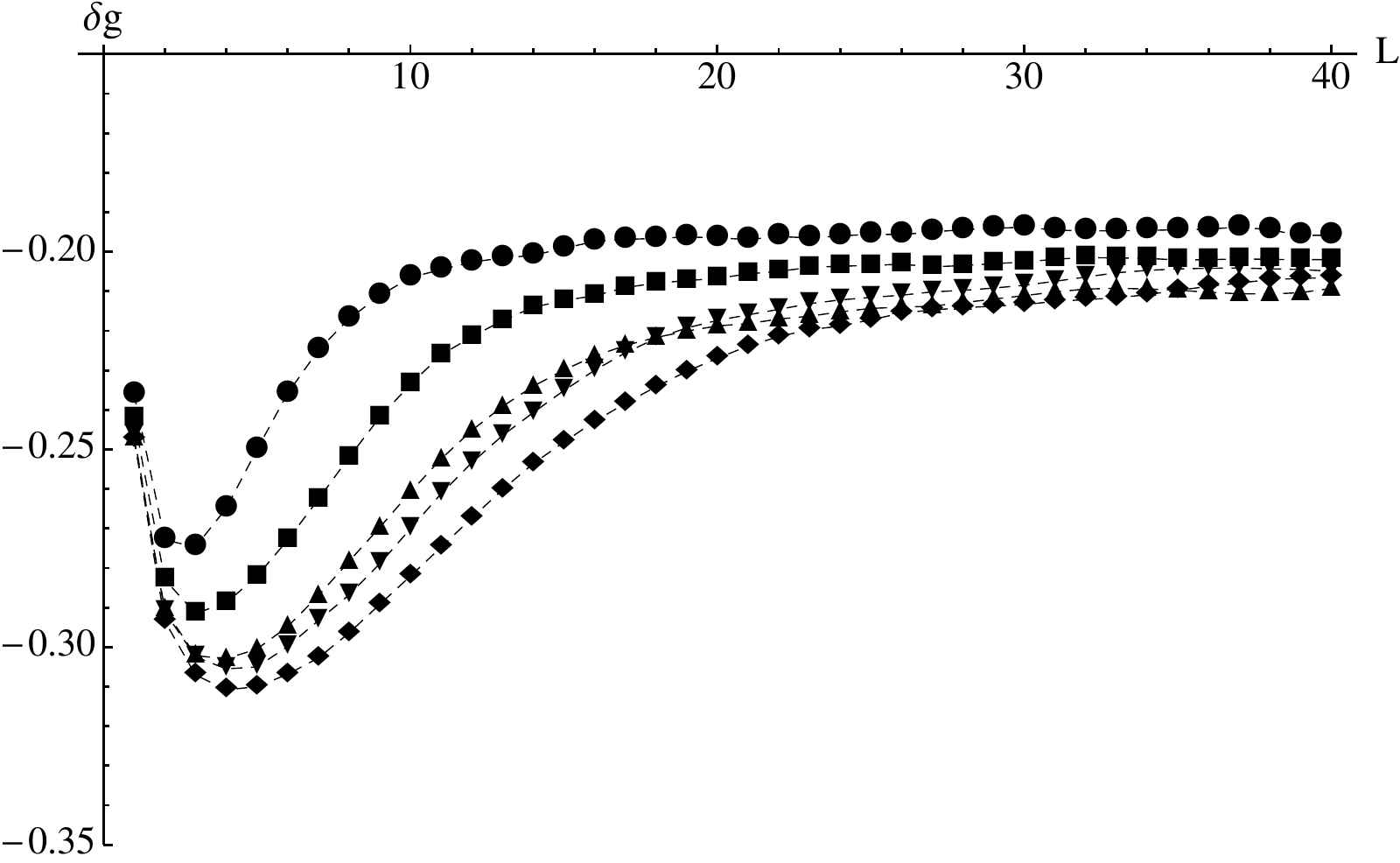}
 \caption[Weak localization in a RMT periodic chain]{Plots of $\delta g = \langle g(L)\rangle_{\rm COE} - \langle g(L)\rangle_{\rm CUE}$ for $N = 10, 20, 30, 40, 50$ (dots, squares, tringles up, tringles down, diamonds) showing the weak localization correction in a RMT periodic chain as a function of its length $L$.}
 \label{fig-dg} 
\end{figure}

\section{Weak localization}\label{section-weak-loc-g}

As we discussed in section \ref{section-dmpk}, the average conductance in a disordered wire or quantum dot (a single cell chain) differs for systems with and without time reversal symmetry, namely $\delta g = \langle g\rangle_{\rm COE} -  \langle g\rangle_{\rm CUE}$ is given by
\begin{equation}
\delta g = \left\{
\begin{array}{ll}
  \displaystyle - \frac{1}{4} & \;\mbox{for a quantum dot}   \\
  &\\
  \displaystyle - \frac{1}{3} & \;\mbox{for a disordered wire.} 
\end{array}\right.
\end{equation}
This difference, called the weak localization correction (WLC), is a purely quantum effect which stems from the destructive interference of time-reversed trajectories present in time-reversal symmetric systems (corresponding to the COE). 

On the other hand, as we saw in section \ref{subsection-nb-rmt}, the number of propagating Bloch modes in the RMT periodic chain model also displays a weak-localization-like effect shown by equation \eref{nb-rmt}, this is $\delta N_B = \langle N_B\rangle_{\rm COE} - \langle N_B\rangle_{\rm CUE} = -0.2$. Since in long periodic chains $\langle g \rangle \sim \langle N_B\rangle$, then we also expect a WLC in the Bloch-ballistic regime. This is shown in figure \ref{fig-dg} where $\delta g$ is plotted as a function of $L$ for the RMT periodic chain model. By construction, we expected to recover $\delta g (L=1) = -1/4$ (the quantum dot WLC) as shown. Then, for $1<L\lesssim \sqrt{N}$, i.e. during the ohmic regime, $\delta g$ undergoes a transient stage before approaching a constant plateau close to $-0.2$ for $L\gg \sqrt{N}$, i.e. in the Bloch-ballistic regime, as expected. We note that the magnitude of $\delta g$ is largest during the ohmic (metallic) regime taking a value $\sim - 0.3$, which is close to the WLC in a disordered wire. 

In conclusion, although the WLC is usually associated with the metallic regime, here we see that a small difference $\delta g = - 0.2$ also persists in the Bloch-ballistic regime for long periodic chains. Recently, the WLC for a periodic system was studied \cite{tian2009} assuming that the Ehrenfest time is larger than the ergodic time. In our case, where RMT is a good model for the periodic waveguide, the Ehrenfest time is smaller than the ergodic time.

\section{Conductance fluctuations}\label{section-conductance-fluct}

Conductance fluctuations in chaotic periodic chains, characterized by the variance $\mbox{Var}[g]$, behave quite differently to the disordered wire. For the latter, in the metallic (ohmic) regime $\mbox{Var}[g] \sim \mathcal{O}(1)$ independent of $N$, $L$ and disorder properties, which is the reason they are called universal conductance fluctuations. In a quantum dot $\mbox{Var}[g]= 1/8\,\beta^{-1}$ whereas for a disordered wire $\mbox{Var}[g]= 2/15\,\beta^{-1}$, the discrepancy being a result of non-geometric correlations of the eigenvalues $T_n$ in the DMPK equation\cite{beenakker1997}. For a long wire, once the localized regime is reached the conductance fluctuations diverge, growing such that $\mbox{Var}[\log{g}] \sim L$. 

\begin{figure}[h!]
 \centering
 \includegraphics[width=.85\columnwidth]{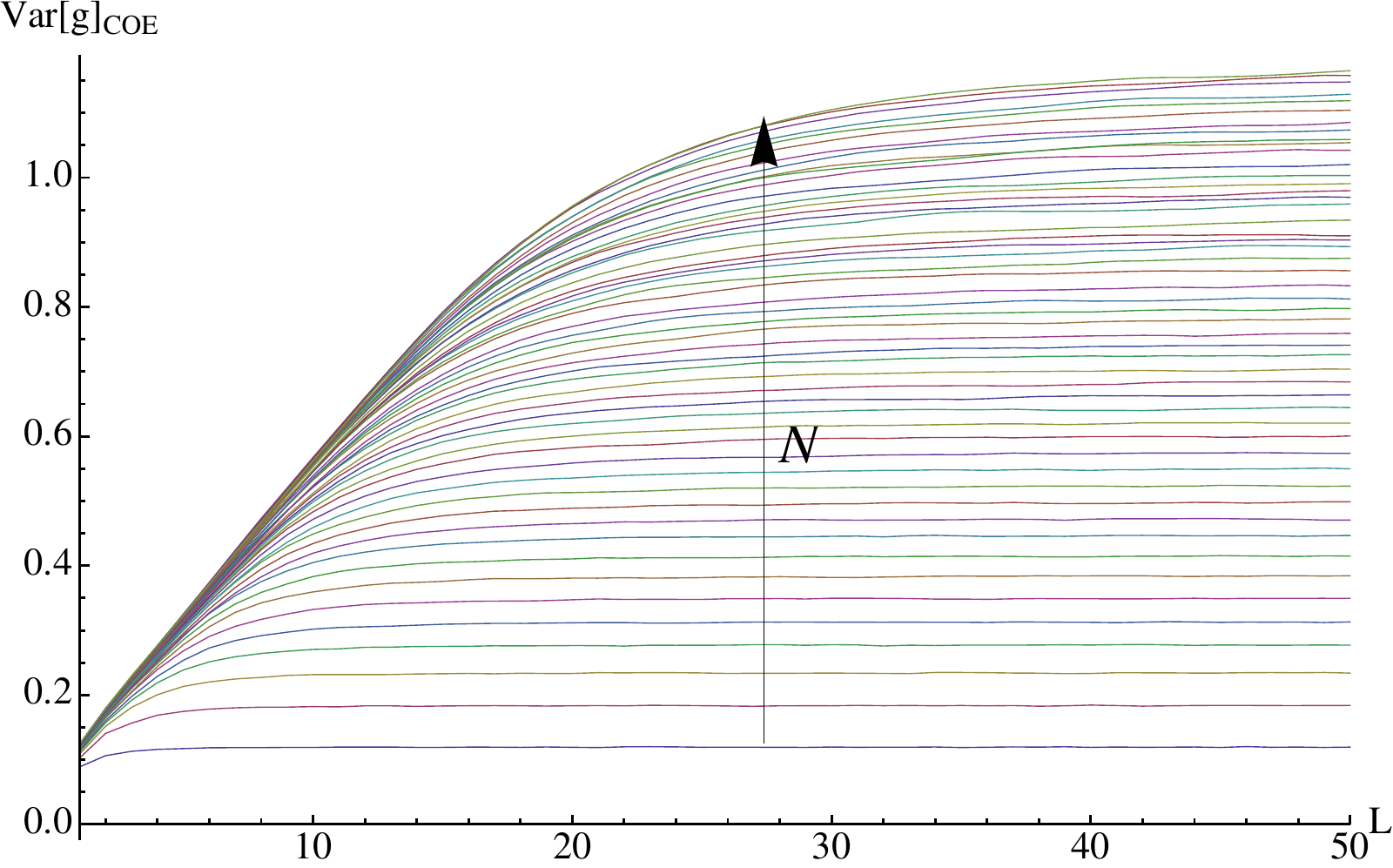}\\
 \vspace{0.5cm}
 \includegraphics[width=.85\columnwidth]{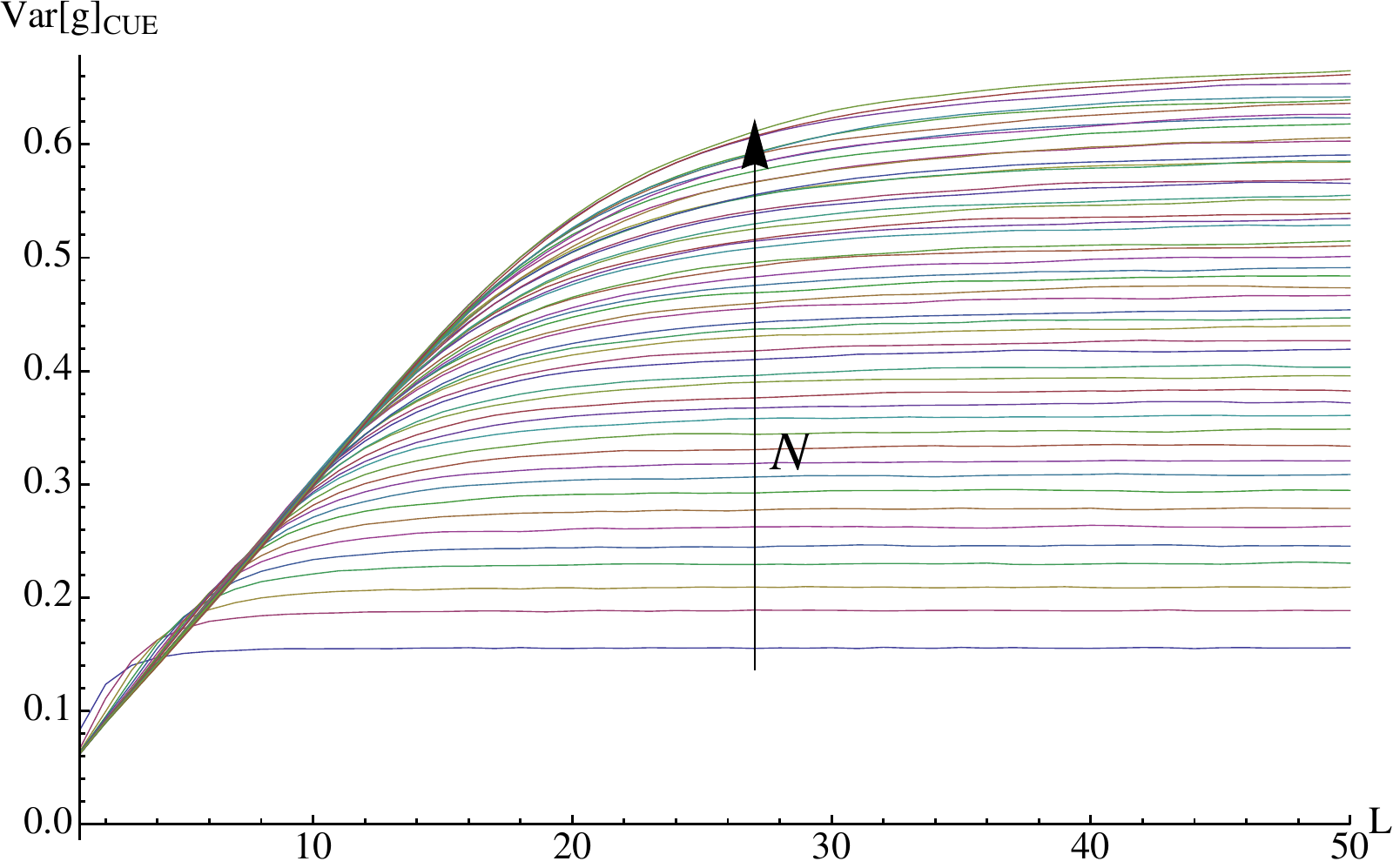}
 \caption[Conductance variance for the COE and CUE periodic chains as a function of $L$]{Conductance variance for the COE and CUE periodic chains as a function of $L$. The plots show ensembles with $N=1$ to 50. For $L\lesssim\sqrt{N}$ the variance grows linear with $L$ [see \eref{var-lin-reg}] and reaches an asymptotic value $\sim\sqrt{N}$ [see \eref{var-large-L}].}
 \label{fig-var-nb} 
\end{figure}

\begin{figure}[h!]
 \centering
 \includegraphics[width=.85\columnwidth]{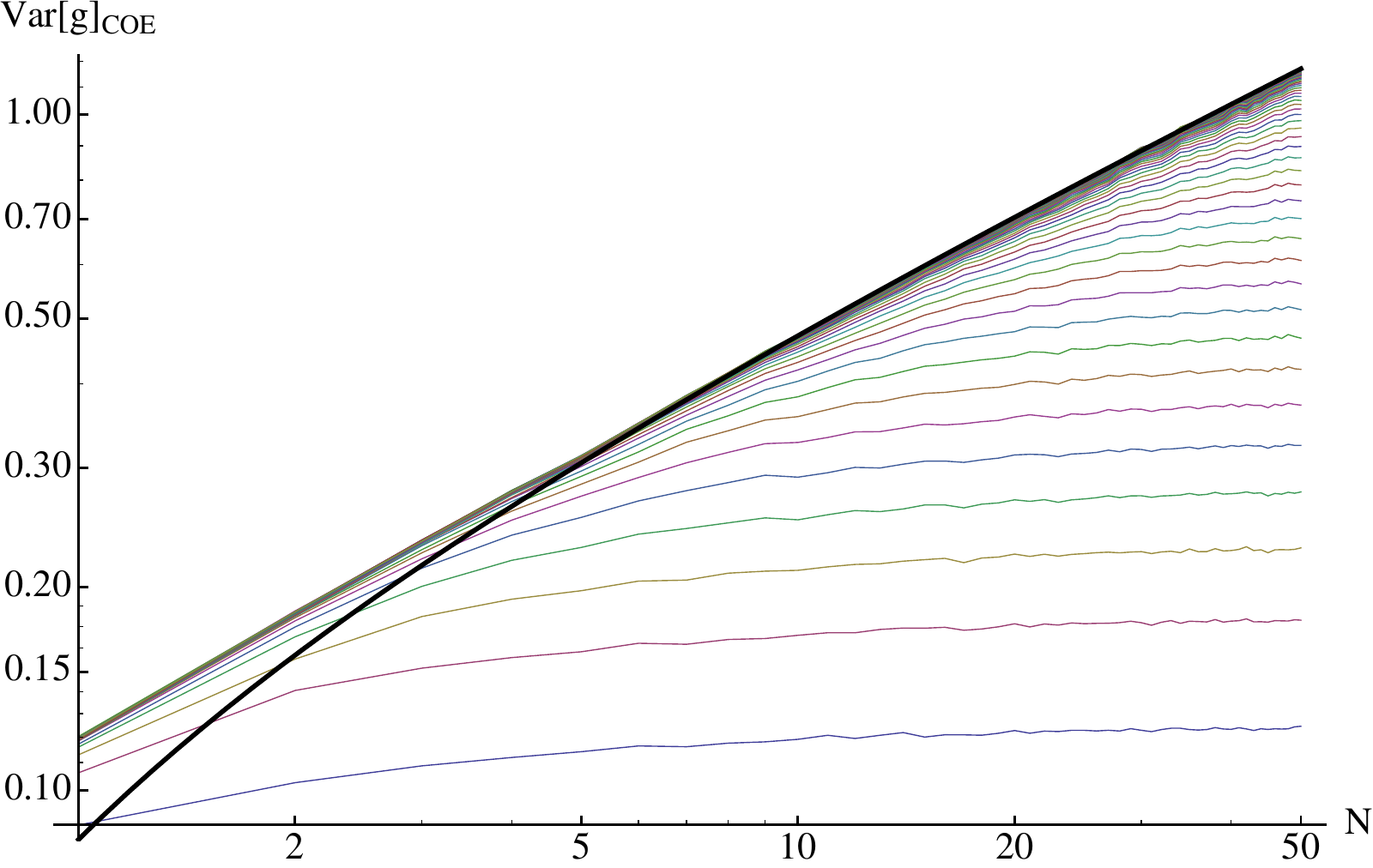}\\
 \vspace{0.5cm}
 \includegraphics[width=.85\columnwidth]{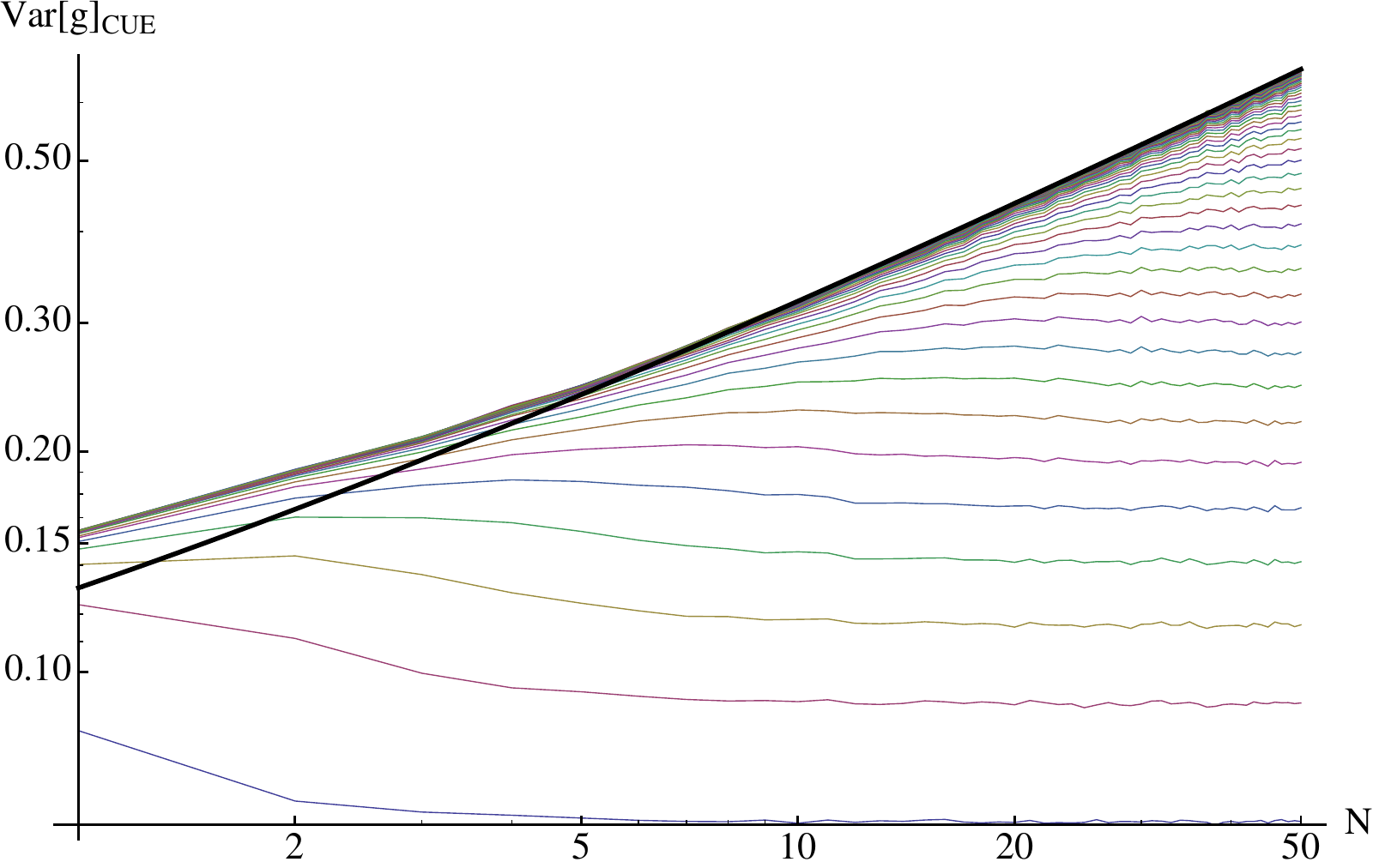}
 \caption[Conductance variance for the RMT periodic chains as a function of $N$]{Log-log plot of the conductance variance for the COE and CUE periodic chains as a function of $N$. The plots show the variance for different values of $L$ (color lines). For large $N$ and $L\gtrsim\sqrt{N}$ the variance reaches the Bloch-ballistic regime with $\mbox{Var}[g]\sim\sqrt{N}$ (black line) [see equation \eref{var-large-L}].}
 \label{fig-var-nb-large-L} 
\end{figure}

Alternatively, as depicted in figure \ref{fig-var-nb}, the conductance variance in a chaotic periodic chain increases linearly with $L$ during the metallic regime and is well described in the range $2<L\lesssim\mathcal{O}(\sqrt{N})$ by
\begin{equation}\label{var-lin-reg}
\mbox{Var}[g(L)] = \left\{
\begin{array}{ll}
0.05\,L + 0.09  & \;\mbox{for COE}   \\
0.03\,L + 0.04 & \;\mbox{for CUE}   .  
\end{array}
\right.
\end{equation}
By construction, for $L=1$ the quantum dot value is obtained. The linear regime given by  (\ref{var-lin-reg}) breaks at $L\sim\mathcal{O}(\sqrt{N})$ signaling the diffusive to Bloch-ballistic transition and is followed by $\mbox{Var}[g(L)]$ reaching an asymptotic constant value of the same order. This asymptotic value can be understood using decomposition \eref{g_inf} and the law of total variance to obtain
\begin{align}\label{g_var_bound}
\mbox{Var}[g(L)] 	&= \langle N_B\mbox{Var}[A_{N_B}|N_B]\rangle +  \mbox{Var}[N_B\langle A_{N_B}|N_B\rangle] \nonumber \\			
				&\leq \frac{1}{4}\langle N_B\rangle + \mbox{Var}[N_B],
\end{align}
where we used that $\mbox{Var}[A_{N_B}|N_B]<1/4$ since $A_{N_B}$ has its support in $[0,1]$\footnote{In general, the variance of a random variable with compact support on $[a,b]$ is bounded above by $(b-a)^2/4$.}. We know from chapter \ref{chapter-nb} that $\langle N_B\rangle \sim  \mbox{Var}[N_B] \sim \sqrt{N}$ for large $N$ [see  \eref{nb_kk_diff},  \eref{nb-rmt}, \eref{var-nb}]. As a result we have that 
\begin{equation}\label{var-large-L}
\lim_{L\to\infty}\mbox{Var}[g(L)] \sim \sqrt{N} .
\end{equation}
This is shown in figure \ref{fig-var-nb-large-L} for the COE and CUE periodic chain models.
\newpage
\newpage

\section{Conductance distribution}\label{section-g-dist}

In this section we summarize some qualitative properties of the conductance pdf $P(g)$ in periodic chains. For short systems in the metallic regime, before the asymptotic Bloch-ballistic state is reached at $L\sim\sqrt{N}$, $P(g)$ is Gaussian --like the metallic regime of a disordered wire-- with mean and variance given by \eref{g-ohm-reg} and \eref{var-lin-reg}, respectively. Although the variance grows with $L$, there is a range where it is small enough so that $\langle 1/g\rangle \sim 1/\langle g\rangle$. In fact, consider the expansion 
\begin{equation}\label{g-moments-exp}
\langle g^{-1}\rangle = \langle g\rangle^{-1} - \mbox{Var}[g]\, \langle g\rangle^{-3}  + \mathcal{O}\left(\langle g\rangle^{-4} \right) .
\end{equation}
Since $\langle g\rangle = N/(L+1)$ we have that deep in the ohmic regime, where $L\ll\sqrt{N}<N$, $\mbox{Var}[g]\ll L$ and $ \mbox{Var}[g]\, \langle g\rangle^{-3} \ll \langle g\rangle^{-1}$. Hence, keeping the dominant term in \eref{g-moments-exp} we obtain $\langle 1/g\rangle \sim 1/\langle g\rangle$. Note that at the diffusive to Bloch-ballistic threshold $L\sim \sqrt{N}$,  $\mbox{Var}[g] \sim \sqrt{N}$ therefore apparently the first term is still dominating becasue $\mbox{Var}[g]\, \langle g\rangle^{-3} \sim 1/N < \langle g\rangle^{-1} \sim 1/\sqrt{N}$. However, the approximation breaks down because $P(g)$ losses its Gaussian shape so the higher order moments in \eref{g-moments-exp} cannot be neglected.

\begin{figure}[h]%
 \centering
 \includegraphics[width=0.48\columnwidth]{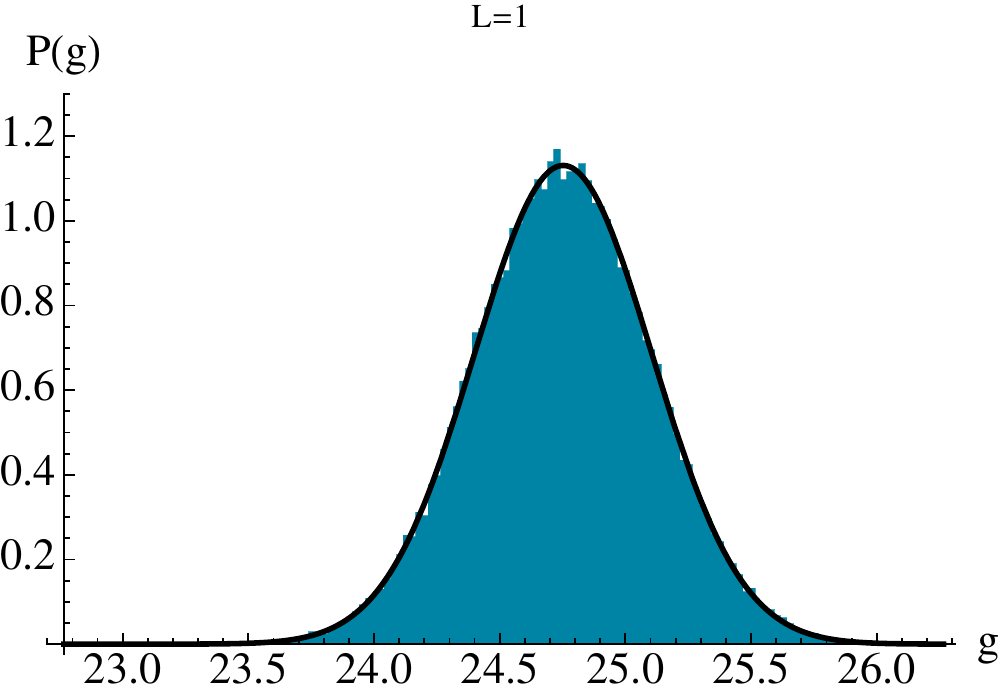}
 \includegraphics[width=0.48\columnwidth]{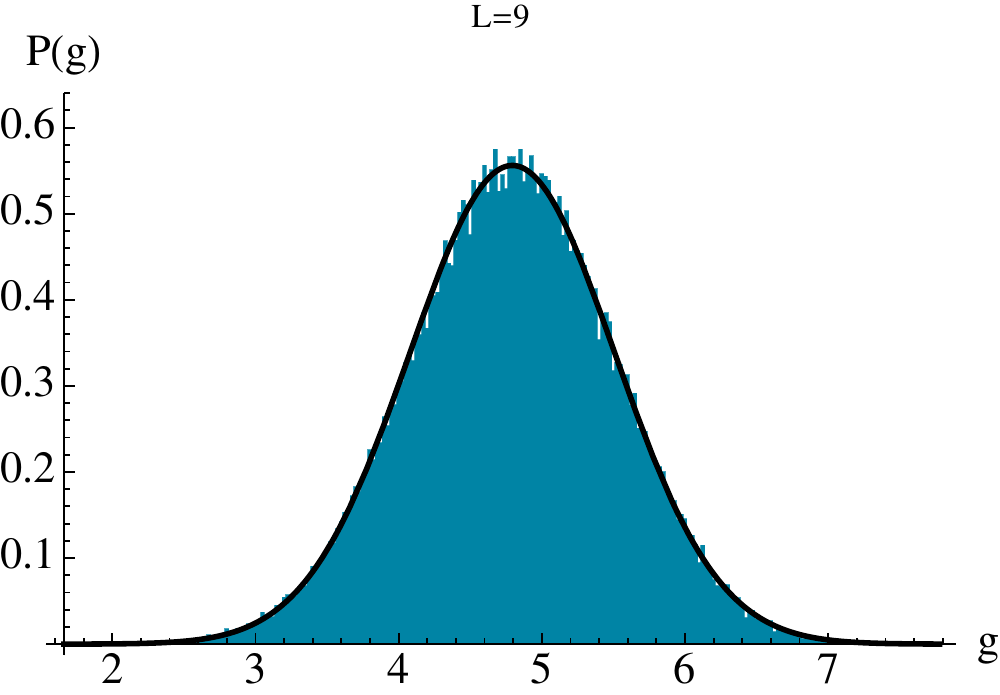}\\
 \vspace{0.1cm}
 \includegraphics[width=0.48\columnwidth]{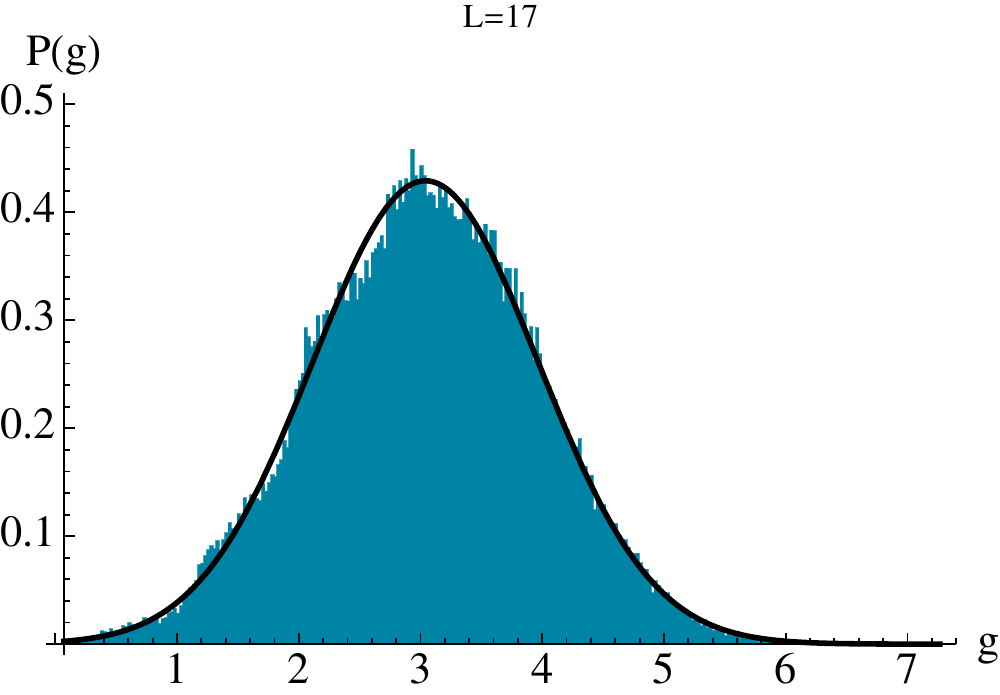}
 \includegraphics[width=0.48\columnwidth]{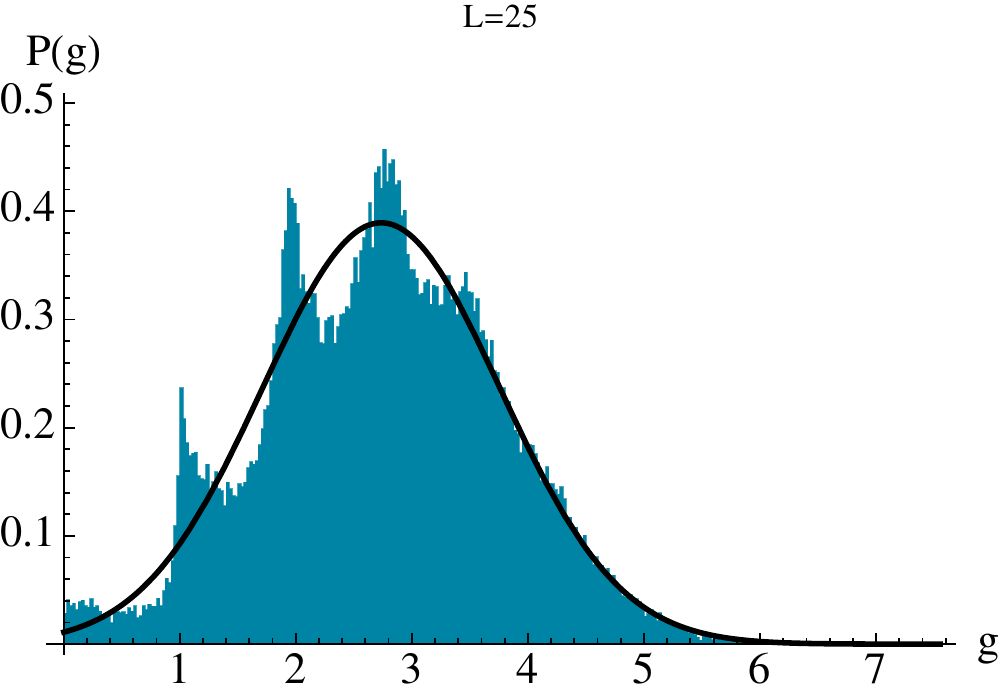}\\
  \vspace{0.1cm}
 \includegraphics[width=0.48\columnwidth]{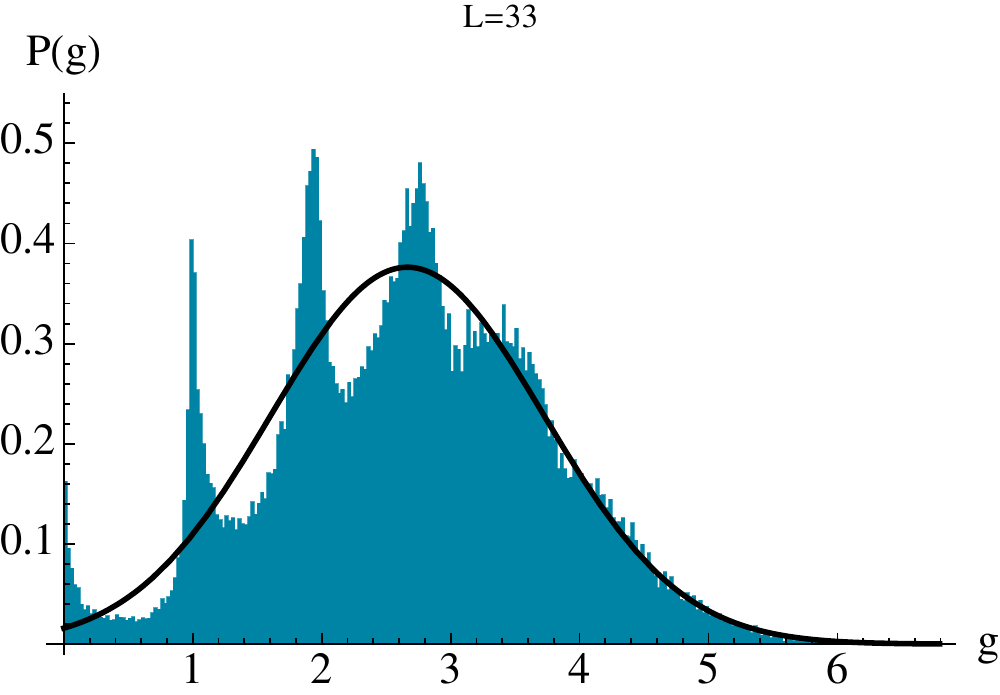}
 \includegraphics[width=0.48\columnwidth]{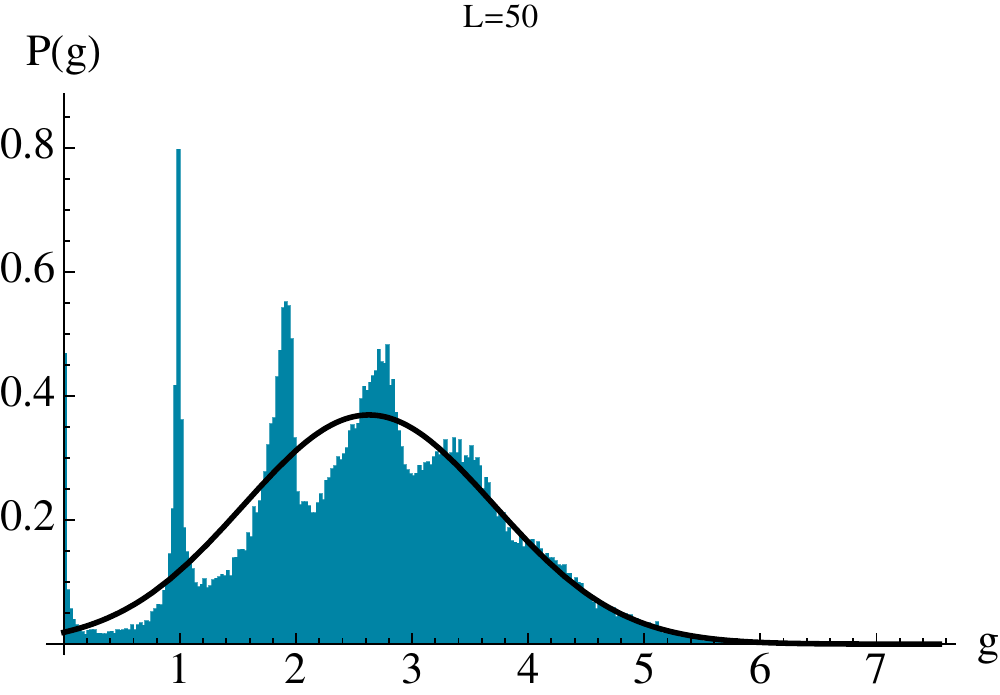}
 \caption[Conductance pdf of a COE periodic chain model as a function of $L$]{Conductance pdf of a COE periodic chain model with $N=50$ for several values of $L$. We see that up to $L=9$ the distribution is clearly gaussian (black line) and at $L=17$ peaks start to develop. At $L=50$ the distribution has reached its long-chain stationary form consisting in a multimodal structure which arises from the asymptotic quasi-periodic behavior of $g$ around its average $\sim N_B$. Given an integer $n$ the peak just below $g=n$ is due to realizations with $N_B=n$. Note that for $N_B=1$ the peak is very narrow and its center close to one but for bigger $N_B$ the peaks start to spread and locate further away from its respective integer upper boundary. This can be understood from the behavior of $\langle A_{N_B}  | N_B\rangle$ as a function of $N_B$ [see \eref{g_inf} and figure \ref{fig-Anb}].}
 \label{fig-p_g} 
\end{figure}

In figure \ref{fig-p_g}, the distribution $P(g)$ of a COE periodic chain is plotted for different values of $L$. $P(g)$ starts being a Gaussian with its center $\langle g \rangle = N/(L+1)$ decreasing with $L$ until $\langle g \rangle \sim \langle N_B\rangle\sim\sqrt{N}$ at $L\sim\sqrt{N}$, where the pdf begins to spread and develop a multimodal structure signaling the diffusive to Bloch-ballistic transition. The asymptotic long-chain distribution converges to this shape with multiple peaks which arise as a consequence of the Bloch modes ballistic propagation. These peaks correspond to realizations of the conductance $g=N_B A_{N_B}$ with different $N_B$. As we have discussed, $\langle A_{N_B}|N_B\rangle$ is close to 1 for small $N_B$ and tends to $1/2$ for larger $N_B$, hence the better defined peaks at lower $N_B$. This is similar to what is observed in CUE chains.

\chapter{Conclusions}\label{chapter-conclusion}

This thesis has dealt with semiclassical transport in periodic chains of chaotic cavities. Our analysis has been based on the study of Landauer conductance as a function of the chain length and energy. For this purpose, we have employed a realistic quasi-one-dimensional waveguide model were we performed direct numerical calculations as well as an RMT periodic chain model. The link of RMT to chaotic systems is well known\cite{bohigas1984} and has been mathematically proved for closed systems\cite{haake2004, haake2007}; this connection explains the famous universality observed in the quantum dynamics of chaotic systems as predicted by the Gaussian ensembles of Hamiltonians. The application of RMT to open chaotic systems is more recent\cite{blumel1990,jalabert1994} and known to be appropriate in cases with strongly chaotic unit cells where the Ehrenfest time is much smaller than the ergodic time. In the opposite case, non-universal behaviour is expected. We always assumed that systems were in the RMT regime in this work. The connection between RMT and semiclassical averages is given by taking an ensemble of energy realizations over several mean level spacings, which is a large interval for quantum mechanics but classically infinitesimal. We defined our RMT periodic chain model by taking scattering matrices from one of the circular ensembles (depending on the symmetries of the system being modeled) and composing it $L$ times to obtain a finite periodic chain.

The system under study consisted in a quasi-one-dimensional waveguide with a scattering region made of a chain of $L$ identical cavities connected to two semi-infinite plane leads at its extremes. We assumed the underlying wave equation was the Helmholtz equation, i.e. waves (and their corresponding classical particles) were free inside the waveguide and only scattered by the geometry of the boundaries. We focused on cavities whose classical dynamic was chaotic and strongly diffusive. The correspondence of the observed semiclassical statistics in this chaotic systems and RMT means that the details of their dynamics are not relevant for the universal properties we have found, therefore, we expect them to hold for different underlying wave equations as long as they exhibit strong chaotic dynamics. 

We have found that the average conductance as a function of the chain length displays two regimes. First, for chains of length $L\lesssim\sqrt{N}$, where $N$ is the number of propagating modes in the leads, the dynamics is diffusive just like in a disordered wire in the metallic regime and an ohmic scaling for the conductance $\langle g(L)\rangle =N/(L+1)$ is observed. In this regime the conductance distribution is Gaussian and its variance grows linearly with $L$, but is small enough such that $\langle g^{-1} \rangle \approx 1/\langle g\rangle$ holds similarly to a disorder wire. Comparing the COE and CUE ensembles we observed a weak localization effect of magnitude similar to the usual one in the disordered wire. We note that the existence of a semiclassical diffusive regime is usually associated with disordered systems but in fact is a consequence of an underlying chaotic dynamics and can also be observed in periodic systems as we have shown. 

For long periodic chains, an important quantity for transport is the number of propagating Bloch $N_B $ of the associated infinite system. Faure\cite{faure2002} provided a semiclassical explicit form for $\langle N_B \rangle$ in diffusive systems [see equation \eref{nb_k_diff}]; we have checked this result in a physically realistic waveguide and extended its validity when the unit cell possess anti-unitary symmetries [see equation \eref{nb_k_diff}]. We have also linked Faure's result to the universal parametric correlation function \cite{altshuler1993, dittrich1997, dittrich1998}. Additionally, using the Machta-Zwanzig approximation for the diffusion coefficient, we conjectured the RMT value for $\langle N_B \rangle$ [see equation \eref{nb-rmt}] in the COE and CUE ensembles, which we checked numerically. The difference observed $\delta N_B =\langle N_B \rangle_{\rm COE} - \langle N_B \rangle_{\rm CUE} = -0.2$ shows the existence of a novel weak localization effect for the conductance $\delta g\approx  \delta N_B $ which extends into the asymptotic Bloch-ballistic regime [see figure \ref{fig-dg}] of a periodic chain of chaotic cavities. 

The ohmic regime breaks down at $L\sim \sqrt{N}$ and for $L\gg\sqrt{N}$ the asymptotic regime is reached where transport in the periodic chain is ballistic and quasi-periodic in $L$.  This is a consequence of the strong resonances owing to the Bloch modes present in the associated infinite chain. The resonances approach the real axis as $L\to\infty$ and cluster together forming the bands of the infinite periodic system, thus creating $N_B$ effectively propagative channels in the chain. In this regime, the transmission through these $N_B$ channels is quasi-periodic in $L$. We have shown that the average conductance asymptotic value is $\langle g\rangle \approx 0.75 \langle N_B\rangle$. Due to the discrete nature of $N_B$ the conductance distribution shows a multimodal structure in the Bloch-ballistic regime, with multiple peaks related to different integer realizations of $N_B$ [see figure \ref{fig-p_g}]. We showed that the conductance variance approaches an asymptotic value of order $\sqrt{N}$ as $L\to\infty$. On the other hand, the distribution of $N_B$ resembles a Poisson distribution where the \textit{events} correspond to the Bloch modes; however, the non-null correlation of Bloch spectra breaks down the exact analogy. Still, a remarkable feature is observed, namely that $\langle N_B\rangle / \mbox{Var}[N_B]$ approach a constant value as $N\to\infty$, which reminded us of shot noise.  

The Bloch-ballistic to diffusive transition occurring at $L \sim \sqrt{N}$ can be understood from several arguments. In section \ref{section-ohms-law}, we noted that the distribution of the inverse decay length spectrum $\ell_n^{-1}$ possess a linear region in $n$ for $\langle N_B\rangle \leq n \ll N$ and for $n<\langle N_B\rangle$ the inverse decay length is close to zero [see figure \ref{fig-coe-lambdas}] due to the propagating nature of Bloch modes. The linear dependence of $\ell_n^{-1} = n/N$ inferred from the DMPK equation is the reason for the ohmic regime in the disordered wire. Likewise, in diffusive periodic systems, this property holds but with the previously named constraint which implies the ohmic regime $\langle 1/g\rangle \sim L$ holds only for $L\lesssim N/\langle N_B\rangle \sim \sqrt{N}$. An alternative argument is that the ohmic regime should be observed only for chains whose diffusion time $t_D$ is smaller than the unit cell Heisenberg time $t_H $ at which Bloch states are resolved, which leads to the same result. For $L\gtrsim \sqrt{N}$, the conductance can be decomposed as $g=g_\infty + g_{\rm np}$, with $g_\infty$ the quasi-periodic asymptotic conductance which is a sum of the $N_B$ Bloch modes and $g_{\rm np} = 4 e^{-2L/\ell}$ the slowest to decay non-propagating mode. By means of our RMT model, we have studied the pdf of $\ell$ as $N\to\infty$ and shown that under an appropriate scaling [see equation \eref{l*-pdf}] it converges to a limit distribution $\hat{P}(\ell)$ possessing an algebraic tail $\hat{P}(\ell)\sim\ell^{-3}$ for $\ell\to\infty$. This allowed us to conjecture the decay $\langle g_{\rm np}\rangle \sim L^{-2}$ for the non-propagating conductance which we observed with excellent agreement in the cosine waveguide.

\section{Outlook}

There remain several open topics which are natural extensions of the work presented in this thesis. In this section we briefly discuss two of them that seem interesting for future research.  

\subsection{Waveguides with Infinite Horizon}

In this work we have explicitly excluded waveguides with infinite-horizon, i.e. geometries allowing classical particle trajectories with free flights of infinite length. Open systems with strong chaotic properties possessing this property display marginally anomalous diffusion such that $\langle x_t ^2 \rangle = \tilde{D} t\log{t}$ for $t\to\infty$ with $x_t$ the particle position at time $t$ and $\tilde{D}$ the anomalous diffusion coefficient. Hence, it is obvious that Faure's result \eref{nb_k_diff} for $\langle N_B\rangle$ must be modified since the normal diffusion coefficient diverge to infinity. Classically, the asymptotic billiard dynamics in infinite horizon systems is well understood thanks to the work of Bleher\cite{bleher1992} and Szazs\cite{szasz2007}. They showed that the decay of correlation in the infinite horizon Lorentz Gas systems corresponds to the marginal case $C(t) \to t^{-1}$ and that the random variable 
\begin{equation}
\eta_t = \frac{x_t - x_0}{\sqrt{\tilde{D}\, t\log{t}}} 
\end{equation}
converge in distribution to a standard Gaussian law. They also gave an explicit expression for $\tilde{D}$ in terms of the infinite horizon corridors geometry. However, the convergence of $\eta_t$ is weak and do not imply convergence of the moments unlike the finite horizon case where even though most of the convergence theorems are also proved in distribution in practice it is observed that the finite time moments --like the finite time approximations for the diffusion coefficient-- converge to the Gaussian moments [see discussion around equation \eref{moments-relation}]. Hence, the relevance of the classical anomalous diffusion constant $\tilde{D}$ (which can be computed analytically in some cases) to the semiclassical dynamics remains unclear. 

On the other hand, coming back to the derivation of Faure's expression for $\langle N_B\rangle$, there is a direct extension for the infinite horizon case replacing equation \eref{eckhardt_equiv} appropriately. In fact, in the same work where \eref{eckhardt_equiv} was dereived, Eckhardt\cite{eckhardt1995} gave the asymptotic dependance for the semiclassical variance when the classical correlation decays as $C(t) \to t^{-1}$, namely
\begin{equation}
 \sigma_\hbar^2 \sim \frac{\log{ t_H}}{t_H}.
\end{equation}
Using $t_H = \nu_E/h = \nu_E k/(2\pi)$ and conjecture \eref{faure_conjecture}, we obtain the dominant term in the semiclassical limit 
\begin{equation}\label{nb-ih}
\langle N_B\rangle \sim \sqrt{k \log{\left(\frac{\nu_E}{2\pi} k\right)}} \quad \mbox{for $k\to\infty$}.
\end{equation}
The difference between this expression and \eref{nb_kk_diff} is numerically small in the range of energies we explored in this thesis. Employing a infinite horizon version of the cosine billiard,  we calculated the average number of propagating Bloch modes but it was not possible to statistically distinguish \eref{nb-ih} from \eref{nb_kk_diff} because of the large fluctuations displayed by $\langle N_B\rangle$. In order to test the validity of \eref{nb-ih}, a much deeper semiclassical region (higher energies) must be considered. An interesting alternative to explore would be to test \eref{nb-ih} in an RMT model for periodic chains with infinite horizon, which could be done adding a \emph{direct process} to the Circular ensembles of scattering matrices we have considered. This can be accomplished using the Poisson kernel, which is a generalization of the Circular ensembles given by the pdf
\begin{equation}\label{poissonkernel}
P(\bm S) \propto |\mbox{Det}(1 - \bar{\bm S}^\dagger\bm S)|^{-\beta(2N - 1) - 2},
\end{equation}
where $2N=N_1+N_2$ with $N_1$, $N_2$ the number of propagating modes on the right and left leads, respectively. The Poisson kernel given by \eref{poissonkernel} is the maximum entropy pdf satisfying the Circular ensembles symmetry constraints plus the condition $\langle \bm S^p\rangle = \langle \bm S\rangle^p = \bar{\bm S}^{p}$, where the average scattering matrix $\bar{\bm S}$ is a given sub-unitary\footnote{A sub-unitary matrix $\bar{\bm S}$ is such that the eigenvalues of $\bar{\bm S} \bar{\bm S}^\dagger$ are $\leq 1$.} matrix. In case $\bar{\bm{S}}=0$ one recovers the Circular ensemble with $P(\bm S) \propto \mbox{constant}$. The addition of a direct process modeling an infinite horizon configuration can be done defining the unit cell scattering matrix as an element of the Poisson kernel with
\begin{equation}
\bar{\bm S} =\left(
\begin{array}{cc}
0 & \bar{\bm t}  \\
\bar{\bm t}  &  0      
\end{array}
\right),
\end{equation}
where $\bar{\bm t}$ is the transmission matrix associated to the direct process of the infinite horizon unit cell. 

\subsection{Weakly disordered periodic chains}

In practice, any experimental realization of a periodic chain in the laboratory will always possess some degree of disorder because fabrication of trully identical unit cells is not possible. Thus, it would be interesting to study how weak disorder breaks down the dynamics we have discovered for perfect periodic chains. We have started working in this direction by considering an RMT model of weakly disordered time-reversal symmetric periodic chains, consisting in a chain of very similar but not identical cavities. We consider the case where all the cavities are taken from the COE ensemble; the chain ensemble is composed of cavities whose orthogonal scattering matrices are
\begin{equation}
\bm{S}_i = \bm{U}^T \displaystyle e^{ i\epsilon \bm{H}_i} \bm{U} \quad i=1,\ldots,L, 
\end{equation}
where $L\in\mathbb{N}$ is the chain length, $\bm U \in U(N)$ is fixed for each realization and $\{\bm H_i\}_{i=1}^N$ is a realization-dependent sequence of Gaussian random real symmetric matrices. The parameter $\epsilon$ is related to the disorder strength and define a particular ensemble (i.e., it does not change between realization of a given ensemble). The chain scattering matrix is obtained composing the $L$ matrices $\bm S_i$ by \eref{t-r-comp}. It is clear that in the limit $\epsilon = 0$ we recover the perfect periodic chain. Let $\bm S = \bm U^T \bm U$ be an unperturbed unit cell scattering matrix. Then, all scattering matrices in the chain are perturbations of this original matrix, and for small $\epsilon$ we have
\begin{equation}
\bm S_i = \bm S + d\bm S_i. 
\end{equation}
Hence, according to \eref{coe-ds} and \eref{mu_rmt_orth}, the diference between the perturbed unit cell scattering matrix and the original one can be measured by $\mu_{\rm O}(\bm S - \bm S_i) = \mu_{\rm O} (d\bm S_i)  \sim \epsilon^N$. 

\begin{figure}[h]
 \centering
 \includegraphics[width=1\columnwidth]{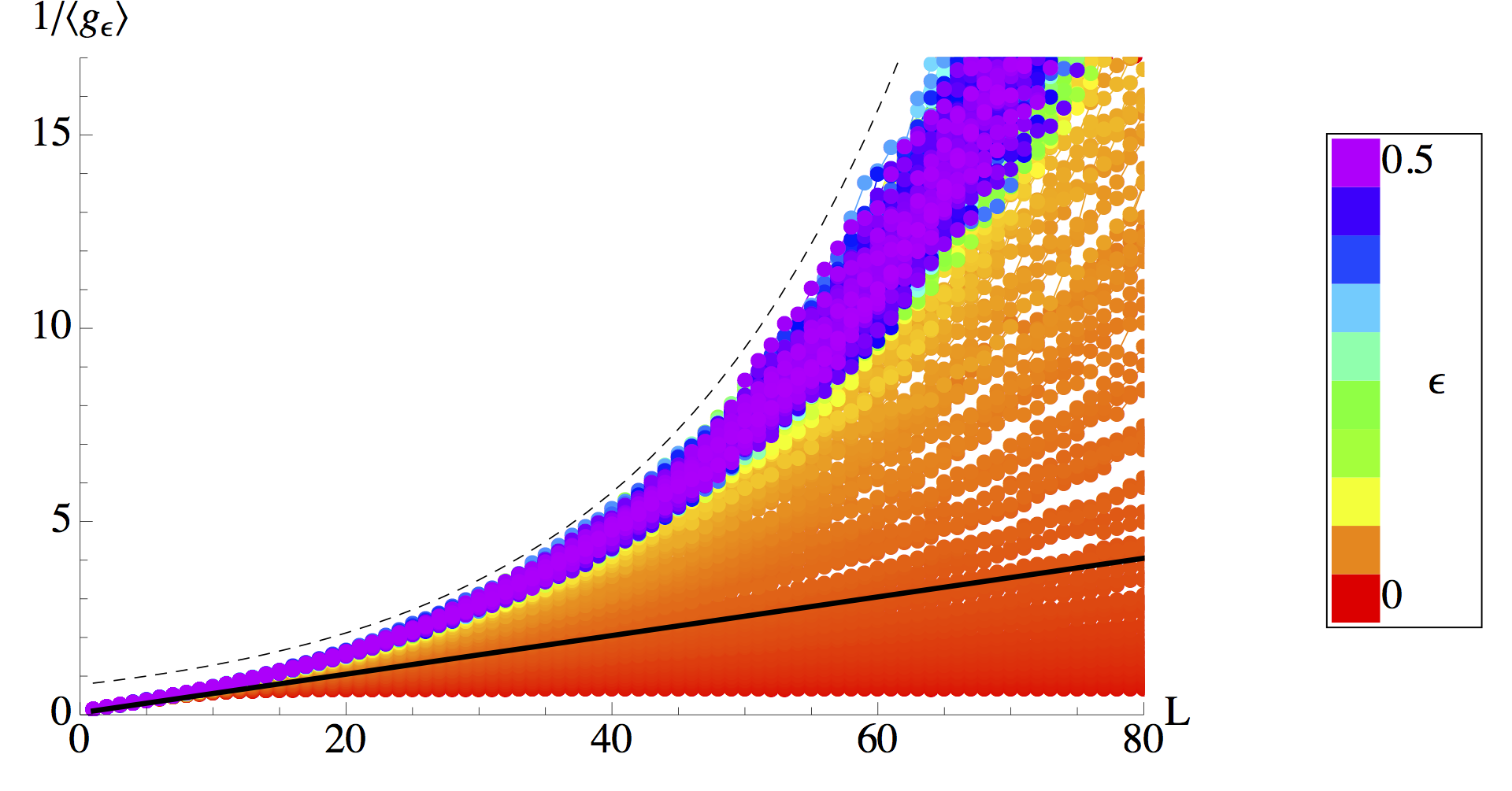}
 \caption[Inverse average conductance of a weakly disordered chain]{Inverse of the average conductance $1/\langle g_\epsilon(L)\rangle$ as a function of $L$ for several values of $\epsilon\in[0, 0.5]$ and $N=20$. The black line is $(L+1)/N$ which describes the ohmic regime and the dashed line is an exponential $e^{(L-5)/N}$ shown here as a guide to the eye. For $\epsilon=0$ the asymptotic value of $1/\langle g_\epsilon(L)\rangle$ is a finite constant and for any $\epsilon>0$ is infinite since localization makes conductance null for $L$ sufficiently large. In the latter case, it is possible to distinguish two different cases: for $\epsilon$ small enough, there is still a region where $1/\langle g_\epsilon(L)\rangle < (L+1)/N$ with negative curvature (see figure \ref{fig-disordered-chain-2}) and for $\epsilon>\bar{\epsilon}$ this is not observed and pure exponential divergence after the ohmic regime is observed. We have found numerically that $\bar{\epsilon}\approx 0.4$ for $N=20$.
 }
 \label{fig-disordered-chain} 
\end{figure}

\begin{figure}[h]
 \centering
 \includegraphics[width=0.95\columnwidth]{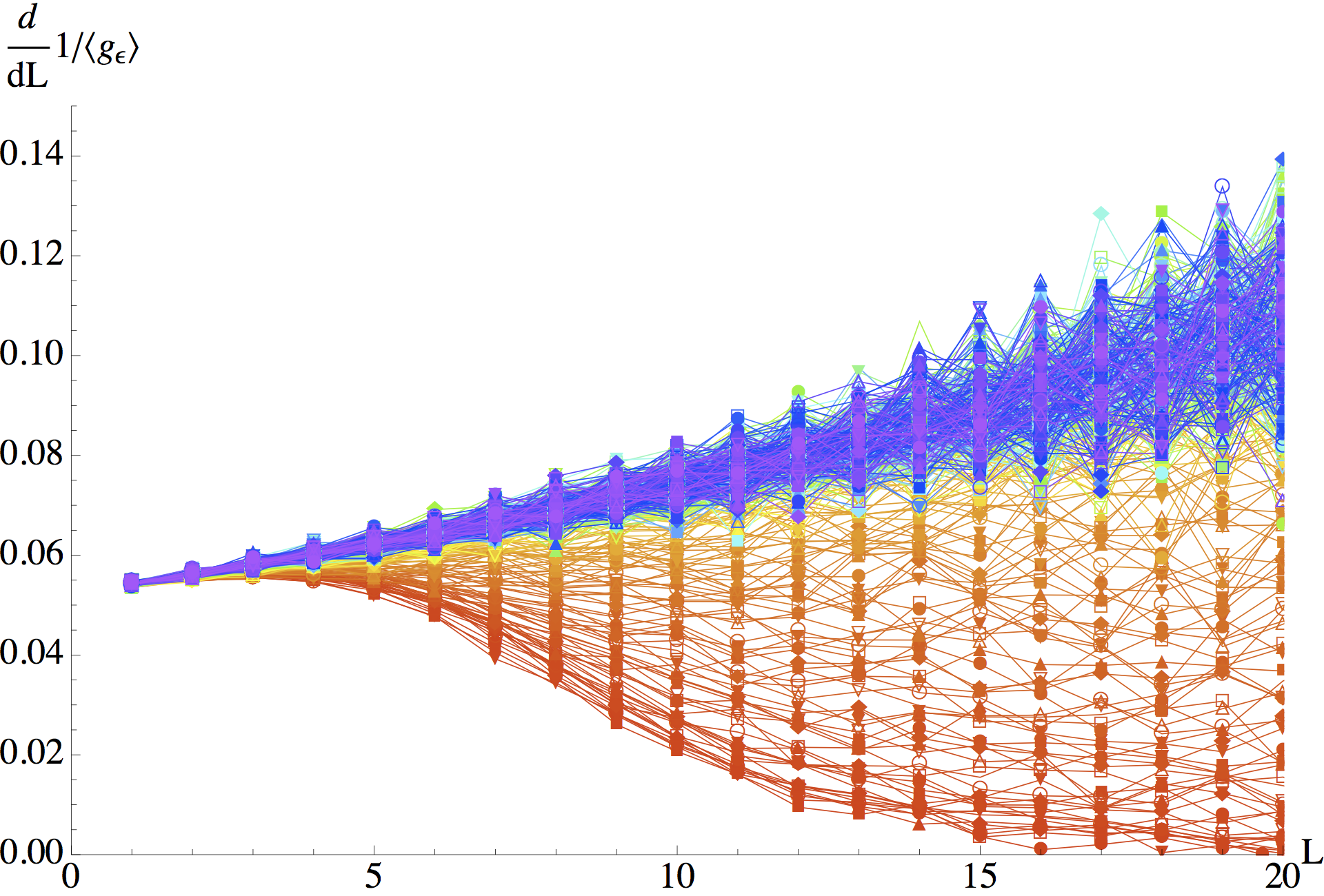}
 \caption[Inverse average conductance slope of a weakly disordered chain]{Derivative of the inverse average conductance as a function of $L$ for the same $\epsilon$-realizations than figure \ref{fig-disordered-chain} (the color map is the same of that figure). It can be seen that for small enough $\epsilon$ there is a region with negative second derivative, i.e. where $1/\langle g_\epsilon(L)\rangle$ is convex (negative curvature). 
 }
 \label{fig-disordered-chain-2} 
\end{figure}

In order to study how disorder breaks down the transport properties of the periodic chain we have focused on the conductance as a function of $L$ and $\epsilon$, which we denote $g_\epsilon(L)$. It is known that in quasi-one-dimensional systems the presence of disorder of any strength will trigger Anderson localization, therefore for any $\epsilon>0$ we have that $\langle g_\epsilon(L)\rangle \to 0$ as $L\to\infty$. However, for sufficiently small $\epsilon$, we expect that in a range of $L$ some of the periodic chain characteristics will still be observed. In figure \ref{fig-disordered-chain}, we show $1/\langle g_\epsilon(L) \rangle$ for several values of $\epsilon\in[0, 0.5]$ and $N=20$. In all cases we see the ohmic linear regime for $L<L_{\rm max}$; as $\epsilon$ increases $L_{\rm max}$ grows from $\sim\sqrt{N}$ as we have shown for a periodic chain to $\sim N$ like in a disordered wire. Following the ohmic regime we observe different behaviors depending on the value of $\epsilon$. For $\epsilon=0$ we have that $1/\langle g_\epsilon(L) \rangle \to \mbox{constant}$ for $L\to\infty$ as expected. As $\epsilon>0$ grows, we see how $1/\langle g_\epsilon(L)\rangle$ starts diverging as $L\to\infty$. We note that for small $\epsilon$ a negative curvature region is still observed for $L>L_{\rm max}$ before reaching exponential divergence signaling Anderson localization [see figures \ref{fig-disordered-chain} and \ref{fig-disordered-chain-2}]. This sub-exponential regime disappears for larger $\epsilon$ suggesting that in those cases disorder is strong enough to erase any trace of periodicity.  We have found numerically that this transition from weak disorder to localization occurs near $\epsilon=\bar{\epsilon}\approx0.4$ in the case considered here (N=20). Since $\mu_{\rm O} (d\bm S_i)  \sim \epsilon^N$, we expect that as $N\to\infty$, this transition point moves to $\bar{\epsilon}\to1$.

The previous argument shows that for weak disorder near a periodic chain there is a regime before localization where some reminiscent transport properties of the original system survive. Many interesting questions remain open, in particular about the nature of the transition from weak into full disorder which might be related to a percolation problem\cite{zhang1994, islam2008}.

\begin{appendices}
\appendix
\chapter{Scattering matrix composition using the Bloch basis}\label{appendix-s-matrix-comp}

In this appendix we show an alternative method to obtain the scattering matrix of a periodic chain given the unit cell scattering matrix. The method consists in writing the $\bm S$ matrix using the transfer matrix Bloch spectrum. Its advantage is that it allows us to keep evanescent modes in the composition, hence it is more precise when the coupling between adjacent unit cells is not completely negligible. 

Let the eigenvectors of the unit cell transfer matrix $\bm M_1$ be given by $\bm F_n^{+}$, $\bm F_n^{-}$ with associated eigenvalues $\lambda_n^{+}$, $\lambda_n^{-}$ where $n=1, M=\ldots, N+N_e$, with $N$ and $N_e$ the number of open and evanescent modes in the plane leads, respectively. We choose the eigenvalues $\lambda^+$ ($\lambda^-$) such that $|\lambda^+|>1$ ($|\lambda^-|$<1) or $|\lambda^\pm|=1$  and its associated eigenstate energy flux is right-going (left-going). Now, consider a periodic chain with $L$ unit cells and let the wavefunction at the chain left end be
\begin{equation}\label{left-end-wf}
\left(
\begin{array}{l}
\bm A_1     \\
\bm B_1
\end{array}
\right) = \sum_{n=1}^M \alpha_n \bm F_n^{+} + \frac{\beta_n}{(\lambda_n^-)^L} \bm F_n^{-}
\end{equation}
with $\bm \alpha = (\alpha_1, \ldots, \alpha_M)^T$ and $\bm \beta = (\beta_1, \ldots, \beta_M)^T$ complex vectors. Therefore, applying the transfer matrix $\bm M =  (\bm M_1)^L$ to \eref{left-end-wf}, we obtain the wavefunction at the right end of the chain,
\begin{equation}\label{right-end-wf}
\left(
\begin{array}{l}
\bm A_2     \\
\bm B_2
\end{array}
\right) = \sum_{n=1}^M \alpha_n (\lambda_n^{+})^L \bm F_n^{+} + \beta_n \bm F_n^{-} . 
\end{equation}
We write the Bloch eigenvectors as
\begin{equation}
\bm F_n^\pm = \left(
\begin{array}{ll}
\bm f^\pm_n     \\
\bm{\bar{f}}^\pm_n
\end{array}
\right),
\end{equation}
where $\bm f^\pm_n$ and $\bm{\bar{f}}^\pm_n$ are vectors of dimension $M$. We define the $M\times M$ matrices $\bm f^\pm$ and $\bm{\bar{f}}^\pm$ with these vectors as their columns, respectively. Then, we have that
\begin{equation}\label{eq-s-1}
\left(
\begin{array}{l}
\bm A_1     \\
\bm B_2
\end{array}
\right) =
\left(
\begin{array}{cc}
\bm f^{+}  & (\bm\lambda_{-})^{-L}\, \bm f^{-}    \\
(\bm \lambda_{+})^L\, \bm{\bar{f}}^+  & \bm{\bar{f}}^-
\end{array}
\right) \left(
\begin{array}{l}
\bm \alpha     \\
\bm \beta
\end{array}
\right) 
\equiv 
\bm R_1 \left( \begin{array}{l}
\bm \alpha     \\
\bm \beta
\end{array}
\right) 
\end{equation} 
and
\begin{equation}\label{eq-s-2}
\left(
\begin{array}{l}
\bm B_1     \\
\bm A_2
\end{array}
\right) =
\left(
\begin{array}{cc}
\bm{\bar{f}}^+ & (\bm\lambda_{-})^{-L}\, \bm{\bar{f}}^{-}    \\
(\bm \lambda_{+})^L\, \bm f^{-}   & \bm f^-
\end{array}
\right) \left( \begin{array}{l}
\bm \alpha     \\
\bm \beta
\end{array}
\right) 
\equiv 
\bm R_2 \left( \begin{array}{l}
\bm \alpha     \\
\bm \beta
\end{array}
\right) 
\end{equation}
with $\bm \lambda_{\pm}$ the $M\times M$ diagonal matrix with the eigenvalues $\lambda^\pm_n$ in its diagonal. Thus, since 
\begin{equation}
\psi_{\rm in} = \left(
\begin{array}{l}
\bm A_1     \\
\bm B_2
\end{array}
\right) 
\qquad \mbox{and}\qquad 
\psi_{\rm out} = \left(
\begin{array}{l}
\bm B_1     \\
\bm A_2
\end{array}
\right),
\end{equation}
we can eliminate $(\bm \alpha, \bm \beta)$ from \eref{eq-s-1} and \eref{eq-s-2} to obtain 
\begin{equation}
\psi_{\rm out} = \bm R_2 \bm R_1^{-1} \psi_{\rm in} ,
\end{equation}
which means the scattering matrix of the length $L$ chain is given by
\begin{equation}
\bm S = \left(
\begin{array}{cc}
\bm{\bar{f}}^+ & (\bm\lambda_{-})^{-L}\, \bm{\bar{f}}^{-}    \\
(\bm \lambda_{+})^L\, \bm f^{-}   & \bm f^-
\end{array}
\right) 
\left(
\begin{array}{cc}
\bm f^{+}  & (\bm\lambda_{-})^{-L}\, \bm f^{-}    \\
(\bm \lambda_{+})^L\, \bm{\bar{f}}^+  & \bm{\bar{f}}^-
\end{array}
\right)^{-1}.
\end{equation}

%

\chapter{Perturbative analysis near a normal transfer matrix}\label{appendix_perturbation}

In this appendix we study the relation between the eigenvalues of $\bm{M}^L$ and $\bm{M}^L\bm{M}^{L\dagger}$ when the transfer matrix $\bm{M}$ is close to be a \textit{normal} matrix\footnote{A matrix is said to be \textit{normal} if it commutes with its Hermitic conjugate or equivalently if its eigenvectors form an orthogonal set.}. Let $\{ \lambda_i, 1/\lambda_i^* \}_{i=1}^{N}$ be the spectrum of $\bm M$ and $\{\Lambda_i^{1}, \Lambda_i^{-1}\}_{i=1}^{N}$ be the spectrum of $\bm{M}^L\bm{M}^{L\dagger}$. If $\bm{M}$ were normal we could map these spectrums such that $\Lambda_i = |\lambda_i|^{2L}$, hence the transfer eigenvalues $T_i = 4\,(2 + \Lambda_i + \Lambda_i^{-1})^{-1}$ associated to propagating Bloch modes would be not dependent on $L$ and equal to one, so for long chains $g=N_B$. Generically, the transfer matrix is not normal, leading to a more involved relation between the sets of $\lambda_i$ and $\Lambda_i$  as a function of $L$ [see discussion around  (\ref{oseledet-limit})].\\

We consider a general unit cell transfer matrix $\bm{M}$ constrained only by  (\ref{M-identity}), which implies $\bm\Sigma$-pseudo-unitarity and no further symmetries. Without loss of generality we can always write the $L$-unit-cells transfer matrix in the form
\begin{equation}
\bm{M}^L = \bm{P} e^{i \bm{D} L} \bm{P}^{-1} ,
\end{equation}
where the matrix $\bm{D}$ is diagonal and composed of $N$ complex conjugate pairs $\{\theta_i$, $\theta_i^*\}_{i=1}^{N}$. Of course, $e^{i \bm{D} L}$ is the diagonal matrix with $\bm M$ eigenvalues $\lambda_i = e^{i\theta_i L}$. The $2N_B$ propagating Bloch modes eigenvalues ($|\lambda_i|=1$) are given by real $\bm{D}$ elements. The matrix $\bm{M}$ is normal if and only if its eigenvectors matrix $\bm{P}$ is unitary. We start the perturbative expansion assuming that 
\begin{equation}
\bm{P} = \bm{P}_0 + \epsilon \bm{P}_1
\end{equation}
with $\bm{P}_0$ unitary and $\epsilon \ll 1$, i.e. $\bm M$ is close to be normal. We now proceed to expand 
\begin{equation}
\bm{M}^L\bm{M}^{L\dagger} = \bm{P}e^{i \bm{D} L}\bm{P}^{-1}\bm{P}^{-1\dagger}e^{-i \bm{D}^* L}\bm{P}^\dagger
\end{equation}
to order $\epsilon$. First, we take the inverse of $\bm{P}$,
\begin{equation}
\bm{P}^{-1} = ( \bm{P}_0 + \epsilon \bm{P}_1 )^{-1} = \bm{P}_0^\dagger - \epsilon \bm{P}_0^\dagger\bm{P}_1\bm{P}_0^\dagger + \mathcal{O}(\epsilon^2)
\end{equation}
Then,
\begin{equation}
\bm{P}^{-1} \bm{P}^{-1\dagger}  =  \bm{1} - \epsilon \left( \bm{P}_1^\dagger\bm{P}_0 + \bm{P}_0^\dagger\bm{P}_1  \right) + \mathcal{O}(\epsilon^2) 
\end{equation}
where we have used that $\bm{P}_0^\dagger\bm{P}_0 = \bm{1}$. Finally, we obtain
\begin{equation}\label{MM-epsilon}
\bm{M}^L\bm{M}^{L\dagger} = \bm{P}_0 \left[ \bm{H}_0 + \epsilon\bm{H}_1 \right]\bm{P}_0^\dagger + \mathcal{O}(\epsilon^2) 
\end{equation}
with $\bm{H}_0 = e^{i(\bm{D}-\bm{D}^*)L} = e^{ -2\, \mathfrak{Im}({\bm{D}})L}$ and 
\begin{equation}
\bm{H}_1 =\bm{H}_0\bm{P}_1^\dagger\bm{P}_0  +  \bm{P}_0^\dagger\bm{P}_1\bm{H}_0 
  -  e^{i\bm{D}L} \left( \bm{P}_1^\dagger\bm{P}_0 + \bm{P}_0^\dagger\bm{P}_1  \right) e^{-i\bm{D}^*L}  .
\end{equation}
Note that $ \bm{H}= \bm{H}_0 + \epsilon \bm{H}_1 $ is Hermitic so the eigenvalues of $\bm{M}^L\bm{M}^{L\dagger}$ are equal to the eigenvalues of $\bm{H}$. Hence, in order to obtain this spectrum  we can use the well known perturbation theory for time-independent Hamiltonians\cite{sakurai}. We want to calculate the spectrum given by the eigenvalue problem 
\begin{equation}
\bm{H}\bm{\phi} = \Lambda\bm{\phi} .
\end{equation}
As usual, we set $\Lambda = \Lambda^0 + \epsilon \Lambda^1$. Since our unperturbed Hamiltonian $\bm{H}_0$ is already diagonal, we have that $\Lambda^0_i = e^{-2\,\mathfrak{Im}(\theta_i)L}$, for $i=1,\ldots,2N$, with $\theta_{2i} = \theta_{2i-1}^*$. Note that the unperturbed spectrum $\{\Lambda^0_i, (\Lambda^0_i)^{-1}\}_{i=1}^N$ is degenerate with the $2N_B$ eigenvalues related to propagating Bloch modes equal to one.  Also, it is clear that the transmission eigenvalues $T_i$ associated to non-propagating modes decays exponentially so we are interested in calculating only the correction to the degenerate set of Bloch transmission eigenvalues.\\

The first order correction $\Lambda^1$ to the $2N_B$-degenerate eigenvalues $\Lambda^0_i = 1$ is given by the solutions of the eigenvalue problem
\begin{equation}\label{deg-evp}
\tilde{\bm{H}}_1^{(2 N_B)} \,\tilde{\bm\phi}_i = \Lambda^1_i \,\tilde{\bm\phi}_i
\end{equation}
with $\tilde{\bm{H}}_1^{(2 N_B)}$ the projection of $\bm{H}_1$ to the degenerate eigenspace. If we choose the $2N_B$ unit eigenvalues at $i=1,\ldots,2N_B$ then $\tilde{\bm{H}}_1^{(2 N_B)}$ is the upper left $2N_B \times 2N_B$ block of $\bm{H}_1$. The transmission eigenvalues associated to the degenerate Bloch subspace are given to dominant order in $\epsilon$ by
\begin{equation}
T_i = 1 - \frac{\epsilon^2}{4} (\Lambda_i^1)^2 + \mathcal{O}(\epsilon^3) .
\end{equation}

We can illustrate this result by means of the simple case $N_B=1$. When there is only one propagating Bloch mode, it is trivial to obtain the solutions to  (\ref{deg-evp}), which gives 
\begin{equation}
\Lambda_\pm =1 \pm 2 | \bm{v}_1^\dagger\bm{v}_2\, \sin{[(\theta_1 - \theta_2)L]} |,
\end{equation}
where $\bm{P} = [\bm{v}_1\; \bm{v}_2]$ with $\bm{v}_1^\dagger\bm{v}_2 \sim \mathcal{O}(\epsilon)$. In this case, the transmission eigenvalue is
\begin{equation}
T_1 = 1 -  | \bm{v}_1^\dagger\bm{v}_2\, \sin{[(\theta_1 - \theta_2)L]}|^2.
\end{equation}
This shows how the non-orthogonality of $\bm M$ eigenvectors breaks $\Lambda$ degeneracy creating a periodic dependance of $T_1$ on the chain length $L$.

\chapter{Analytical calculation of \texorpdfstring{$\langle N_B\rangle$}{<NB>}  for $N=1$}

In this appendix we calculate analitically the average number of propagating Bloch modes $\langle N_B\rangle$ for a COE and CUE periodic chain model with one open mode in the plane leads, this is $N=1$. In this case the scattering and transmission matrices are $2\times2$ allowing an explicit analitycal solution of $\langle N_B\rangle$. 

From definition \eref{M(S)}, of the transfer matrix $\bm M$ as a function of the scattering matrix $\bm S$, we have that for $N=1$,
\begin{equation}
   \bm{M} =\left(\begin{array}{cc}
   t - rr'/t' & r'/t'\\
   -r/t' & 1/t'
 \end{array}\right) ,
\end{equation}
where $t, r$ ($t', r'$) are the left-to-right (right-to-left) transmission and reflection coefficients, respectively. They are complex numbers and define the scattering matrix [see section \ref{section-s-matrix}],
\begin{equation}\label{S-matrix-dim1}
   \bm{S} =\left(\begin{array}{cc}
    r & t'\\
    t & r'
 \end{array}\right).
\end{equation}
As we have shown in section \ref{waveguide-bloch-spec}, the number of propagating Bloch modes $N_B$ in a periodic chain with unit cell $\bm S$-matrix given by \eref{S-matrix-dim1} is equal to the number of $\bm M$-matrix eigenvalues $\lambda_\pm$ with unit modulus. In the present case, we can solve the $\bm M$ eigenvalue problem algebraically to obtain
\begin{equation}\label{lambda-M-pm}
\lambda_\pm = \left( \frac{1-rr'+tt'}{2t'} \right) \pm \sqrt{-\frac{t}{t'} + \left( \frac{1-rr'+tt'}{2t'}\right)^2 }.
\end{equation}
Since there is only one open channel in the leads, the Bloch spectrum consists of a pair of states corresponding to modes traveling in both chain directions. As we will see explicitly later, $|\lambda_{-}|^{-1}=|\lambda_{+}|$. Therefore, we have that
\begin{equation}\label{nb-n1}
\langle N_B(N=1) \rangle = P[\lambda_\pm \in S^1] ,
\end{equation}
where $P[\lambda_\pm\in S^1]$ is the probability that $|\lambda_\pm| = 1$ under the appropriate RMT measure. We start with the CUE periodic chain ensemble. As discussed in section \ref{section-circular-ensembles}, the ensemble of scattering matrices $\bm S$ taken from the CUE ensemble are distributed according to the Haar measure of the unitary group ${\rm U}(2N)$. In general, any matrix of ${\rm U}(2N)$ can be represented using Euler angles as a composition of rotations in ${\rm U}(2)$ \cite{zyczkowski1994}. For $N=1$, a general $\bm S$ matrix is described by four independent parameters as
\begin{equation}\label{S-euler}
   \bm{S}(\xi, \phi_1, \phi_2, \phi_3) =\left(\begin{array}{cc}
    \sqrt{\xi}\,e^{i \phi_1} & -\sqrt{1-\xi}\,e^{i \phi_2} \\
    \sqrt{1-\xi}\,e^{- i \phi_2} & \sqrt{\xi}\,e^{- i \phi_1}
 \end{array}\right) e^{i\phi_3}
\end{equation} 
with $\xi\in[0,1]$ and $\phi_1, \phi_2, \phi_3\in[0,2\pi)$. The Haar measure in this parametrization is simply
\begin{equation}\label{measure-cue-dim1}
\mu_{U(2)}(d\bm S) = \frac{1}{(2\pi)^3}d\xi\, d\phi_1\, d\phi_2\, d\phi_3 .
\end{equation}
 Now, in order to compute $P[\lambda_\pm\in S^1]$, we rewrite \eref{lambda-M-pm} using \eref{S-euler} as
\begin{equation}
\lambda_{\pm} = e^{i(\pi/2 - \phi_2)} \left( \alpha \pm \sqrt{-1+\alpha^2} \right)
\end{equation}
with
\begin{equation}
\alpha = \frac{1-rr'+tt'}{2\sqrt{tt'}} = \frac{-\sin{\phi_3}}{\sqrt{1-\xi}}. 
\end{equation}
Thus, since $\alpha\in\mathbb{R}$, $|\lambda_\pm|=1$ if and only if $|\alpha| \leq 1$. Hence, we obtain
\begin{align}
P[\lambda_\pm\in S^1]  &=  \int_{|\sin{\phi_3}|^2\leq 1-\xi}  \mu_{U(2)}(d\bm S) \\
			& = \frac{1}{(2\pi)}\int_0^{2\pi}\int_0^{1-|\sin{\phi_3}|^2}\,d\xi\,d\phi_3 . 	\label{P-l-int}		
\end{align}
The integral in \eref{P-l-int} can be easily computed and by \eref{nb-n1} leads to
\begin{equation}
\langle N_B(N=1)\rangle_{\rm CUE} = \frac{1}{2},
\end{equation}
which is an exact result. This means that in a CUE periodic chain with one open channel the events of a Bloch state being propagative or evanescent are equiprobable (and complementary). We note that the asymptotic result obtained in section \ref{section-nb-rmt}, namely $\langle N_B(N)\rangle_{\rm CUE} = \sqrt{N/\pi}$ for $N\to\infty$, in this case gives $\sqrt{1/\pi}=0.56$.

We now turn to the COE periodic chain. As we have shown in section \ref{section-circular-ensembles}, an $\bm S$ matrix in this ensemble can be constructed by taking $\bm U\in {\rm U(2N)}$ distributed according to the Haar measure $\mu_{\rm U(2N)}$ and setting $\bm S =\bm{U}^T\bm{U}$. Since for $N=1$ we have an explicit parametrization for $\bm U$ given by \eref{S-euler}, we can write a general symmetric unitary matrix in COE as
\begin{equation}\label{S-euler-coe}
   \bm{S}(\xi, \phi_1, \phi_2, \phi_3) =\left(\begin{array}{cc}
    \xi\,e^{2i\phi_1} + (1-\xi)\,e^{-2i\phi_2}     &     -2i\sqrt{\xi(1-\xi)}\sin{(\phi_1+\phi_2)}\\
    -2i\sqrt{\xi(1-\xi)}\sin{(\phi_1+\phi_2)}        &       \xi\,e^{-2i\phi_1} + (1-\xi)\,e^{2i\phi_2} 
 \end{array}\right) e^{2i\phi_3} ,
\end{equation} 
where again $\xi\in[0,1]$ and $\phi_1, \phi_2, \phi_3\in[0,2\pi)$ are distributed according to the uniform measure \eref{measure-cue-dim1}. In this case, the eigenvalues of $\bm M$ are 
\begin{equation}
\lambda_\pm  = \alpha' \pm \sqrt{-1+\alpha'^2} 
\end{equation}
with 
\begin{equation}
\alpha' = \frac{\sin{(2\phi_3)}}{2\sqrt{\xi(1-\xi)} } \csc{(\phi_1 + \phi_2)}
\end{equation}
a real function of the parameters. Therefore, we have again that $|\lambda_\pm|=1$ if and only if $|\alpha'| \leq 1$, which implies
\begin{align}
P[\lambda_\pm\in S^1] &= \int_{|\sin{2 \phi_3}|^2\leq 2\sqrt{\xi(1-\xi)}|\sin{(\phi_1+\phi_2)}   }  \mu_{U(2)}(d\bm S) \\
		& = \frac{1}{2\pi^2}\int_0^1 \int_0^{4\pi} \arcsin{\left(2\sqrt{\xi(1-\xi)}\, |\sin{\theta}| \right)} \,d\theta\, d\xi .
\end{align}
This integral can be evaluated numerically to obtain
\begin{equation}
\langle N_B(N=1)\rangle_{\rm COE} = 0.3634.
\end{equation}
This result is quite close to the value given by the asymptotic expression [see section \ref{section-nb-rmt}] $\langle N_B(N)\rangle_{\rm COE} = \sqrt{N/\pi} - 0.2$ for $N\to\infty$, which in this case gives $\sqrt{1/\pi}-0.2 = 0.3642$. 

The present procedure can in principle be generalized to arbitrary $N$, however for larger dimensions the method becomes increasingly involved and impractical, since eigenvalue characteristic polynomials are not solvable analytically and, in addition, the direct evaluation of integral expressions is not feasible in high dimensional spaces.

\end{appendices}

\bibliographystyle{amsalpha}
\bibliography{thesis_bibliography}

\end{document}